\documentclass[10pt]{article}
\usepackage{geometry}
\usepackage{amsfonts}
\usepackage{amsmath}
\usepackage{slashed}
\usepackage{amssymb}
\usepackage{cancel}
\usepackage{mathrsfs}
\usepackage{wasysym}
\usepackage{cite}
\usepackage{bbm}
\usepackage{bm}
\usepackage{graphicx}
\usepackage[labelfont=small]{caption}
\usepackage{hyperref}
\usepackage[usenames,dvipsnames]{color}
\usepackage{wrapfig}
\usepackage{comment}
\usepackage{multicol}

\usepackage[LGR,T1]{fontenc}
\usepackage[utf8]{inputenc}   

\hypersetup{
colorlinks=true,         
linkcolor=Red,          
citecolor=blue,           
urlcolor=Cyan          
}
\title{\vspace{-1 cm} \textbf{T-Minkowski noncommutative spacetimes II:\\
classical field theory}}
\author{Flavio Mercati\footnote{flavio.mercati@gmail.com}
\vspace{12pt}
\\
Departamento de F\'isica, Universidad de Burgos, 09001 Burgos, Spain.
}

\newcommand{\sfrac}[2]{ {\textstyle \frac{#1}{#2}}}
\newcommand{\st}[1]{\text{\tiny \rm #1}}

\newcommand{\1}{\hat{1}}
\newcommand{\Q}{4}

\newcommand{\x}{\hat{x}}
\newcommand{\A}{\hat{a}}
\newcommand{\La}{\hat{\Lambda}}

\newcommand{\E}{\hat{E}}

\newcommand{\X}{\xi}

\newcommand{\mP}{\mathbbm{P}}
\newcommand{\mM}{\mathbbm{M}}

\usepackage{tocloft}

\begin{document}

\maketitle

\begin{abstract}
This paper is the second part of a series that develops the mathematical framework necessary for studying field theories on ``T-Minkowski'' noncommutative spacetimes. These spacetimes constitute a class of noncommutative geometries, introduced in Part I, that are each invariant under distinct quantum group deformations of the Poincaré group.  All these noncommutative geometries possess certain  physically desirable characteristics, which allow me to develop all the tools of  differential geometry and functional analysis, that are necessary in order to build consistent and T-Poincar\'e invariant noncommutative classical field theories.
\end{abstract}

\vspace{.5cm}

\tableofcontents

\newpage

\subsection*{Notations used in the paper}

\begin{itemize}
\item The Einstein summation convention is assumed. Greek indices $\alpha, \beta , \dots \mu,\nu,\dots$ will run from $0$ to $3$. Lowercase Latin indices from the middle of the alphabet, $i,j,k,l,\dots$ will run from $1$ to $3$. The ones from the beginning of the alphabet, $a,b,c,\dots$ will be natural numbers. Uppercase Latin indices $A,B,C, \dots, I,J,K,\dots$ will run from $0$ to $4$.

\item Symmetrization and antisymmetrization of indices will be done with a $1/2$ weight:
$$
T^{(\mu\nu)} = {\sfrac 1 2} (T^{\mu\nu} + T^{\nu\mu} ) \,,
\qquad
T^{[\mu\nu]} = {\sfrac 1 2} (T^{\mu\nu} - T^{\nu\mu} ) \,.
$$

\item A hat on a symbol, $\hat f$, indicates that $\hat f$ is an element of a not-necessarily-commutative algebra.

\item $\eta_{\mu\nu}$ will indicate a \textit{numerical,} constant symmetric metric. The Minkowski metric in Cartesian coordinates, \textit{i.e.} $\eta_{\mu\nu} = \text{diag}(-1,1,1,1)$ will be assumed unless specified otherwise. The signature is the mostly-positive one.

\item Complex conjugates will be indicated with a bar, $\overline{x+iy} = x-iy$, $x,y \in \mathbbm{R}$. The involution of our noncommutative algebras (an antilinear antihomomorphism) will be represented by a $(\,\cdot \,)^*$ symbol, to distinguish it from the Hermitian conjugate symbol $(\,\cdot \,)^\dagger$, which I prefer to reserve to operators on Hilbert space (the two symbols will be related to each other if a representation of the algebra is introduced).

\end{itemize}

\section{Introduction}

In Part I of this series~\cite{Mercati:2023apu}, I introduced the concept of T-Minkowski noncommutative spacetimes, a set of 17 classes of noncommutative geometries that are invariant under  quantum-group deformations of the Poincar\'e group, characterized by a series of physically and mathematically desirable properties.  The starting point is the algebra of continuous functions on Minkowski space, $C[\mathbbm{R}^{3,1}]$, equipped with the (commutative) pointwise product between functions, $f \dot g(x) = f(x) g(x) = g \dot f (x)$. This algebra can be promoted to a non Abelian algebra $C_\ell[\mathbbm{R}^{3,1}]$, where $\ell$ is a length scale characterizing the scale of fuzziness/noncommutativity, by relaxing the commutativity axiom of the product. This could be achieved in a number of possible ways, but an important assumption can make our life a lot easier: suppose that four coordinate functions $\x^\mu$, which reduce to ordinary Cartesian coordinates on Minkowski space in the commutative $\ell \to 0$ limit, form a basis for $C_\ell[\mathbbm{R}^{3,1}]$, and any element of the algebra can be obtained as a (limit) of \textit{ordered} polynomials in these coordinates. This statement, known as \textit{Poincar\'e--Birkhoff--Witt} property~\cite{wess1999qdeformed}, makes our algebras of functions much more amenable to be studied with techniques that resemble closely those of functional analysis and differential geometry that are familiar in the study of commutative spaces, keeping noncommutativity under control.
A crucial property, for the purposes of field theory, of the algebra $C[\mathbbm{R}^{3,1}]$, is that it transforms covariantly under Poincar\'e transformations, as a consequence of the fact that Minkowski space is a homogeneous space of the Poincar\'e group. To formulate algebraically this notion of covariance, we introduce the algebra of functions on the Poincar\'e group $C[ISO(3,1)]$,  so that a Poincar\'e-transformed function on Minkowski space can be represented as a function on the Cartesian product of Minkowski space and the Poincar\'e group, $C[ISO(3,1)\times \mathbbm{R}^{3,1}]$. This, under the canonical isomorphism, is an element of $C[ISO(3,1)]\otimes C[\mathbbm{R}^{3,1}]$. In terms of algebras of functions, therefore, Poincar\'e transformations are formulated as a map from $C[\mathbbm{R}^{3,1}]$ to $C[ISO(3,1)]\otimes C[\mathbbm{R}^{3,1}]$. This carries over to the noncommutative case: a \textit{coaction}   $C_\ell[\mathbbm{R}^{3,1}] \to C_\ell[ISO(3,1)]\otimes C_\ell[\mathbbm{R}^{3,1}]$ defines the transformation rule of functions on the noncommutative Minkowski space. In the noncommutative case, however, $C_\ell[\mathbbm{R}^{3,1}]$ has additional structure: namely, the noncommutative product. The statement that the space is covariant under Poincar\'e transformations, then, constrains the structure of the quantum group $ C_\ell[\mathbbm{R}^{3,1}]$, because we do not want that the two sides of some identities implied by the commutation relations end up transforming differently under Poincar\'e transformations. This is codified algebraically by the requirement that the coaction is a homomorphism for the product of $C_\ell[\mathbbm{R}^{3,1}]$. Physically, I interpret this as stating that different observers, in different reference frames, will observe the same commutation relations between coordinates (whatever the physical implications of such commutation relations may be). Then the noncommutativity scale, $\ell$, which is supposed to measure the distance scales at which noncommutative fuzziness becomes visible, is a relativistic invariant, like the speed of light. A theory based on the construction that I outlined would possess a notion of minimal distance, without breaking Lorentz invariance: a \textit{doubly special relativity} theory~\cite{AmelinoCamelia:2000mn}.

The possible quantum-group deformations of the Poincar\'e group are limited in number. A way to ensure that the algebra of functions on the group, $C_\ell[ISO(3,1)]$ possesses the Poincar\'e--Birkhoff--Witt property is to demand that the commutation relations between the coordinates on the group can be expressed in terms of \textit{RTT} relations:
\begin{equation}
   R^{AB}_{EF} ~ \hat T^E{}_C \, \hat T^F{}_D  ~=~  \hat T^B{}_G \, \hat T^A{}_H  ~  R^{HG}_{CD} \,,     \label{RTT_relations}
\end{equation}
where $\hat{T}^A{}_B$ are the matrix elements of the group elements in a finite-dimensional representation of the group (in the case of the Poincar\'e group, this is five-dimensional). Unless the group is $GL(N)$, such matrix elements are redundant coordinates on the group, and they need to be supplemented with algebraic relations, like \textit{e.g.} orthogonality.
The numerical coefficients $ R^{AB}_{EF}$ form the so-called \textit{R matrix,} which satisfies the \textit{Quantum Yang--Baxter Equation} (QYBE):
\begin{equation}\label{QYBE}
    R^{AB}_{IJ} \, 
    R^{IC}_{LK} \, 
    R^{JK}_{MN} 
    =
    R^{BC}_{JK} \, 
    R^{AK}_{IN} \, 
    R^{IJ}_{LM} \,,
\end{equation}
which essentially ensures the associativity of the noncommutative product between functions.
The Lie group structure is encoded, in terms of the algebra of functions on the group, into a \textit{coproduct,} an \textit{antipode} and a \textit{counit} map, respectively:
\begin{equation}
\label{T_matrix_coproduct_counits_antipodes}
    \Delta( \hat T^A{}_B  ) = \hat T^A{}_C  \otimes \hat T^C{}_B \,,  \qquad S(\hat T^A{}_B) = (\hat T^{-1})^A{}_B  \qquad \epsilon(\hat T^A{}_B) = \delta^A{}_B  \,, \,.
\end{equation}
Given a function $f$ on the group, its coproduct $\Delta(f)$ is a function of two group elements $g$ and $h$, whose value is $f$ calculated on $g\,h$. The antipode of $f$ is a function whose value on $g$ is $f(g^{-1})$, and the counit of $f$ gives the value of $f$ at the identity:
\begin{equation}
f(g \, h) = \Delta(f)(g,h) \,, \qquad f(g^{-1}) = S(f)[g] \,, \qquad f(e) = \epsilon(f) \,.
\end{equation}
Eq.~\eqref{T_matrix_coproduct_counits_antipodes} shows how the three maps act on the coordinate functions $\hat{T}^A{}_B$.

The algebraic relations satisfied by the matrix $\hat{T}^A{}_B$ in the case of the Poincar\'e group are
\begin{equation} \label{T_matrix_decomposition}
\hat T^{\mu}{}_\nu = \La^\mu{}_\nu \,, \qquad \hat T^\mu{}_\Q = \A^\mu \,,
    \qquad
\hat T^\Q{}_\mu = 0 \,,\qquad \hat T^\Q{}_\Q = \1 \,,
\end{equation}
where the $\La^\mu{}_\nu$ matrices satisfy the additional constraint of being Lorentz matrices, $SO(3,1)$:
\begin{equation}
\label{MetricInvariance}
    \eta_{\mu\nu}  \,   \La^\mu{}_\rho \, \La^\nu{}_\sigma 
= \eta_{\rho\sigma} 
\,, \qquad
\eta^{\rho\sigma}  \,   \La^\mu{}_\rho \, \La^\nu{}_\sigma 
= \eta^{\mu\nu} \,,
\end{equation}
where $\eta^{\mu\nu}$ is any (commutative) real nondegenerate symmetric matrix with signature $(-,+,+,+)$. For the present paper I will assume the form $\eta_{\mu\nu} = \text{diag}(-1,1,1,1)$ for the metric.

The noncommutative space $C_\ell[\mathbbm{R}^{3,1}]$, invariant under the quantum group described above, is defined by a product that is expressed in terms of \textit{RXX} relations: 
\begin{equation}
   \x^A \, \x^B  =   \,  R^{BA}_{CD}  \,  \x^C \, \x^D  \,, \qquad \x^A = (\x^\mu ,\1)\,, \label{RXX_relations}
\end{equation}
notice the algebraic relation $\x^4 = \1$ satisfied by the quintuplet $\x^A$.
The invariance is expressed by a coaction map $( \, \cdot \, )' : C_\ell[\mathbbm{R}^{3,1}] \to  C_\ell[ISO(3,1)] \otimes C_\ell[\mathbbm{R}^{3,1}] $, which essentially encodes the usual row-by-column product between a matrix and a vector:
\begin{equation}\label{Coaction_5D_coordinates}
\x'^A = \hat T^A{}_B \otimes \x^B \,,  
\end{equation}
in terms of the four independent Cartesian coordinates, the coaction reproduces the usual Poincar\'e transformation of Minkowski coordinates:
\begin{equation}\label{Coaction_spacetime_coordinates}
 \x'^\mu = \La^\mu{}_\nu \otimes \x^\nu + \A^\mu \otimes \1 \,.
\end{equation}
This coaction leaves the relations~\eqref{RXX_relations} invariant, in the sense that
\begin{equation}\label{PoincareInvariance_RXX}
\x'^A \, \x'^B  =   \,  R^{BA}_{CD}  \,  \x'^C \, \x'^D \,. 
\end{equation}
The fact that the coordinates of the noncommutative space and those of the quantum group commute, as implied by the tensor product structure in~\eqref{Coaction_spacetime_coordinates}, seems like a physically reasonable assumption, as one can independently specify a state on the algebra  $C_\ell[\mathbbm{R}^{3,1}]$ (the spacetime), and one on $C_\ell[ISO(3,1)]$ (the Poincare\'e group), which together give a state on $C_\ell[ISO(3,1)] \otimes C_\ell[\mathbbm{R}^{3,1}]$. The physical meaning of this would be that one is free to independently specify ``points'' (or, more precisely, localizations) in spacetime in one reference frame, and the relations between that frame and a second one. The transformed coordinates~\eqref{PoincareInvariance_RXX} close an algebra that is isomorphic to $C_\ell[\mathbbm{R}^{3,1}]$, and, therefore, states on $C_\ell[ISO(3,1)] \otimes C_\ell[\mathbbm{R}^{3,1}]$ admit the interpretation of states on the algebra of coordinates on the noncommutative Minkowski space, seen from a second reference frame.

The existence of an R-matrix allows also one to define a 4-dimensional bicovariant differential calculus in the sense of Woronowicz~\cite{Woronowicz:1989}, as\footnote{If the R-matrix is triangular,   $R^{EF}_{AB} R^{BA}_{CD} = \delta^F{}_C \, \delta^E{}_D$, Eqs.~\eqref{RXdX_relations} can be equivalently written as $   d \x^A \, \x^B   =   R^{BA}_{CD}   \, d \x^C \, \x^D   $.}
\begin{equation}\label{RXdX_relations}
  \x^A \, d\x^B   =   R^{BA}_{CD} \,  \x^C \, d\x^D        \,,  \qquad  d\x^A = (d \x^\mu ,0) \,,
\end{equation}
which is covariant under the simultaneous coaction~\eqref{Coaction_spacetime_coordinates} and $d\x'^A = \hat T^A{}_B \otimes d \x^B$. I will call $\Gamma_\ell^1$ the bimodule generated by $d \x^\mu$ and refer to~\cite{Woronowicz:1989} for the theorem stating that the most general element of $\Gamma_\ell^1$ takes the form
\begin{equation}
\omega = \omega_\mu \, d \x^\mu \,, \qquad  \omega_\mu \in C_\ell[\mathbbm{R}^{3,1}] \,.
\end{equation}
Finally, the R-matrix also allows one to introduce higher-dimensional covariant comodules with a braiding construction:
\begin{equation}
\x_a^A \,\x_b^B  =   R^{BA}_{CD} \,\x_b^C \, \x_a^D   \,,\qquad \x_a^A = (\x_a^\mu ,\1) \,, \qquad a = 1,\dots, N, \label{RXY_relations}
\end{equation}
which are covariant under the coaction $\x'^A_a = \hat T^A{}_B \otimes \x^B_a $. These enable the description of the \textit{braided tensor product} algebra $C_\ell[\mathbbm{R}^{3,1}]^{\otimes_\ell N}$, generalizing to the noncommutative case the algebra $C[\mathbbm{R}^{3,1}]^{\otimes N}$ of functions of $N$ different points, which is a crucial ingredient of Quantum Field Theory (QFT), the physical content of the theory being encoded in the form of the $N$-point functions.

The frameword I described involves only commutation relations that are quadratic in the (group or spacetime) coordinates [Eqs.~\eqref{RTT_relations} and~\eqref{RXX_relations}], however the algebraic relations satisfied by the $\hat{T}^A{}_B$ matrix and the coordinates $\hat{x}^A$, in particular~\eqref{T_matrix_decomposition}, set some components equal to the identity, and this introduces in the commutation relations terms that are proportional to the identity, and terms that are linear in the coordinates. Their general form will be:
\begin{equation}\label{CommutatorPlanckLenghExpansion}
\begin{aligned}   
[\x^\mu , \x^\nu ] =&  i \, \ell^2  \, \theta^{\mu\nu} \, \1 + i \, \ell \, c^{\mu\nu}{}_\rho \,  \x^\rho + i \, d^{\mu\nu}{}_{\rho\sigma} \, \x^\rho \, \x^\sigma   \,,
\\
[\A^\mu , \A^\nu] =& i \, \ell^2  \left( \theta_1^{\mu\nu} \, \1  +  (\theta_2)^{\mu\nu\alpha}_{\beta} \, \La^\beta{}_\alpha   +  (\theta_3)^{\mu\nu\alpha\gamma}_{\beta\delta} \, \La^\beta{}_\alpha\, \La^\delta{}_\gamma \right) 
\\
& 
+ i \, \ell \, \, \A^\rho  \left(  c_1^{\mu\nu}{}_\rho + (c_2)^{\mu\nu\alpha}_{\rho\gamma} \,  \La^\gamma{}_\alpha \right)+ i \, (d_1)^{\mu\nu}{}_{\rho\sigma} \, \A^\rho \, \A^\sigma \,,
\\
[\A^\mu , \La^\nu{}_\rho] =&  i \, \ell \left(
 (\theta_4)^{\mu\nu}_\rho  \, \1 +(\theta_5)^{\mu\nu\alpha}_{\rho\beta} \, \La^\beta{}_\alpha 
+   (\theta_6)^{\mu\nu\alpha\gamma}_{\rho\beta \delta} \, \La^\beta{}_\alpha  \, \La^\delta{}_\gamma
\right) +
\\
& i \, \A^\sigma \left( (c_3)^{\mu\nu}{}_{\rho\sigma} + (c_4)^{\mu\nu\alpha}_{\rho\sigma\gamma} \, \La^\gamma{}_\alpha \right)
+ {\sfrac i \ell} \, (d_2)^{\mu\nu}{}_{\rho\sigma\lambda} \, \A^\sigma \, \A^\lambda  \,,
\\
[\La^\mu{}_\rho, \La^\nu{}_\sigma] =&   
i \, \left(
 (\theta_7)^{\mu\nu}_{\rho\sigma} \, \1 +(\theta_8)^{\mu\nu\alpha}_{\rho\beta\sigma} \, \La^\beta{}_\alpha 
+   (\theta_9)^{\mu\nu\alpha\gamma}_{\rho\beta \delta\sigma} \, \La^\beta{}_\alpha  \, \La^\delta{}_\gamma
\right) +
\\
& {\sfrac i \ell} \, \A^\lambda \left( (c_5)^{\mu\nu}{}_{\rho\sigma\lambda} + (c_6)^{\mu\nu\alpha}_{\rho\sigma\gamma\lambda} \, \La^\gamma{}_\alpha \right)
+ {\sfrac i {\ell^2}} \, (d_3)^{\mu\nu}{}_{\rho\sigma\lambda\kappa} \, \A^\lambda \, \A^\kappa\,,
\,,
\end{aligned}
\end{equation} 
where I expressed all the components of the R-matrix (except the ones proportional to the identity) in terms  of dimensionless coefficients, multiplied by appropriate powers of the noncommutativity scale $\ell$ that gives them the correct dimensions. A reasonable physical request on the coefficient above is that the expressions on the right-hand side admit a regular limit as $\ell \to 0$, which corresponds to states on the algebra in which the coordinates $\x^\mu$ and the translation parameters $\A^\mu$ are macroscopic. The terms involving negative powers of $\ell$ are excluded by this assumption. The presence of these terms would be hard to reconcile with the observation of a commutative spacetime at large scales, and commutative relations between inertial reference frames, because the larger the distances involved, the larger the contributions to the uncertainty relations of the terms of this type.
In Part I of this series,~\cite{Mercati:2023apu}, I presented further physical arguments that constrain the commutation relations above.
First, all terms that do not depend on $\ell$ have no dependence on scale, and therefore the uncertainty relations that they imply will appear the same at all scales. Compatibility with the large-scale commutative limit then imposes that those coefficients are extremely small dimensionless numbers, and they can be ignored in first approximation, because then at small scales, whenever the noncommutative effects will be measurable, the other terms with a dependence on $\ell$ will completely dominate over them. This is true, of course, unless all coefficient that depend on $\ell$ are zero. Then we have a class of models with quadratic noncommutativity of the coordinates, and noncommutative Lorentz matrices. These have been studied recently in~\cite{Meier:2023kzt,Meier:2023lku}, and will not be the concern of this series of articles. Here, therefore, I will be assuming that the commutation relations of the coordinates and of the translation parameters are at most linear in  $\x^\mu$  and, respectively, in $\A^\mu$. Moreover the commutation relations of the translations with the Lorentz matrices will only depend on $\La^\mu{}_\nu$, and, finally, the Lorentz matrices will close a commutative subalgebra.
The last assumption I made in Part I~\cite{Mercati:2023apu}, is that the coefficient $c_2$ in $[\A^\mu , \A^\nu]$ vanishes. This was motivated by the structure of the so-called \textit{lightlike $\kappa$-Minkowski}~\cite{Ballesteros_1995,BALLESTEROS_1996,Ballesteros:1996awf},  \textit{$\rho$-Minkowski}  and the simpler $\theta$-Minkowski spacetimes, which our class of theories is modeled after. In these three models, the coefficients  $(c_2)^{\mu\nu\alpha}_{\rho\sigma\gamma}$  are zero, and the commutator $[\A^\mu , \La^\nu{}_\rho]$ depends only on $\La^\mu{}_\nu$. These two properties allow us to interpret the operators $\A^\mu$ as vector fields on the Lorentz group (without the $c_2$ term, the commutator $[\A^\mu , \A^\nu]$ as the Lie brackets between the four vector fields $\A^\mu$). In Part I, I called the quantum groups defined by these assumptions T-Poincar\'e models, and their quantum homogeneous spaces T-Minkowski. The T in the name refers to the special role of the translation parameters, which are the only noncommutative part of the group, and can be represented as vector fields on the Lorentz group. As shown in~\cite{Mercati:2023apu}, once the assumptions I listed are imposed, one is left with an R-matrix with 64 free parameters, and the QYBE should be imposed on those. Instead, it is convenient to impose the necessary condition that all the commutators found thus far satisfy the Jacobi identities, and the following consistency conditions:
\begin{itemize}
\item The invariance of the metric~\eqref{MetricInvariance}, \textit{i.e.} the compatibility of the commutators with $\La^\mu{}_\nu$ being a Lorentz matrix.

\item The existence of a quantum Homogeneous space, whose commutation relations are expressed in terms of RXX relations~\eqref{RXX_relations}.

\item The existence of a differential calculus and of a braiding, expressed in terms of RXY relations~\eqref{RXX_relations}. These two conditions are equivalent.
\end{itemize}
These conditions, together, are sufficient to satisfy the QYBE, and to fix the R-matrix to be \textit{triangular}~\cite{majid_1995}. Whether they are necessary or not has not been checked, so one could have other solutions of the QYBE that do not satisfy those requirements. Even if they existed, however, those solutions would be  less physically interesting, because they would not meet the minimal physical requirement that I consider necessary in order to develop consistent quantum field theories. The final result takes the following form: 
\begin{equation}\label{T-PoincareGroup_commutators}
\left\{\begin{aligned}
  &[ \A^\mu , \A^\nu ]  =  i \, \theta^{\rho\sigma} \,  \left( \delta^{[\mu}{}_\rho \, \delta^{\nu]}{}_\sigma - \La^{[\mu}{}_\rho \, \La^{\nu]}{}_\sigma    \right)  
  + i \, \left( f^{\nu\mu}{}_\rho - f^{\mu\nu}{}_\rho \right) \, \A^\rho  \,,
\\
&[ \La^{\mu}{}_\rho  , \A^{\nu} ]   =
  i \,  f^{\alpha\beta}{}_\gamma
\left(  \La^{\mu}{}_\alpha \, \La^{\nu}{}_\beta \, \delta^\gamma{}_\rho  - \delta^{\mu}{}_\alpha \, \delta^{\nu}{}_\beta \, \La^\gamma{}_\rho \right) 
 \,, 
\\
&[ \La^{\mu}{}_\rho ,  \La^{\nu}{}_\sigma ] = 0   \,,
 \end{aligned}\right.
\end{equation}
where the coefficients $ f^{\alpha\beta}{}_\gamma$ are real ``structure constants'', not necessarily antisymmetric in $\alpha$ and $\beta$. $\theta^{\alpha\beta}$ is, in turn, antisymmetric in $\alpha$ and $\beta$, and  it is real as well. Together they satisfy the following constraints
\begin{equation}  \label{StructureConstants_constraints}
     \begin{gathered}
      c^{\mu\nu}{}_\lambda \, c^{\lambda\rho}{}_\sigma - 
        \, c^{\mu\rho}{}_\lambda c^{\lambda\nu}{}_\sigma 
        - c^{\nu\rho}{}_\lambda \, c^{\mu\lambda}{}_\sigma  = 0\,,
        \qquad 
 c^{\mu\nu}{}_\lambda \, \theta^{\rho \lambda}   + c^{\rho\mu}{}_\lambda \, \theta^{\nu \lambda} + c^{\nu\rho}{}_\lambda \, \theta^{\mu \lambda} =  0\,,
\\
    f^{\alpha\mu}{}_\lambda  \, f^{\lambda\nu}{}_\beta -      f^{\alpha\nu}{}_\lambda  \, f^{\lambda\mu}{}_\beta 
    = - c^{\mu\nu}{}_\rho \, f^{\alpha\rho}{}_\beta \,,
    \qquad  
    \eta^{\theta (\sigma}  \,  f^{\varphi ) \lambda }{}_\theta =  0 \,,
     \end{gathered}
\end{equation}
where $c^{\alpha\beta}{}_\gamma$ is (minus) twice the antisymmetric part of $f^{\alpha\beta}{}_\gamma$
\begin{equation}
c^{\alpha\beta}{}_\gamma  = f^{\beta\alpha}{}_\gamma - f^{\alpha\beta}{}_\gamma \,,
\qquad \Rightarrow \qquad
 f^{[\alpha\beta]}{}_\gamma = - 
    {\sfrac 1 2 } c^{\alpha\beta}{}_\gamma \,.
\end{equation}
The constraints~\eqref{StructureConstants_constraints} can be solved as follows: consider the four matrices
\begin{equation}\label{K_matrices_definition}
    (K^\mu)^\alpha{}_\beta = f^{\alpha\mu}{}_\beta \,,
\end{equation}
Eqs.~\eqref{StructureConstants_constraints} imply that $K^\mu$ are  (possibly degenerate) linear combinations of four-dimensional Lorentz algebra matrices,
\begin{equation}\label{K_as_M}
    K^\mu =  {\sfrac 1 2}   \, \omega^\mu{}_{\rho\sigma} \, M^{\rho\sigma} \,, \qquad  (M^{\mu\nu})^\alpha{}_\beta = \eta^{\mu \alpha} \, \delta^\nu{}_\beta - \eta^{\nu \alpha} \, \delta^\mu{}_\beta \,, \qquad  \omega^\mu{}_{\rho\sigma} = - \omega^\mu{}_{\sigma\rho} \in \mathbbm{R} \,,
\end{equation}
which close a Lie algebra with structure constants $-c^{\mu\nu}{}_\rho$:
\begin{equation}\label{K_commutation_relations}
    [K^\mu , K^\nu] = - c^{\mu\nu}{}_\rho \, K^\rho \,,
\end{equation}
then, if $c^{\mu\nu}{}_\rho$ satisfy the Jacobi identities, all of the constraints above are satisfied, except the one involving $\theta^{\mu\nu}$, which needs to be solved after we have all of the components of $f^{\mu\nu}{}_\rho$. The equations for $\theta^{\mu\nu}$ are the conditions for the existence of a central extension of the Lie algebra with structure constants $c^{\mu\nu}{}_\rho$.
The relation between the Lorentz algebra coefficients $\omega^\mu{}_{\rho\sigma}$ and the structure constants can be written
\begin{equation}
\omega^\mu{}_{\sigma\rho}  = \eta_{\nu\sigma}  \, f^{\nu\mu}{}_\rho  \qquad \Rightarrow \qquad
f^{\mu\nu}{}_\rho =\eta^{\mu\sigma}  \, \omega^\nu{}_{\sigma\rho}  \,.
\end{equation}
The noncommutative spacetime that is covariant under the coaction~\eqref{Coaction_spacetime_coordinates} of a T-Poincar\'e quantum group with commutation relations~\eqref{T-PoincareGroup_commutators}, satisfies the following commutation relations:
\begin{equation}
 \label{x_comm_rel}
  [\x^{\mu} , \x^{\nu}] 
    = i \, \theta^{\rho\sigma}   \, \1 
  + i \, \left( f^{\nu\mu}{}_\rho - f^{\mu\nu}{}_\rho \right) \, \x^\rho    \,,
\end{equation}
which are completely fixed once one solves the constraints~\eqref{StructureConstants_constraints}.

The 5-dimensional R-matrix one finds after solving all the constraints takes the following form:
\begin{equation}\label{R-matrix_r-matrix}
 R = 1 \otimes 1 + i \, r \,, \qquad
r =   - {\sfrac 1 2} \, \theta^{\mu\nu} \,  \mP_\mu \wedge \mP_\nu - {\sfrac 1 2}   \,   \omega^{\mu\sigma\rho} \,  \mP_\mu \wedge \mM_{\sigma\rho}  \,,
\end{equation}
where 
\begin{equation}\label{Eq:5DrepPoincare}
 (\mP_\mu)^A{}_B = \delta^{A}{}_\mu \, \delta^\Q{}_{B}\,, \qquad
(\mM_{\mu\nu})^A{}_B  = \delta^A{}_\mu\, \eta_{\nu\beta} \, \delta^\beta{}_B - \delta^A{}_\nu \, \eta_{\mu\beta} \, \delta^\beta_B \,, 
\end{equation}
are 5-dimensional representations of the Poincar\'e algebra generators, and $r$ is a \textit{triangular classical r-matrix,}\footnote{Notice that
this makes the coordinate algebras~\eqref{RXX_relations} and~\eqref{RXY_relations} into $H$-module algebras, with $H$ a triangular Hopf algebra. These structures are the basis of the approach developed in~\cite{Aschieri:2020ifa}, in turn inspired by the older~\cite{Aschieri:2005zs}. It would be interesting to explore the formulation of T-Minkowski models in this language, in particular in connection to their Riemannian geometry.} satisfying the \textit{Classical Yang--Baxter equation}~\cite{Chari}:
\begin{equation}
[ r_{12} , r_{13}]  + [r_{12} , r_{23} ] + [r_{13} , r_{23}] =  0  \,.
\end{equation}
The solutions to the above equation (in the case of the Poincar\'e algebra) have been classified by Zakrzewski~\cite{Zakrzewski_1997}, and leads to 17 classes (Cases \textbf{7}, \textbf{8}, \textbf{9}$^{\bm{(-1)}}$, \textbf{9}$^{\bm{(0)}}$,\textbf{9}$^{\bm{(+1)}}$, \textbf{10}--\textbf{21} in~\cite{Zakrzewski_1997}) of inequivalent models which I will list here, using the original numbering of Zakrzewski's classification,  lumping together some models that are variation of the same theme.

\subsection{List of the T-Poincar\'e models}

for each model, I will list the r-matrix, the corresponding expressions of the $\theta^{\mu\nu}$ and $\omega^{\mu\nu\rho}$ coefficients, the structure constants $f^{\mu\nu}{}_\rho  = \omega^{\nu\mu\sigma} \, \eta_{\sigma\rho} $, and the commutation relations of the homogeneous space $[\x^\mu, \x^\nu] = i \, \theta^{\mu\nu} + i \, c^{\mu\nu}{}_\rho \, \x^\rho$ where $ c^{\mu\nu}{}_\rho  = 2 \, f^{[\nu\mu]}{}_\rho$.

\begin{itemize}

\item \textbf{Case 7} 
When $\zeta^{(7)} =0$, this reduces to the $\kappa$-lightlike model, discovered in~\cite{Ballesteros_1995,BALLESTEROS_1996,Ballesteros:1996awf} and recently developed up to the point of free quantum scalar fields in~\cite{Lizzi:2021rlb,DiLuca:2022idu,Fabiano:2023xke}.
\begin{equation}
\begin{aligned}
    r_7 =&   \eta^{\mu\nu} \, \mP_\mu \wedge \mM_{\nu 0} + \eta^{\mu\nu} \, \mP_\mu \wedge  \mM_{\nu 3} + \zeta^{(7)} \, (\mP_0 + \mP_3) \wedge \mM_{12}  \,,
\\
\theta^{\mu \nu} =& 0\,,
\\
\omega^{\mu \nu \rho } =&\zeta ^{(7)} \left(\delta ^{\mu }{}_0+\delta ^{\mu }{}_3\right) \left(\delta ^{\rho }{}_1 \delta ^{\nu }{}_2 -\delta ^{\rho }{}_2\delta ^{\nu }{}_1 \right) +\left(\delta ^{\nu }{}_0+\delta ^{\nu }{}_3\right) \eta^{\mu \rho }-\left(\delta ^{\rho }{}_0+\delta ^{\rho }{}_3\right) \eta^{\mu \nu }\,,
\\
f^{\mu \nu }{}_{\rho } =&\zeta ^{(7)} \left(\delta ^{\nu }{}_0+\delta ^{\nu }{}_3\right)  \left(\eta _{\rho 1} \delta ^{\mu }{}_2-\eta _{\rho 2} \delta ^{\mu }{}_1\right)+
\left(\delta ^{\mu }{}_0+\delta ^{\mu }{}_3\right) \delta^\nu{}_\rho
-\left(\eta_{\rho 0}+\eta_{\rho 3} \right) \eta^{\mu\nu }\,,
\end{aligned}
\end{equation}
\begin{equation}
   \begin{aligned}
    & [\x^0,\x^1]=-i \left(\zeta ^{(7)} \, \x^2 +\x^1\right)\,,& &[\x^0,\x^2]=i \left(\zeta ^{(7)} \, \x^1 -\x^2\right)\,,& &[\x^1,\x^2]=0\,,
 \\ 
 & [\x^0,\x^3]=i \left(\x^0-\x^3\right)\,,& &[\x^1,\x^3]=i \left(\zeta ^{(7)} \, \x^2 +\x^1\right)\,,& &[\x^2,\x^3]=i \left(\x^2-\zeta ^{(7)} \, \x^1 \right)\,.
   \end{aligned}
\end{equation}

\item \textbf{Case 8}
\begin{equation}
\begin{aligned}
    r_8 =&  \eta^{\mu\nu} \, \mP_\mu \wedge \mM_{\nu 0} + \eta^{\mu\nu} \,\mP_\mu \wedge \mM_{\nu 3} + \zeta^{(8)}\, (\mP_0 + \mP_3) \wedge (\mM_{01}-\mM_{13}) \,,
\\
\theta^{\mu \nu} =& 0\,,
\\
\omega^{\mu \nu \rho } =&2 \, \zeta ^{(8)} \,\left(\delta ^{\mu }{}_0+\delta ^{\mu }{}_3\right)  \delta ^{[\nu }{}_1 \left(\delta ^{\rho]}{}_0 +\delta ^{\rho]}{}_3\right)  +\left(\delta ^{\nu }{}_0+\delta ^{\nu }{}_3\right) \eta^{\mu \rho }-\left(\delta ^{\rho }{}_0+\delta ^{\rho }{}_3\right) \eta^{\mu \nu }\,,
\\
f^{\mu \nu }{}_{\rho } =&2 \,\zeta ^{(8)} \, \left(\delta ^{\nu }{}_0+\delta ^{\nu }{}_3\right)  \left( 
\delta ^{\mu }{}_{[1} \eta _{0] \rho} + \delta ^{\mu }{}_{[1} \eta _{3]\rho} \right)+\left(\delta ^{\mu }{}_0+\delta ^{\mu }{}_3\right) \delta^\nu{}_\rho
-\left(\eta _{\rho 0}+\eta _{\rho 3}\right)  \eta^{\mu\nu} \,,
\end{aligned}
\end{equation}
\begin{equation}
   \begin{aligned}
& [\x^0,\x^1]=i \left(\zeta ^{(8)} \, \left(\x^3-\x^0\right) - \x^1\right)\,,& &[\x^0,\x^2]=-i \,  \x^2\,,& &[\x^1,\x^2]=0\,,
 \\ 
 & [\x^0,\x^3]=i \left(\x^0-\x^3\right)\,,& &[\x^1,\x^3]=i \left(\zeta ^{(8)} \,\left(\x^0-\x^3\right) +\x^1\right)\,,& &[\x^2,\x^3]=i \, \x^2\,.
   \end{aligned}
\end{equation}
\\
~

\item \textbf{Cases 9$^{\bm{(s)}}$},\footnote{In Table 1 of Ref.~\cite{Zakrzewski_1997}, the r-matrices are wrong unless $s \neq 0$, as they do not satisfy the CYBE.  This mistake was corrected in~\cite{tolstoy2007twisted} (pag.~13). I reproduce here the correct r-matrices from~\cite{tolstoy2007twisted}.} $s= (-1,0,1)$ 
\begin{equation}
\begin{aligned}
    r_{9,s} =~& \theta^{(9)}  \, (\mP_0 + \mP_3) \wedge \mP_2 + [ \mP_1 + s \, (\mP_0 + \mP_3) ] \wedge [\zeta^{(9)}_1( \mM_{01}-\mM_{13} ) +  \zeta^{(9)}_2 (\mM_{23} - \mM_{02}) ]
    + \\
    &\zeta^{(9)}_1 \, (\mP_0 + \mP_3) \wedge  \mM_{03} \,,
\\
\theta^{\mu \nu} =& \theta^{(9)} \left(\delta ^{\mu }{}_2 \left(\delta ^{\nu }{}_0+\delta ^{\nu }{}_3\right)-\left(\delta ^{\mu }{}_0+\delta ^{\mu }{}_3\right) \delta ^{\nu }{}_2\right) \,,
\\
\omega^{\mu \nu \rho } =&\zeta _1^{(9)} \left(s (\delta ^{\mu }{}_0+ \delta ^{\mu }{}_3)+\delta ^{\mu }{}_1\right) 
\left[  
 \left(\delta ^{\rho }{}_0+\delta ^{\rho }{}_3 \right) \delta ^{\nu }{}_1
-\left(\delta ^{\nu }{}_0+\delta ^{\nu }{}_3 \right) \delta ^{\rho }{}_1  
\right]+
\\
&\zeta _2^{(9)} \left(s (\delta ^{\mu }{}_0+ \delta ^{\mu }{}_3)+\delta ^{\mu }{}_1\right) \left[
\left(\delta ^{\nu }{}_0+\delta ^{\nu }{}_3 \right) \delta ^{\rho }{}_2 
-\left(\delta ^{\rho }{}_0 
+\delta ^{\rho }{}_3   \right)\delta ^{\nu }{}_2 
\right]+
 \,,
\\
&\zeta _1^{(9)} \left(\delta ^{\mu }{}_0+\delta ^{\mu }{}_3\right) \left(\delta ^{\nu }{}_3 \delta ^{\rho }{}_0-\delta ^{\nu }{}_0 \delta ^{\rho }{}_3\right) \,,
\\
f^{\mu \nu }{}_{\rho } =&
\zeta _1^{(9)} \left(s \left(\delta ^{\nu }{}_0+\delta ^{\nu }{}_3\right)+\delta ^{\nu }{}_1\right) 
\left[
\left(\eta _{\rho 0}+\eta _{\rho 3}\right) \delta ^{\mu }{}_1 
-\left(\delta ^{\mu }{}_0+\delta ^{\mu }{}_3\right)
\eta _{\rho 1} \right] +
\\
& \zeta _2^{(9)}\left(s \left(\delta ^{\nu }{}_0+\delta ^{\nu }{}_3\right)+\delta ^{\nu }{}_1\right)  \left[ 
\left(\delta ^{\mu }{}_0+\delta ^{\mu }{}_3\right) \eta _{\rho 2} - \left(\eta _{\rho 0}+\eta _{\rho 3}\right) \delta ^{\mu }{}_2  \right]+
\\
&\zeta _1^{(9)}\left(\delta ^{\nu }{}_0+\delta ^{\nu }{}_3\right) \left(\delta ^{\mu }{}_3\eta _{\rho 0} -\delta ^{\mu }{}_0\eta _{\rho 3} \right)  \,,
\end{aligned}
\end{equation}
\begin{equation}
   \begin{aligned}
& [\x^0,\x^1]=i\, \zeta _1^{(9)} \left[ s \left(\x^3-\x^0\right) +\x^1 \right] - i\, \zeta _2^{(9)}  \, \x^2 \,,& &[\x^0,\x^2]=- i\, \theta ^{(9)} + i \, s  \, \zeta _2^{(9)} \left(\x^0-\x^3\right) \,,
\\ 
 &[\x^1,\x^2]=i\,\zeta _2^{(9)} \left(\x^0-\x^3\right) \,,& &[\x^0,\x^3]=i \,\zeta _1^{(9)}\left(\x^3-\x^0\right) \,,
 \\ 
 &[\x^1,\x^3]=i \, \zeta _1^{(9)} \left[  s \left(\x^0-\x^3\right) -\x^1 \right] + i \,\zeta _2^{(9)}\, \x^2  \,, & &[\x^2,\x^3]=i \, \theta ^{(9)} - i \, s \, \zeta _2^{(9)} \left(\x^0-\x^3\right) \,.
   \end{aligned}
\end{equation}

\item \textbf{Case 10}
\begin{equation}
    \begin{aligned}
    r_{10} =~& \theta^{(10)}_1 \,  \, (\mP_0 - \mP_3) \wedge \mP_1 + \theta^{(10)}_2 \,  (\mP_0 + \mP_3) \wedge \mP_2 + \mP_1 \wedge (\mM_{23}-\mM_{02}) \\
    &+(\mP_0 + \mP_3) \wedge (\mM_{01}-\mM_{13}) \,,
\\
\theta^{\mu \nu} =&  2 \, \theta _1{}^{(10)}  \, \delta ^{[\mu}{}_1 \left(\delta ^{\nu]}{}_0-\delta ^{\nu]}{}_3\right)
+2 \,\theta _2{}^{(10)} \, \delta ^{[\mu}{}_2 \left(\delta ^{\nu]}{}_0+\delta ^{\nu]}{}_3\right) \,,
\\
\omega^{\mu \nu \rho } =& 2 \, \left(\delta^{\mu }{}_0+\delta ^{\mu }{}_3\right) \left(   \delta ^{\nu }{}_{[1} \delta ^{\rho }{}_{0]} +  \delta ^{\nu }{}_{[1} \delta ^{\rho }{}_{3]}  \right)
- \delta ^{\mu }{}_1 \left(  \delta ^{\nu }{}_{[2} \delta ^{\rho }{}_{0]}  + \delta ^{\nu }{}_{[2} \delta ^{\rho }{}_{3]} \right)\,,
\\
f^{\mu \nu }{}_{\rho } =&
2 \, \left(\delta^{\nu }{}_0+\delta ^{\nu }{}_3\right) \left(   \delta ^{\mu }{}_{[1} \eta_{0]\rho} +  \delta ^{\mu }{}_{[1} \eta_{3]\rho}  \right)
- \delta ^{\nu }{}_1 \left(  \delta ^{\mu }{}_{[2} \eta_{0]\rho}  + \delta ^{\mu }{}_{[2} \eta_{3]\rho} \right)\,,
\end{aligned}
\end{equation}
\begin{equation}
   \begin{aligned}
& [\x^0,\x^1]=i \left(\x^3-\x^0-\x^2-\theta _1{}^{(10)}\right)\,,& &[\x^0,\x^2]=-i\, \theta _2{}^{(10)}\,,
 \\ 
 & [\x^1,\x^2]=i \left(\x^0-\x^3\right)\,,& 
 & [\x^0,\x^3]=0 \,,
 \\ 
 & [\x^1,\x^3]=i \left(\x^0+\x^2-\x^3-\theta _1{}^{(10)}\right)\,,& &[\x^2,\x^3]=i \,\theta _2{}^{(10)}\,.
   \end{aligned}
\end{equation}

\item \textbf{Case 11}
\begin{equation}
    \begin{aligned}
    r_{11} =&\theta^{(11)}_1 \,  \, (\mP_0+\mP_3) \wedge \mP_1 + \theta^{(11)}_2 \, (\mP_0-\mP_3)\wedge \mP_2 + \mP_2 \wedge ( \mM_{01} - \mM_{13} )   \,,
\\
\theta^{\mu \nu} =&  2 \, \theta _1{}^{(11)}  \, \delta ^{[\mu}{}_1 \left(\delta ^{\nu]}{}_0+\delta ^{\nu]}{}_3\right)  + \theta_2{}^{(11)} \,\delta ^{[\mu}{}_2\left(\delta ^{\nu]}{}_0-\delta ^{\nu]}{}_3\right)  \,,
\\
\omega^{\mu \nu \rho } =&  \delta ^{\mu }{}_2 \left( \delta ^{\nu }{}_1 \delta ^{\rho }{}_0-\delta ^{\nu }{}_0 \delta ^{\rho }{}_1 + \delta ^{\nu }{}_1 \delta ^{\rho }{}_3-\delta ^{\nu }{}_3 \delta ^{\rho }{}_1 \right)\,,
\\
f^{\mu \nu }{}_{\rho } =& \delta ^{\nu }{}_2 \left( \delta ^{\mu }{}_1 \eta_{\rho 0}-\delta ^{\mu }{}_0 \eta_{\rho 1} + \delta ^{\mu }{}_1 \eta_{\rho 3}-\delta ^{\mu }{}_3 \eta_{\rho 1} \right)\,,
\end{aligned}
\end{equation}
\begin{equation}
   \begin{aligned}
& [\x^0,\x^1]=-i \,\theta _1{}^{(11)}\,,& &[\x^0,\x^2]=i \left(\x^1-\theta _2{}^{(11)}\right)\,,& &[\x^1,\x^2]=i \left(\x^0-\x^3\right)\,,
 \\ 
 & [\x^0,\x^3]=0\,,& &[\x^1,\x^3]=i \, \theta _1{}^{(11)}\,,& &[\x^2,\x^3]=-i \left(\theta _2{}^{(11)}+\x^1\right)\,.
   \end{aligned}
\end{equation}

$$
~
$$

\item \textbf{Case 12}\footnote{There is another  mistake in~\cite{Zakrzewski_1997}. Case 12 in Table 1 has an additional term, of the form $(\mP_0-\mP_3) \wedge \mP_2$, which does not satisfy the CYBE. This term has been removed here. This mistake was pointed out in~\cite{tolstoy2007twisted} (pag.~15) first, and more recently in~\cite{Meier:2023kzt,Meier:2023lku}.}
\begin{equation}
\begin{aligned}
    r_{12} =&   (\mP_0-\mP_3) \wedge \left[ \theta^{(12)}_1 \, (\mP_0+\mP_3) + \theta^{(12)}_2 \, \mP_1  \right] 
+  (\mP_0+\mP_3) \wedge  \left[ \theta^{(12)}_3 \,  \mP_2  +  ( \mM_{01} - \mM_{13} )   \right] \,,
\\
\theta^{\mu \nu} =&  2\left[ \left( \theta _1{}^{(12)} \left(\delta ^{[\mu}{}_0+\delta ^{[\mu}{}_3\right)
+\theta _2{}^{(12)} \, \delta ^{[\mu}{}_1\right) \left(\delta ^{\nu]}{}_0-\delta ^{\nu]}{}_3\right) +\theta _3{}^{(12)}   \delta ^{[\mu }{}_2 \left(\delta ^{\nu]}{}_0+\delta ^{\nu]}{}_3\right) \right] \,,
\\
\omega^{\mu \nu \rho } =&  \left(\delta ^{\mu }{}_0+\delta ^{\mu }{}_3\right) \left(  \delta ^{\nu }{}_1 \delta ^{\rho }{}_0-\delta ^{\nu }{}_0 \delta ^{\rho }{}_1 +   \delta ^{\nu }{}_1 \delta ^{\rho }{}_3-\delta ^{\nu }{}_3 \delta ^{\rho }{}_1  \right)\,,
\\
f^{\mu \nu }{}_{\rho } =&\left(\delta ^{\mu }{}_0+\delta ^{\mu }{}_3\right) \left(  \delta ^{\nu }{}_1 \eta_{\rho 0}-\delta ^{\nu }{}_0 \eta_{\rho 1} +   \delta ^{\nu }{}_1 \eta_{\rho 3}-\delta ^{\nu }{}_3 \eta_{\rho 1}  \right)\,,
\end{aligned}
\end{equation}
\begin{equation}
   \begin{aligned}
        & [\x^0,\x^1]=i \left(\x^3-\x^0 - \theta _2{}^{(12)}\right)\,,& &[\x^0,\x^2]=-i\, \theta _3{}^{(12)}\,,& &[\x^1,\x^2]=0\,,
 \\ 
 & [\x^0,\x^3]=-2 \,i\, \theta _1{}^{(12)}\,,& &[\x^1,\x^3]=i \left(\x^0-\x^3 -\theta _2{}^{(12)}\right)\,,& &[\x^2,\x^3]=i \,\theta _3{}^{(12)}\,.
   \end{aligned}
\end{equation}
\item \textbf{Case 13, 14 \& 15}

Assume $\zeta^{(\varrho)}_1 = \zeta^{(\varrho)}_2 = 0$. When $\zeta^{(\varrho)}_\mu = \delta^0_\mu$, $\zeta^{(\varrho)}_\mu = \delta^0_\mu + \delta^3{}_\mu$ and  $\zeta^{(\varrho)}_\mu = \delta^3{}_\mu$ respectively, these are the timelike, lightlike and spacelike version of $\varrho$-Poincar\'e~\cite{Lizzi:2022hcq,Fabiano:2023uhg}. This class of models was introduced in~\cite{Lukierski:2005fc}.
\begin{equation}
\begin{aligned}
    r_{\varrho} =&  \theta^{(\varrho)}_1 \, \mP_0 \wedge \mP_3 +  \theta^{(\varrho)}_2 \, \mP_1 \wedge \mP_2  +  \zeta^{(\varrho)}_\mu \, \eta^{\mu\nu} \, \mP_\nu \wedge  \mM_{12}    \,,
\\
 \theta^{\mu \nu} =& \delta ^{\mu }{}_3 \delta ^{\nu }{}_0 \theta _1{}^{(\varrho )}-\delta ^{\mu }{}_0 \delta ^{\nu }{}_3 \theta _1{}^{(\varrho )}+\left(\delta ^{\mu }{}_2 \delta ^{\nu }{}_1-\delta ^{\mu }{}_1 \delta ^{\nu }{}_2\right) \theta _2{}^{(\varrho )}\,,
\\
\omega^{\mu \nu \rho } =&\left(\delta ^{\nu }{}_2 \delta ^{\rho }{}_1-\delta ^{\nu }{}_1 \delta ^{\rho }{}_2\right) \left(\eta^{\mu 0} \zeta _0{}^{(\varrho )}+\eta^{\mu 3} \zeta _3{}^{(\varrho )}\right)\,,
\\
f^{\mu \nu }{}_{\rho } =&\left( \delta ^{\mu }{}_2\eta _{\rho 1} -\delta ^{\mu }{}_1\eta _{\rho 2}  \right) \left(\eta^{\nu 0} \zeta _0{}^{(\varrho )}+\eta^{\nu 3} \zeta _3{}^{(\varrho )}\right)\,,
\end{aligned}
\end{equation}
\begin{equation}
   \begin{aligned}
& [\x^0,\x^1]=i \, \zeta _0{}^{(\varrho )} \,\x^2 \,,& &[\x^0,\x^2]=-i \,\zeta _0{}^{(\varrho )}\, \x^1\,,& &[\x^1,\x^2]=-i \, \theta _2{}^{(\varrho )}\,,
 \\ 
 & [\x^0,\x^3]=-i \, \theta _1{}^{(\varrho )}\,,& &[\x^1,\x^3]=i\, \zeta _3{}^{(\varrho )}\, \x^2 \,,& &[\x^2,\x^3]=-i \,  \zeta _3{}^{(\varrho )}\,\x^1\,.
   \end{aligned}
\end{equation}

\item \textbf{Case 16}
\begin{equation}
\begin{aligned}
    r_{16} =&  \theta^{(16)}_1 \, \mP_0 \wedge \mP_3 +  \theta^{(16)}_2 \, \mP_1 \wedge \mP_2  +  \mP_1 \wedge  \mM_{03}    \,,
\\
\theta^{\mu \nu} =&  \theta _1{}^{(16)} \,  \left( \delta ^{\mu }{}_3 \delta ^{\nu }{}_0-\delta ^{\mu }{}_0 \delta ^{\nu }{}_3 \right) +\theta _2{}^{(16)} \left(\delta ^{\mu }{}_2 \delta ^{\nu }{}_1-\delta ^{\mu }{}_1 \delta ^{\nu }{}_2\right) \,,
\\
\omega^{\mu \nu \rho } =&\delta^\mu{}_1 \left(\delta ^{\nu }{}_3 \delta ^{\rho }{}_0-\delta ^{\nu }{}_0 \delta ^{\rho }{}_3\right)\,,
\\
f^{\mu \nu }{}_{\rho } =&\delta^\nu{}_1 \left(\delta ^{\mu }{}_3 \eta _{\rho 0} - \delta ^{\mu }{}_0 \eta _{\rho 3}\right)\,,
\end{aligned}
\end{equation}
\begin{equation}
   \begin{aligned}
& [\x^0,\x^1]=i \, \x^3\,,& &[\x^0,\x^2]=0\,,& &[\x^1,\x^2]=-i \, \theta _2{}^{(16)}\,,
 \\ 
 & [\x^0,\x^3]=-i \, \theta _1{}^{(16)}\,,& &[\x^1,\x^3]=-i \, \x^0\,,& &[\x^2,\x^3]=0\,.
   \end{aligned}
\end{equation}

\item \textbf{Case 17}
\begin{equation}
\begin{aligned}
    r_{17} =&  \theta^{(17)}_1 \, \mP_1 \wedge \mP_2 +  \theta^{(17)}_2 \, (\mP_0 + \mP_3) \wedge \mP_1  +  (\mP_0 + \mP_3) \wedge  \mM_{03}    \,,
\\
\theta^{\mu \nu} =& \delta ^{\mu }{}_2 \delta ^{\nu }{}_1 \theta _1{}^{(17)}-\left(\delta ^{\mu }{}_0+\delta ^{\mu }{}_3\right) \delta ^{\nu }{}_1 \theta _2{}^{(17)}+\delta ^{\mu }{}_1 \left(\left(\delta ^{\nu }{}_0+\delta ^{\nu }{}_3\right) \theta _2{}^{(17)}-\delta ^{\nu }{}_2 \theta _1{}^{(17)}\right)\,,
\\
\omega^{\mu \nu \rho } =&\left(\delta ^{\mu }{}_0+\delta ^{\mu }{}_3\right) \left(\delta ^{\nu }{}_3 \delta ^{\rho }{}_0-\delta ^{\nu }{}_0 \delta ^{\rho }{}_3\right)\,,
\\
f^{\mu \nu }{}_{\rho } =&\left(\delta ^{\nu }{}_0+\delta ^{\nu }{}_3\right) \left(\delta ^{\mu }{}_3 \eta _{\rho 0} - \delta ^{\mu }{}_0 \eta _{\rho 3}\right)\,,
\end{aligned}
\end{equation}
\begin{equation}
   \begin{aligned}
& [\x^0,\x^1]=-i \, \theta _2{}^{(17)}\,,& &[\x^0,\x^2]=0\,,& &[\x^1,\x^2]=-i \, \theta _1{}^{(17)}\,,
 \\ 
 & [\x^0,\x^3]=i \left(\x^3-\x^0\right)\,,& &[\x^1,\x^3]=i \, \theta _2{}^{(17)}\,,& &[\x^2,\x^3]=0\,.
   \end{aligned}
\end{equation}

\item \textbf{Case 18}
\begin{equation}
\begin{aligned}
    r_{18} =&  \theta^{(18)} \, \mP_1 \wedge \mP_2 
     +  (\mP_0 + \mP_3) \wedge \left(  \mM_{03}  + \zeta^{(18)} \,  \mM_{12} \right)   \,.
\\
\theta^{\mu \nu} =& \left(\delta ^{\mu }{}_2 \delta ^{\nu }{}_1-\delta ^{\mu }{}_1 \delta ^{\nu }{}_2\right) \theta ^{(18)}\,,
\\
\omega^{\mu \nu \rho } =&\left(\delta ^{\mu }{}_0+\delta ^{\mu }{}_3\right) \left(\zeta ^{(18)}\left(\delta ^{\nu }{}_2 \delta ^{\rho }{}_1-\delta ^{\nu }{}_1 \delta ^{\rho }{}_2\right) +\delta ^{\nu }{}_3 \delta ^{\rho }{}_0-\delta ^{\nu }{}_0 \delta ^{\rho }{}_3\right)\,,
\\
f^{\mu \nu }{}_{\rho } =&\left(\delta ^{\mu }{}_0+\delta ^{\mu }{}_3\right) \left(\zeta ^{(18)} \left(\delta ^{\nu }{}_2 \eta_{\rho 1}-\delta ^{\nu }{}_1  \eta_{\rho 2}\right) +\delta ^{\nu }{}_3  \eta_{\rho 0} -\delta ^{\nu }{}_0  \eta_{\rho 3}\right)\,,
    \end{aligned}
\end{equation}
\begin{equation}
   \begin{aligned}
& [\x^0,\x^1]=-i \,\zeta ^{(18)} \, \x^2 \,,& &[\x^0,\x^2]=i\, \zeta ^{(18)} \, \x^1 \,,& &[\x^1,\x^2]=-i \,\theta ^{(18)}\,,
 \\ 
 & [\x^0,\x^3]=i \left(\x^3-\x^0\right)\,,& &[\x^1,\x^3]=i \, \zeta ^{(18)} \,\x^2  \,,& &[\x^2,\x^3]=-i \,\zeta ^{(18)} \,\x^1\,.
   \end{aligned}
\end{equation}

\item \textbf{Cases 19, 20 \& 21}

These are the three classes of the $\theta$-Poincar\'e group, corresponding to choices of the $\theta^{\mu\nu}$ matrix that are inequivalent under automorphisms of the Poincar\'e group.
\begin{equation}
    r_{\theta} =  - {\sfrac 1 2} \, \theta^{\mu\nu} \,  \mP_\mu \wedge \mP_\nu      \,,
\qquad
 \theta^{\mu\nu} =   \text{\textit{arbitrary}} \,,
\qquad
    \omega^{\mu\nu\rho} = f^{\mu\nu}{}_\rho  = 0 \,,
\end{equation}
\begin{equation}
[\x^\mu ,\x^\nu ]  = i \, \theta^{\mu\nu} \,.
\end{equation}
\end{itemize}

Of the 17 classes listed above, for nomenclature purposes, we can lump together the three \textbf{9}$^{\bm{(s)}}$ classes, classes \textbf{13},\textbf{14} \& \textbf{15} (which we can collectively name $\varrho$-deformations), and \textbf{19}, \textbf{20} \& \textbf{21}  (which are commonly all called $\theta$-deformations). This leaves us with 11 different models. As these models get studied in detail, my hope is that the community will keep the tradition of assigning lowercase greek letters to identify them, of which there is enough unused ones. The models corresponding to classes \textbf{10} to \textbf{16} are special, because the Lie group generated by the spacetime coordinates are \textit{unimodular,} which is an important property that leads to significant simplifications, as I will show below. Furthermore, classes \textbf{7} and \textbf{8} are special too, in that the Lie algebra generated by the spacetime coordinates does not admit a central extension, and therefore $\theta^{\mu\nu}$ can only be zero.

\subsection{Bicrossproduct and Drinfel'd Twist structures}

When the central extension parameters are set to zero, $\theta^{\mu\nu}=0$, all the models above have a bicrossproduct structure~\cite{Majid:1994cy}, in which the Hopf algebra $C_\ell[ISO(3,1)]$  can be written as
\begin{equation}
C_\ell[ISO(3,1] = C_\ell[\mathbbm{R}^{3,1}] \vartriangleright \! \blacktriangleleft C[SO(3,1)] \,,
\end{equation}
where $C_\ell[\mathbbm{R}^{3,1}]$ is extended to a Hopf algebra with primitive coproduct, antipode and counit:
\begin{equation}
  [\x^{\mu} , \x^{\nu}]  = i \, \left( f^{\nu\mu}{}_\rho - f^{\mu\nu}{}_\rho \right) \, \x^\rho    \,, \qquad \Delta[\x^\mu] = \x^\mu \otimes \1 + \1 \otimes \x^\mu \,, \qquad S(\x^\mu) = - \x^\mu \,, \qquad \epsilon(\x^\mu) = 0\,,
\end{equation}
and $C[SO(3,1)]$ is the undeformed Hopf-algebra formulation of the algebra of functions on the Lorentz group, \textit{i.e.}
\begin{equation}
[\La^\mu{}_\nu,\La^\rho{}_\sigma]= 0 \,, \qquad \Delta( \La^\mu{}_\nu  ) = \La^\mu{}_\rho  \otimes \La^\rho{}_\nu  \,, \qquad S(\La^\mu{}_\nu) = (\La^{-1})^\mu{}_\nu \,, \qquad \epsilon(\La^\mu{}_\nu) = \delta^\mu{}_\nu   \,,
\end{equation}
supplemented by relations~\eqref{MetricInvariance}. The two algebras have a covariant right action of $ C_\ell[\mathbbm{R}^{3,1}]$ on $C[SO(3,1)]$ and a covariant left coaction of   $C[SO(3,1)]$ on $ C_\ell[\mathbbm{R}^{3,1}]$:
\begin{equation}
\begin{gathered}
\La^\mu{}_\rho \triangleleft \x^\nu  =  i \,  f^{\alpha\beta}{}_\gamma
\left(  \La^{\mu}{}_\alpha \, \La^{\nu}{}_\beta \, \delta^\gamma{}_\rho  - \delta^{\mu}{}_\alpha \, \delta^{\nu}{}_\beta \, \La^\gamma{}_\rho \right) , 
\\
\La^\mu{}_\rho \triangleleft \1 = \La^\mu{}_\rho  \,, \qquad \1 \triangleleft \x^\nu  = 0 \,, \qquad \1 \triangleleft \1 = \1 \,,
\\
\beta (\x^\mu) = \La^\mu{}_\nu \otimes \x^\nu  \,, \qquad
\beta (\1) = \1 \otimes \1 \,.
\end{gathered}
\end{equation}
The above maps satisfy the axioms of a bicrossproduct Hopf algebra (here I am using Sweedler's notation $\Delta(X) = X_{(1)} \otimes  X_{(2)}  = \sum_i  X^i_{(1)} \otimes  X^i_{(2)} $):
\begin{equation}
\begin{gathered}
\lambda \triangleleft ( x \, y) = (\lambda \triangleleft  x ) \triangleleft y \,,
\qquad
1 \triangleleft x = \epsilon(x) \, \1 \,,
\qquad
( \lambda \, \sigma ) \triangleleft x = (\lambda \triangleleft x_{(1)} ) (\sigma \triangleleft x_{(2)} ) \,,
\\
( \text{id} \otimes \beta ) \circ \beta = (\Delta \otimes \text{id}) \circ \beta \,,
\qquad
(\epsilon \otimes \text{id} ) \circ \beta = \text{id} \,,
\qquad
\beta (\lambda \, \sigma ) = \beta(\lambda) \beta(\sigma) \,, \qquad
\beta (\1) = \1 \otimes \1 \,,
\end{gathered}
\end{equation}
for any $x, y \in C_\ell[\mathbbm{R}^{3,1}]$ and  $\sigma, \lambda \in  C[SO(3,1)]$. These are all trivial to verify except the first, which holds only if $\theta^{\mu\nu} =0$ (hence the assumption above), and when the third of Eqs.~\eqref{StructureConstants_constraints} is satisfied (this is proven in Sec. 4.1 of~\cite{Mercati:2023apu}, when checking the Jacobi identities for two translation parameters and one Lorentz matrix). The third property on the first line follows from the fact that the coproduct of $\x^\mu$ is primitive, meaning that it acts as a derivation on $C[SO(3,1)]$ (satisfying the Leibniz rule), and the fact that the algebra $C[SO(3,1)]$ is commutative, which is compatible with a trivial coproduct for $\x^\mu$.

For the bicrossproduct to be well-defined, a set of compatibility conditions need to be verified. The counit needs to satisfy $\epsilon(\lambda \triangleleft x ) = \epsilon(\lambda ) \epsilon(x)$, which is trivially verified. The coproduct needs to be such that
$\Delta (\lambda \triangleleft x) = \left( \lambda_{(1)} \triangleleft x_{(1)} \right) \beta^{(1)}(x_{(2)}) \otimes \left( \lambda_{(2)} \triangleleft \beta^{(2)}(x_{(2)}) \right)   $, where $\beta(X) = \beta^{(1)}(X) \otimes \beta^{(2)}(X)$. This can be verified by direct calculation:
\begin{equation}
\begin{aligned}
&\left( (\La^\mu{}_\rho)_{(1)} \triangleleft \x^\nu_{(1)} \right) \beta^{(1)}[\x^\nu_{(2)}] \otimes \left( (\La^\mu{}_\rho)_{(2)} \triangleleft \beta^{(2)}[\x^\nu_{(2)}] \right)   
 \\
 &=  i \, f^{\alpha\beta}{}_\gamma
\left(  \La^{\mu}{}_\alpha \, \La^{\nu}{}_\beta \, \delta^\gamma{}_\sigma  - \delta^{\mu}{}_\alpha \, \delta^{\nu}{}_\beta \, \La^\gamma{}_\sigma \right)  \otimes  \La^\sigma{}_\rho 
+ i \,
 \La^\mu{}_\sigma \,\La^\nu{}_\lambda \otimes  f^{\alpha\beta}{}_\gamma
\left(  \La^{\sigma}{}_\alpha \, \La^{\lambda}{}_\beta \, \delta^\gamma{}_\rho  - \delta^{\sigma}{}_\alpha \, \delta^{\lambda}{}_\beta \, \La^\gamma{}_\rho \right)
 \\
 &=  i \, f^{\alpha\beta}{}_\gamma
\left(  \cancel{\La^{\mu}{}_\alpha \, \La^{\nu}{}_\beta \otimes  \La^\sigma{}_\rho \, \delta^\gamma{}_\sigma }  - \delta^{\mu}{}_\alpha \, \delta^{\nu}{}_\beta \, \La^\gamma{}_\sigma \otimes  \La^\sigma{}_\rho    
+  \La^\mu{}_\sigma \,\La^\nu{}_\lambda \otimes  \La^{\sigma}{}_\alpha \, \La^{\lambda}{}_\beta \, \delta^\gamma{}_\rho  -    \cancel{  \La^\mu{}_\alpha \,\La^\nu{}_\beta \otimes  \La^\gamma{}_\rho} \right) \\
 &=  i \, f^{\alpha\beta}{}_\gamma
\left(  \Delta[\La^\mu{}_\alpha \,\La^\nu{}_\beta]  \, \delta^\gamma{}_\rho - \delta^{\mu}{}_\alpha \, \delta^{\nu}{}_\beta \, \Delta [\La^\gamma{}_\rho]    \right)
= \Delta \left[ \La^\mu{}_\rho \triangleleft \x^\nu \right] \,.
\end{aligned}
\end{equation}
The left coaction needs to satisfy $\beta(x \, y) = \left[ \beta^{(1)}(x) \triangleleft y_{(1)} \right]\beta^{(1)}(y_{(2)}) \otimes \beta^{(2)}(x) \, \beta^{(2)}(y_{(2)})$, which means:
\begin{equation}
\begin{aligned}
&\beta \left( {\sfrac i 2} \, c^{\mu\nu}{}_\rho \, \x^\rho \right) = \beta(\x^{[\mu} \, \x^{\nu]}) = \left[ \La^{[\mu}{}_\rho \triangleleft \x^{\nu]}\right]\beta^{(1)}(\1) \otimes  \x^\rho \, \beta^{(2)}(\1)+
\left[ \La^{[\mu}{}_\rho \triangleleft \1 \right]\beta^{(1)}(\x^{\nu]}) \otimes  \x^\rho \, \beta^{(2)}(\x^{\nu]})
\\
&=\left[ \La^{[\mu}{}_\rho \triangleleft \x^{\nu]} \right]  \otimes  \x^\rho  +
  \La^{[\mu}{}_\rho \, \La^{\nu]}{}_\sigma \otimes  \x^\rho \,\x^\sigma 
  \\
  &=  i \,  f^{\alpha\beta}{}_\gamma
\left(  \La^{[\mu}{}_\alpha \, \La^{\nu]}{}_\beta \, \delta^\gamma{}_\rho  - \delta^{[\mu}{}_\alpha \, \delta^{\nu]}{}_\beta \, \La^\gamma{}_\rho \right)  \otimes  \x^\rho  +
  \La^{[\mu}{}_\rho \, \La^{\nu]}{}_\sigma \otimes  \x^\rho \,\x^\sigma
   \\
  &=  i \,  f^{\alpha\beta}{}_\gamma
\left(  \cancel{\La^{[\mu}{}_\alpha \, \La^{\nu]}{}_\beta \, \delta^\gamma{}_\rho } - \delta^{[\mu}{}_\alpha \, \delta^{\nu]}{}_\beta \, \La^\gamma{}_\rho \right)  \otimes  \x^\rho  + \cancel{{\sfrac i 2} \, c^{\rho\sigma}{}_\lambda \,
  \La^{\mu}{}_\rho \, \La^{\nu}{}_\sigma \otimes  \x^\lambda }
  \\
&= -i \,  f^{[\mu\nu]}{}_\rho  \La^\rho{}_\lambda \otimes  \x^\rho  = {\sfrac i 2} \,  c^{\mu\nu}{}_\rho  \ , \La^\rho{}_\lambda \otimes  \x^\rho =   {\sfrac i 2} \,  c^{\mu\nu}{}_\rho \, \beta(\x^\rho) \,.
\end{aligned}
\end{equation}
Finally, the action and the coaction have to satisfy $\beta^{(1)}(x_{(1)}) (\lambda \triangleleft x_{(2)}) \otimes \beta^{(2)}(x_{(1)}) = (\lambda \triangleleft x_{(1)}) \beta^{(1)}(x_{(2)}) \otimes \beta^{(2)}(x_{(2)})$, which can be verified as follows:
\begin{equation}
\begin{gathered}
\beta^{(1)}(\x^\nu_{(1)}) (\La^\mu{}_\rho \triangleleft \x^\nu_{(2)}) \otimes \beta^{(2)}(\x^\nu_{(1)}) =
\beta^{(1)}(\x^\nu) (\La^\mu{}_\rho \triangleleft \1) \otimes \beta^{(2)}(\x^\nu)
+\beta^{(1)}(\1) (\La^\mu{}_\rho \triangleleft \x^\nu) \otimes \beta^{(2)}(\1)
\\
=
\La^\nu{}_\sigma \, \La^\mu{}_\rho \otimes \x^\sigma
+  (\La^\mu{}_\rho \triangleleft \x^\nu) \otimes \1
=
  (\La^\mu{}_\rho \triangleleft \x^\nu)  \otimes \1
  +
    (\La^\mu{}_\rho \triangleleft \1) \La^\mu{}_\sigma \otimes \x^\sigma =
 \\
  (\La^\mu{}_\rho \triangleleft \x^\nu) \beta^{(1)}(\1) \otimes \beta^{(2)}(\1)
  +
    (\La^\mu{}_\rho \triangleleft \1) \beta^{(1)}(\x^\mu) \otimes \beta^{(2)}(\x^\nu)
 =
  (\La^\mu{}_\rho \triangleleft \x^\nu_{(1)}) \beta^{(1)}(\x^\mu_{(2)}) \otimes \beta^{(2)}(\x^\nu_{(2)}) \,.
\end{gathered}
\end{equation}

The T-Poincar\'e groups listed above all have a Drinfel'd twist structure (also when $\theta^{\mu\nu} \neq 0$). This was proven in~\cite{tolstoy2007twisted} for all models except Case \textbf{12}. The twist element for this last case was found in~\cite{Meier:2023lku}.
The r-matrices of the $\theta$-deformation, Cases~\textbf{19}-\textbf{21} are \textit{Abelian,} in the sense that they are made of wedge products of generators that all commute with each other. This allows the introduction of an \textit{Abelian twist} element of the form $\mathcal{F} = \exp(-r)$. In the other cases, the twist elements found in~\cite{tolstoy2007twisted} and~\cite{Meier:2023lku} take the form of a product of exponentials, $\mathcal{F} = \exp(-r_1) \exp(-r_2) \dots \exp(-r_n)$. The twist element allows to express the whole corresponding Hopf-algebra deformation of the Poincar\'e algebra   as follows: the commutation relations, and counits are undeformed, while the coproducts and antipodes are given by
\begin{equation}\label{Eq:Twisted_Hopf_algebra}
\Delta(X) = \mathcal{F} \Delta_0(X) \mathcal{F}^{-1} \,, \qquad
S(X) =   \mathcal{F}_{(1)} S_0 \big{(}\mathcal{F}_{(2)} \big{)} S_0[X] S_0 \big{(} \mathcal{F}^{-1}_{(1)} \big{)} \mathcal{F}^{-1}_{(2)}  \,,
\end{equation}
where $\Delta_0 (P_\mu) = P_\mu \otimes \1 + \1  \otimes P_\mu$,  $\Delta_0 (M_{\mu\nu}) = M_{\mu\nu} \otimes \1 + \1  \otimes M_{\mu\nu}$, $S_0(P_\mu) = - P_ \mu$ and $S_0(M_{\mu\nu} ) = - M_{\mu\nu} $.
As I will show in Sec.~\ref{Sec:T-Poincare_algebra} below, the dual Hopf algebra to $C_\ell[ISO(3,1)]$, which is the Hopf-algebra analog of the Lie algebra associated to the Poincar\'e group, can be derived by looking at the coaction~\eqref{Coaction_spacetime_coordinates} of an ordered exponential of the T-Minkowski coordinates. What one obtains this way differs from the Hopf algebra~\eqref{Eq:Twisted_Hopf_algebra}: for example, the commutation relations are deformed and nonlinear. This difference is due to the fact that~\eqref{Eq:Twisted_Hopf_algebra} with primitive commutators and counit and the Hopf algebra I will obtain in Sec.~\ref{Sec:T-Poincare_algebra} are the same Hopf algebra, expressed in different bases. The basis~\eqref{Eq:Twisted_Hopf_algebra} is the \textit{twist basis,} which could also be called \textit{classical basis} because the commutation relations are undeformed in this basis. The basis of Sec.~\ref{Sec:T-Poincare_algebra} could be called \textit{bicrossproduct basis} (in a particular ordering), and it has greater practical interest because it allows to describe the first-order action of T-Poincar\'e transformations on noncommutative functions. The two bases are related by nonlinear changes of coordinates, which in particular involve momentum-dependent redefinitions of the Lorentz algebra generators $M_{\mu\nu}$. This has been observed in the $\kappa$-lightlike model~\cite{Ballesteros_1995,BALLESTEROS_1996,Ballesteros:1996awf,Lizzi:2021rlb}, and, more recently, in the $\varrho$-Minkowski model~\cite{Fabiano:2023uhg,Lizzi:2022hcq}.

The twist basis, and its relation to the bicrossproduct bases, remains of interest, because if the twist element is known, it is possible to write the \textit{universal R-matrix} of the quantum group with a closed formula:
\begin{equation}\label{Universal_R-matrix}
\mathcal{R} = \mathcal{F} _\text{op} \, \mathcal{F}^{-1} \,, \qquad  \mathcal{F}_\text{op} =  \mathcal{F}_{(2)} \otimes  \mathcal{F}_{(1)} \,.
\end{equation}
This is a useful object to have, because, for example, it allows to write in a compact form the commutator (or the exchange rule) of two general functions (not necessarily coordinates) on the noncommutative space. The matricial R-matrix I wrote above in Eqs.~\eqref{RTT_relations}, \eqref{QYBE} and then in~\eqref{R-matrix_r-matrix}, is just the operator~\eqref{Universal_R-matrix} in its five-dimensional representation. Since the momentum generators are nilpotent, in this representation, the R-matrix can be written as an object at most quadratic in the generators, as in Eq.~\eqref{R-matrix_r-matrix}. However, the universal R-matrix~\eqref{Universal_R-matrix} depends, in principle, on all powers of the Poincar\'e algebra generators. If we are interested in relations that require the universal R-matrix, like the commutators between two plane waves or between two fields (objects that are of interest in quantum field theory, for example), then we might want to calculate the universal  R-matrix from the twist, and then make a change of basis to a bicrossproduct basis.

At any rate, in the present paper I will only consider classical noncommutative field theories \textit{i.e.} a regime in which $\ell \neq 0$ but $\hbar = 0$. For this, knowledge of the exact form of the universal R-matrix is not necessary.

\subsection{Differential calculus, braiding and involution}

All of the T-Minkowski models admit a bicovariant differential calculus~\cite{Woronowicz1987_1,Woronowicz1987_2,Woronowicz:1989}, defined as a bimodule $\Gamma^\ell$ on $C_\ell[\mathbbm{R}^{3,1}]$ with the following commutation relations:
\begin{equation}
 \label{xdx_comm_rel}
 [ \x^{\mu} , d\x^{\nu} ]  = i \,f^{\nu\mu}{}_\rho \, d \x^\rho  \,.
\end{equation}
In~\cite{Mercati:2023apu} I have shown that commutators of this form are invariant under the simultaneous coaction of the T-Poincar\'e group on coordinates~\eqref{Coaction_spacetime_coordinates}, together with the following transformation law for the differentials:
\begin{equation}
d \x'^\mu = \La^\mu{}_\nu \otimes d \x^\nu \,,
\end{equation}
which is essentially the coaction~\eqref{Coaction_spacetime_coordinates} without the translation part, as should be since the differentials are supposed to represent differences between ``infinitesimally close'' coordinates of different points. In Sec.~\ref{Sec:DifferentialGeometry}, building on these results, I will develop all of the differential-geometric instruments that are necessary to define field theories on the T-Minkowski noncommutative spacetimes, following Woronowicz' work on noncommutative differential geometry~\cite{Woronowicz1987_1,Woronowicz1987_2,Woronowicz:1989}.

I mentioned that the noncommutative differential can be regarded, in a loose sense, as a sort of difference between coordinates of different points. This analogy is actually more literal than one  might think: in fact, as I showed in~\cite{Mercati:2023apu}, it is possible to introduce a notion of algebra of functions of N points. In the commutative case, this is isomorphic to $C[\mathbbm{R}^{3,1}]^{\otimes N}$, but the naive tensor product of N copies of $C_\ell[\mathbbm{R}^{3,1}]$ is inadequate in the noncommutative case, because it is not covariant under T-Poincar\'e transformations. In this case, it is necessary to introduce a \textit{braided tensor product algebra,} $C[\mathbbm{R}^{3,1}]^{\otimes_\ell N}$, in which the coordinates of different points do not commute:
\begin{equation}
  \label{xaxb_comm_rel}
  [\x_a^\mu , \x_b^\nu] 
    = i \, \theta^{\mu\nu}   
     -   i    \,   f^{(\mu\nu)}{}_\gamma \,  \left( \x_a^\rho - \x_b^\rho   \right)   
 - i   \, f^{[\mu\nu]}{}_\rho\left(     \x_a^\rho  +    \x_b^\rho     \right)    \,.
\end{equation}
This is covariant under the following simultaneous transformation of all coordinates:
\begin{equation}
\x'^\mu_a = \La^\mu{}_\nu \otimes \x^\nu_a + \A^\mu \otimes \1 \,.
\end{equation}
The analogy between this braided algebra and the differential calculus consists in the fact that the difference between two different coordinates, \textit{e.g.} $\x^\mu_b - \x^\mu_c$ is a bimodule for the algebra generated by $\x^\mu_a$, with the same commutation relations as $d \x^\mu$ in~\eqref{xdx_comm_rel}:
\begin{equation}
 [ \x^{\mu}_a , \x^{\nu}_b - \x^\nu_c ]  = i \,f^{\nu\mu}{}_\rho \, \left( \x^{\nu}_b - \x^\nu_c  \right) \,,
\end{equation}
 as can be straightforwardly derived from Eqs.~\eqref{xaxb_comm_rel}.

Finally, there is a noncommutative notion of complex conjugate, or rather Hermitian conjugate of the T-Minkowski coordinates. I say Hermitian because, for consistency, it has to invert the order of all noncommutative products. This notion extends also to the T-Poincar\'e group, to the differential calculus and to the braiding. Formally it can be introduced as an anti-linear involution on the algebras $C_\ell[\mathbbm{R}^{3,1}]$, $C_\ell[ISO(3,1)]$, $C_\ell[\mathbbm{R}^{3,1}]^{\otimes_\ell N}$ and on the comodule $\Gamma_\ell^1$ as
\begin{equation}\label{Involutions}
(\x^\mu)^* = \x^\mu \,, \qquad (\A^\mu)^* = \A^\mu \,, \qquad (\La^\mu{}_\nu)^* = \La^\mu{}_\nu \,, \qquad (\x^\mu_a)^* = \x^\mu_a \,,  \qquad (d \x^\mu)^* =  d \x^\mu \,.
\end{equation}
The transformation~\eqref{Involutions} leaves all of the commutation relations~\eqref{x_comm_rel}, \eqref{T-PoincareGroup_commutators}, \eqref{xaxb_comm_rel} and~\eqref{xdx_comm_rel} invariant, if we assume that all products are swapped by the involution.  For example, consider the comodule product of $\Gamma^1_\ell$, \textit{i.e.} 
\begin{equation}\label{CompatibilityInvolution_diffcalc}
\left[ d \x^\mu \, f(\x) \right]^* = f^*(\x) \, (d\x^\mu)^*   =  f^*(\x) \, d\x^\mu \,,
\end{equation}
this assumption makes the involution compatible with the commutation rules~\eqref{xdx_comm_rel} (the structure constants $f^{\nu\mu}{}_\rho$ are real):
\begin{equation}
\left( [ \x^{\mu} , d\x^{\nu} ] \right)^*  = [(d\x^{\nu})^*, (\x^{\mu})^* ] =  [ d\x^{\nu}  , \x^{\mu} ] = - [ \x^{\mu} , d\x^{\nu} ]  = - i \,f^{\nu\mu}{}_\rho \, d \x^\rho  =   \left( i \,f^{\nu\mu}{}_\rho \, d \x^\rho  \right)^* \,.
\end{equation}
The existence of the involution puts constraints on the possible representations of these algebras, as happens in the $\kappa$-lightlike model, where the standard Hermitian conjugate with the $L^2$ inner product that one would naturally define on the simplest infinite-dimensional representation of the algebra does not make the coordinates Hermitian. One then needs to represent the involution as a different operator than the naive Hermitian conjugate~\cite{Fabiano:2023xke}.

\subsection{Plan of the paper}
  
The goal of this Part II is to develop in detail all of the noncommutative functional analysis and differential geometric tools that are necessary to study field theories on T-Minkowski spacetimes. My approach will be model-independent, applying equally to all of the 11 T-Minkowski models (or 17 classes, if we adhere strictly to Zakrzewski's  classification by automorphism classes). However, some of the models have simpler properties, above all the unimodularity possessed by Classes \textbf{10} to \textbf{16}. Also, most models admit a central extension, which complicates the formulas, unless it is turned off by setting $\theta^{\mu\nu}=0$ by hand. The vast majority of my result will hold for the most generic case (non-unimodular, with central extension), with very few exceptions (essentially, the discussion of gauge theories is limited to the unimodular case because it is not clear, at this stage, how to define gauge-invariant actions if the integral is not cyclic). The focus of the present paper is not on model building and studying any particular spacetime, but rather to find the largest possible set of results that hold in general, for all T-Minkowski noncommutative spacetimes. This set, as I will show, is surprisingly large, as one can make things quite explicit without having to specialize to any particular model, reducing most statements to the existence of the solutions to certain partial differential equations, which can only be solved once a concrete model is specified. What this ignores are existence and uniqueness issues, and global issues of the solutions to these differential equations. These issues can have a dramatic effect on the theory, as exemplified by the $\kappa$-lightlike model, in which one has to extend the Fourier expansion of functions onto a whole additional space of exponentials~\cite{DiLuca:2022idu,Fabiano:2023xke}. These issues, however, can only be discussed once a particular model is chosen, and therefore will be postponed to studies that are focalized on one of the T-Minkowski spacetimes, and ignored here.

The paper is organized as follows: In Sec.~\ref{Sec:Fourier}, building on the results of Part I~\cite{Mercati:2023apu}, I will develop functional analysis techniques on any T-Minkowski spacetime, based on noncommutative Fourier theory, which in turn rests upon the Lie group structure of the algebra of coordinates $C_\ell[\mathbbm{R}^{3,1}]$. This will allow me to study the T-Poincar\'e transformation laws of scalar fields  (which are just elements of $C_\ell[\mathbbm{R}^{3,1}]$), and how the newly introduced notion of momentum space transforms under them. Section~\ref{Sec:T-Poincare_algebra} picks up from these results, and focuses on the quantum Poincar\'e \textit{algebras} associated to the T-Poincar\'e group, expressed in a bicrossproduct basis.
The following Section~\ref{Sec:DifferentialGeometry} contains the development, based on Woronowicz' construction~\cite{Woronowicz1987_1,Woronowicz1987_2,Woronowicz:1989}, of the bicovariant differential complex on T-Minkowski spacetimes, the Hodge star operator, the Lie and inner derivatives. With these tools, we will be able to study any classical field theory on T-Minkowski, scalar, vectorial or spinorial. This is done in the following Section, number~\ref{Sec:FieldTheory},  where all of the three types of field are studied, including their equations of motion and solutions, and a notion of Noether currents and conservation law is introduced and studied for the T-Poincar\'e symmetries of the theory (in the case of a scalar field). Finally, Section~\ref{Sec:Conclusions} contains my conclusions, and Appendix~\ref{Appendix:Lorentz_transform_momenta} details an important proof, that allows me to demonstrate the general form of the Lorentz transform of momentum space in a T-Minkowski noncommutative spacetime.

\newpage

\section{Noncommutative Fourier theory and integral calculus}\label{Sec:Fourier}
 
 \subsection{Weyl systems and noncommutative exponentials}
 
 The algebra $C_\ell[\mathbbm{R}^{3,1}]$ is generated by $\1$ and the four coordinates $\x^\mu$, and, since the Poincar\'e--Birkhoff--Witt  property holds, a basis for it is given by all possible monomials with a certain ordering prescription. A general function (that is, a scalar field) can be in principle written as a formal (\textit{i.e.}, without regard for its convergence) power series in these monomials. The same could be done in the commutative case: however, as it turns out, we do not need all (possibly badly divergent) field configurations that arise this way in order to do physics. Most of the results in  QFT require only one to consider Schwarz functions, which vanish at infinity (together with all their derivatives) faster than any power of the coordinates, and are closed under Fourier transform. In the case of noncommutative algebras defined by basic commutators of the form~\eqref{x_comm_rel}, there is a generalization of the Fourier transform that possesses all the desirable properties that make its commutative counterpart useful. This is the notion of \textit{generalized Weyl system,} introduced in~\cite{Agostini:2002de}.
 
One begins with a family of noncommutative exponentials in the spacetime coordinates:
\begin{equation}\label{GeneralOrderedExponentials}
\E[q] = ~ : e^{i \, q_\mu \, \x^\mu} : ~ = \sum_{n=0}^\infty  \frac{i^n}{n!} \, :(q_\mu \, \x^\mu)^n  :  \,, 
\end{equation}
where $: \, . \, :$ is a given ordering prescription, and the exponential is understood as a formal power series. Two examples of ordering prescriptions are the lexicographic ordering:
\begin{equation}
\begin{gathered}
: (\x^0)^{n_0}(\x^1)^{n_1}(\x^2)^{n_2}(\x^3)^{n_3} :_\st{Lexi} ~ = ~ (\x^0)^{n_0}(\x^1)^{n_1}(\x^2)^{n_2}(\x^3)^{n_3} \,, \\ \Downarrow 
\\ : e^{i \, q_\mu \, \x^\mu} :_\st{Lexi}  ~ = ~   e^{i \, q_0 \, \x^0}  e^{i \, q_1 \, \x^1}  e^{i \, q_2 \, \x^2}  e^{i \, q_3 \, \x^3}  \,,
\end{gathered}
\end{equation}
and the \textit{Weyl} (or totally symmetric) ordering:
\begin{equation}
\begin{gathered}
:  \x_{\mu_1} \dots \x_{\mu_n} :_\st{Weyl}  ~ = ~ \frac{1}{n!} \, \sum_{\sigma \in S_n} \x_{\sigma(\mu_1)} \dots \x_{\sigma(\mu_n)} \,, \\ \Downarrow 
\\ : e^{i \, q_\mu \, \x^\mu} :_\st{Weyl} ~ = ~ e^{i \, q_\mu \, \x^\mu} \,,
\end{gathered}
\end{equation}
where $S_n$ is the symmetric group of $n$ elements.

Given an ordering prescription, $\E[q]$ can be considered an element of the Lie group associated to the Lie algebra~\eqref{x_comm_rel}, and the ordering choice corresponds to a choice of \textit{factorization} of the group. For example, the lexicographic ordering corresponds to taking elements of the four one-parameter subgroups generated by the four coordinates, and multiplying them in order. On the other hand, the Weyl ordering corresponds to just considering the one-parameter subgroup generated by the linear combination $q_\mu \, \x^\mu$. Different factorizations of a group are related by local coordinate transformations of the Fourier parameters $q_\mu$, \textit{i.e.}
\begin{equation}\label{OrderingChange}
\E_1 [q] = e^{i \,\theta^{\mu\nu}\varphi^{12}_{\mu\nu}(q')}\, \E_2 [q'] \,, \qquad   q'_\mu = q'_\mu(q) \,,
\end{equation}
where $\theta^{\mu\nu}\varphi^{12}_{\mu\nu}(q')$ is a phase that depends on the two ordering choices. It is important to remark that these coordinate transformations are invertible only in a neighbourhood of the identity, and they might get singular far away from it. Think, for example, of the phenomenon of the \textit{gimbal lock} in $SO(3)$~\cite{enwiki:1214961199}, which is due to the Euler angles failing to be injective in certain regions of the group, and different conventions for the choice of axes (\textit{i.e.} different factorizations of the group) suffer from this phenomenon in different regions of the group. Moreover, in general, the exponential map is surjective only in an open neighbourhood of the identity of a Lie group, so we should be careful when identifying the momentum space of a noncommutative spacetime with a Lie group manifold.

Now is a good time to discuss the fact that I am allowing the commutators~\eqref{x_comm_rel} to include terms that are proportional to the identity. I could simply replace the $\1$ on the right-hand side of~\eqref{x_comm_rel} by a further generator, supplement Eq.~\eqref{x_comm_rel} with five more equations expressing the commutativity of this new generator, and treat $C_\ell[\mathbbm{R}^{3,1}]$ as a five-dimensional Lie algebra. However, it is much more convenient to treat it as a central extension of a four-dimensional Lie algebra with structure constants $c^{\mu\nu}{}_\rho$. As a matter of fact, in each T-Minkowski model, the possible values of the matrix $\theta^{\mu\nu}$ cover \textit{all} possible central extensions of the 4D Lie algebra, because the only constraint on $\theta^{\mu\nu}$ is the cocycle condition [second Eq. in~\eqref{StructureConstants_constraints}], which is precisely the condition for a central extension. From now on, I will call $\mathcal{A}_\ell$ the 4D Lie algebra of which  $C_\ell[\mathbbm{R}^{3,1}]$  is a central extension. This algebra has commutation relations
\begin{equation}
[\hat{t}^\mu , \hat{t}^\nu] = i \, c^{\mu\nu}{}_\rho \, \hat{t}^\rho \,,
\qquad  \hat{t}^\mu  \in \mathcal{A}_\ell \,,
\end{equation}
and I will call $\mathcal{G}_\ell$ the Lie group of $\mathcal{A}_\ell$, and refer to its group structure when describing the properties of the algebra of exponential functions.

For each and every one of our algebras $\mathcal{A}_\ell$, we can explicitly build a faithful 5-dimensional representation as follows
\begin{equation}\label{Eq:t_matrices}
\rho (\hat t^\alpha) = 
\left(
\begin{array}{c|c}
 -i \, (K^\alpha)^\mu{}_\nu &   \eta^{\mu\alpha}
\\
\hline
0 & 0
\end{array}
\right)\,, 
\end{equation}
their commutators are
\begin{equation}
[\rho(\hat{t}^\alpha), \rho(\hat{t}^\beta)] =
\left(
\begin{array}{c|c}
- [K^\alpha, K^\beta]^\mu{}_\nu &   - i \left( \eta^{\rho\beta}(K^{\alpha})^\mu{}_\rho - \eta^{\rho\alpha}(K^{\beta})^\mu{}_\rho\right)
\\
\hline
0 & 0
\end{array}
\right)\,,
\end{equation}
now use Eq.~\eqref{K_commutation_relations}, $[K^\mu , K^\nu] = - c^{\mu\nu}{}_\rho \, K^\rho$, and the fact that
 each $K^\alpha$ is a Lorentz matrix, so that it is antisymmetric when we raise one of its indices, \textit{i.e.} $\eta^{\rho[\beta}(K^{\alpha]})^\mu{}_\rho = -\eta^{\rho\mu}(K^{[\alpha})^{\beta]}{}_\rho $. Then, using the definition~\eqref{K_matrices_definition}, $\eta^{\rho[\beta}(K^{\alpha]})^\mu{}_\rho = 
-\eta^{\rho\mu} \, f^{[\beta\alpha]}{}_\rho = 
- \sfrac 1 2 \eta^{\rho\mu} \, c^{\alpha\beta}{}_\rho $, and
\begin{equation}
[\rho(\hat{t}^\alpha), \rho(\hat{t}^\beta)] 
=\left(
\begin{array}{c|c}
  \,c^{\alpha\beta}{}_\gamma  \, (K^\gamma)^\mu{}_\nu &  i\, \eta^{\rho\mu} \, c^{\alpha\beta}{}_\rho
\\
\hline
0 & 0
\end{array}
\right)
=
 i \,c^{\alpha\beta}{}_\gamma  \left(
\begin{array}{c|c}
 - i \, (K^\gamma)^\mu{}_\nu & \eta^{\mu\gamma}  
\\
\hline
0 & 0
\end{array}
\right) =  i \,c^{\alpha\beta}{}_\gamma  \, \rho(t^\gamma) \,.
\end{equation}
The representation I introduced is  faithful: in fact,
\begin{equation}
k_\mu \, \rho(\hat t^\mu) = q_\mu \, \rho(\hat t^\mu) \qquad \text{\textit{iff}} \qquad k_\mu = q_\mu \,,
\end{equation}
as follows from the fact that the top right corner of the matrix $\rho(\hat t^\mu)$ is a nonzero vector, $-\eta^{\mu\alpha}$, $\alpha = 0,\dots ,3$ (if, for some $\mu$, the vector was all zero, then the metric $\eta^{\mu\nu}$ would be degenerate), and the four vectors $\eta^{\mu\alpha}$, $\mu = 0,\dots ,3$  must be linearly independent for the metric to be nondegenerate.

A closer inspection reveals that I am representing the coordinate $\hat t^\mu $ as an element of (the 5-dimensional representation of) the Poincar\'e algebra:
\begin{equation}
\rho(\hat t^\mu)  =   \eta^{\mu\nu} \, \mP_\nu + {\sfrac i 2} \, \omega^{\mu\rho\sigma} \, \mM_{\rho\sigma} \,,
\end{equation}
where $\mP_\mu$ and $\mM_{\mu\nu}$ are defined in~\eqref{Eq:5DrepPoincare}.

Given an ordering choice for the $\x^\mu$ generators, and a corresponding family of noncommutative exponentials, I will use the same ordering to describe elements of $\mathcal{G}_\ell$ in a neighbourhood of the identity in terms of exponentials of $\hat t^\mu$:
\begin{equation}
: e^{i \, q_\mu \, \hat{t}^\mu}  : \,,
\end{equation}
where $: \, . \, :$  orders $\hat{t}^0$, $\hat{t}^1$, $\hat{t}^2$ and $\hat{t}^3$ in the same way it orders $\x^0$, $\x^1$, $\x^2$ and $\x^3$. 
So now $q_\mu \in \mathbbm{R}^4$ is a coordinate system around the identity of $\mathcal{G}_\ell$. Then, the group multiplication rule of $\mathcal{G}_\ell$ can be expressed in terms of a ``coproduct'' map (or \textit{momentum composition law}) $\Delta : \mathbbm{R}^4 \otimes \mathbbm{R}^4 \to \mathbbm{R}^4$, such that
\begin{equation}\label{Eq:Definition_Coproduct}
: e^{i \, p_\mu \, \hat{t}^\mu}  : \, : e^{i \, q_\mu \, \hat{t}^\mu}  :
~=~ : e^{i \, \Delta_\mu (p,q) \, \hat{t}^\mu}  : \,,
\end{equation}
which is associative, \textit{i.e.}
\begin{equation}
\Delta[p,\Delta(q,k)] = \Delta[\Delta(p,q), k] \,.
\end{equation}
The coordinates of the identity are identified by a ``counit'' $o \in \mathbbm{R}^4  $:
\begin{equation}
: \1 : ~=~ : e^{i \,o_\mu \, \hat{t}^\mu}  : \,, \qquad   o_\mu  = (0,0,0,0) \,,
\end{equation}
(notice that one might have introduced non-exponential coordinates on the group, which do not vanish at the identity, in which case the counit would not be the zero 4-vector, but rather the vector of the four values of the coordinates at the identity).
The counit is the neutral element for the coproduct:
\begin{equation}
\Delta(o,p) = \Delta(p,o) = p \,.
\end{equation}
The inverse is expressed in terms of an ``antipode'' map $S :\mathbbm{R}^4 \to \mathbbm{R}^4$, 
\begin{equation}\label{Eq:Definition_Antipode}
: e^{  i \, q_\mu \, \hat{t}^\mu}  : ~  : e^{i \, S _\mu(q) \, \hat{t}^\mu}  : 
~=~ 
: e^{  i \, S_\mu(q) \, \hat{t}^\mu}  : ~  : e^{i \, q_\mu \, \hat{t}^\mu}  : 
~=~  \1 \,,
\end{equation}
the property of being an inverse can be written in terms of the other two maps
\begin{equation}
\Delta(p,S(p)) = \Delta(S(p), p) = o \,.
\end{equation}
The antipode is a linear anti-homomorphism for the composition law:
\begin{equation}\label{AntipodeCoproductAntipode}
S[\Delta(k,q)] = \Delta[S(q), S(k)] \,,
\end{equation}
and the antipode leaves the $o$ element invariant:
\begin{equation}
S(o) = o \,.
\end{equation}
Moreover, if we introduced an involution on our coordinate algebra as in~\eqref{Involutions}, we can do the same on $\mathcal{A}_\ell$ by assuming $(t^\mu)^* =  t^\mu$, and the exponentials would transform as follows under its action:
\begin{equation}
\left(: e^{  i \, q_\mu \, \hat{t}^\mu}  : \right)^* = ~ : e^{  i \, S(q)_\mu \, \hat{t}^\mu}  : \,.
\end{equation} 

I am now ready to describe the Weyl system built from one ordering choice of the exponential functions~\eqref{GeneralOrderedExponentials}. The product of two exponentials follows the rule
\begin{equation}\label{ProductBetweenExponentials}
\E[p] \, \E[q] =  e^{i \, \theta^{\mu\nu} \, \Phi_{\mu\nu}(p,q)} \, \E[\Delta(p,q)] \,,
\end{equation}
where $\Phi_{\mu\nu}(p,q)$ is a cocycle with respect to the product $\Delta$  of the non-centrally-extended Lie group $\mathcal{G}_\ell$,
\begin{equation}\label{PhiCocycle}
\Phi_{\mu\nu} \left[p,q\right] 
+
\Phi_{\mu\nu} \left[\Delta(p,q),k\right]
 =
\Phi_{\mu\nu} \left[q,k\right]
+
\Phi_{\mu\nu} \left[p,\Delta(q,k)\right] \,, 
\end{equation}
which is equivalent to the cocycle condition~\eqref{StructureConstants_constraints} on $\theta^{\mu\nu}$. Eq.~\eqref{PhiCocycle} ensures that the product between three exponentials is associative, just like~\eqref{StructureConstants_constraints} ensures that the triple products between coordinates of $C_\ell[\mathbbm{R}^{3,1}]$ satisfies the associativity condition. One can deduce perturbatively the form of $\Phi_{\mu\nu}(p,q)$ from the Baker--Campbell--Hausdorff formula~\cite{achilles2012early} for the product~\eqref{ProductBetweenExponentials}, which also reveals that the additional phase due to the central extension has to be linear in $\theta^{\mu\nu}$. For example, in the case of Weyl-ordered exponential, the series expansion of $\Phi_{\mu\nu}(p,q)$ in powers of the structure constants of $\mathcal{A}_\ell$ is:
\begin{equation}
\Phi^\st{Weyl}_{\mu\nu}(p,q)=
{\sfrac 1 2} \, p_\mu \, q_\nu 
+ {\sfrac 1 {12}} \, c^{\rho\sigma}{}_\nu \, p_\rho \, q_\sigma (p_\mu-q_\mu) 
-{\sfrac 1 {24}} \, c^{\rho \lambda}{}_\nu c^{\sigma\kappa}{}_\lambda \, q_\mu \, p_\sigma \, q_\kappa \, p_\rho + \mathcal{O}(c^3) \,.
\end{equation}
The definition~\eqref{ProductBetweenExponentials} of the cocycle $\Phi_{\mu\nu}(p,q)$ implies that:
\begin{equation}
\Phi_{\mu\nu}(o,q) = \Phi_{\mu\nu}(p,o)  = 0 \,, \qquad
 \Phi_{\mu\nu}[p,S(p)]  =    \Phi_{\mu\nu}[S(p),p]  = 0\,,
\end{equation}
the second identity follows from the first through the following sequence, using the cocycle relation~\eqref{PhiCocycle}:
\begin{equation}
\begin{aligned}
0 = \Phi_{\mu\nu}(o,o) = \Phi_{\mu\nu}(\Delta[S(p),p],o) + \Phi_{\mu\nu}(S(p),p) 
\\
= \Phi_{\mu\nu}(S(p),\Delta[p,o]) + \Phi_{\mu\nu}(p,o)
 =\Phi_{\mu\nu}[S(p),p]  \,,
\end{aligned}
\end{equation}
and similarly for $ \Phi_{\mu\nu}[p,S(p)] $.

The last identity implies that the antipode of $\mathcal{A}_\ell$ provides the inverse map for the ordered exponential~\eqref{GeneralOrderedExponentials} too:
\begin{equation}
\E[p] \, \E[S(p)] = \E[S(p)] \, \E[p] = \1 \,,
\end{equation}
and the involution~\eqref{Involutions} acts on the ordered exponentials via the antipode map of $\mathcal{A}_\ell$, and therefore coincides with the inverse of the ordered exponential:
\begin{equation}
\E^*[p] = \E[S(p)] \,. 
\end{equation}

\subsection{Fourier transform and integral calculus}

Having a map $\Delta$ that codifies the product of $\mathcal{G}_\ell$, the components of the left- and right-invariant vector fields on $\mathcal{G}_\ell$ can be written explicitly,
\begin{equation}
(X^\mu_\st{L})_\nu (p) = \left. \frac{\partial \Delta_\nu (p,q)}{\partial q_\mu} \right|_{q=o} \,,
\qquad
(X^\mu_\st{R})_\nu (p) = \left. \frac{\partial \Delta_\nu (q,p)}{\partial q_\mu} \right|_{q=o} \,,
\end{equation}
these are vector fields on momentum space, which close two representations of the Lie algebra $\mathcal{A}_\ell$ under Lie brackets:
\begin{equation}
[ X^\mu_\st{L} , X^\nu_\st{L} ]  = - c^{\mu\nu}{}_\rho \, X^\rho_\st{L} \,,
\qquad
[ X^\mu_\st{R} , X^\nu_\st{R} ] =  c^{\mu\nu}{}_\rho \, X^\rho_\st{R} \,,
\end{equation}
and, moreover, Lie-commute with each other:
\begin{equation}
[ X^\mu_\st{L} , X^\nu_\st{R} ]  = 0 \,.
\end{equation}
We can introduce a basis of dual one-forms to these vector fields, the left- and right-invariant forms:
\begin{equation}
\omega^\st{L}_\mu \left[ X^\nu_\st{L} \right] = (\omega^\st{L}_\mu)^\rho \, ( X^\nu_\st{L} )_\rho = \delta^\nu{}_\mu \,, \qquad
\omega^\st{R}_\mu \left[ X^\nu_\st{R} \right] = (\omega^\st{R}_\mu)^\rho \, ( X^\nu_\st{R} )_\rho = \delta^\nu{}_\mu \,,
\end{equation}
these close the Maurer--Cartan identities,
\begin{equation}
d \omega^\st{L}_\mu = {\frac 1 2} \, c^{\rho\sigma}{}_\mu \, \omega^\st{L}_\rho \wedge \omega^\st{L}_\sigma \,, \qquad 
d \omega^\st{R}_\mu = - {\frac 1 2} \,  c^{\rho\sigma}{}_\mu \, \omega^\st{R}_\rho \wedge \omega^\st{R}_\sigma \,,
\end{equation}
and they can be used to define the left- and right-invariant Haar measures on $\mathcal{G}_\ell$:
\begin{equation}\label{Eq:Invariant_measures_definitions}
\begin{aligned}
&d\mu^\st{L} (p) = \omega^\st{L}_0 \wedge \omega^\st{L}_1 \wedge \omega^\st{L}_2 \wedge \omega^\st{L}_3 = \left| \det  (X^\mu_\st{L})_\nu (p) \right|^{-1} \, d^4 p = 
 \left|  \det \frac{\partial \Delta_\nu (p,q)}{\partial q_\mu} \right|_{q=o}^{-1} \, d^4 p \,,
 \\
 &d\mu^\st{R}  (p) = \omega^\st{R}_0 \wedge \omega^\st{R}_1 \wedge \omega^\st{R}_2 \wedge \omega^\st{R}_3 = \left| \det  (X^\mu_\st{R})_\nu (p) \right|^{-1} \, d^4 p = 
 \left|  \det \frac{\partial \Delta_\nu (q,p)}{\partial q_\mu} \right|_{q=o}^{-1} \, d^4 p \,,
\end{aligned}
\end{equation}
these are  respectively left- and right-invariant under the group multiplication
\begin{equation}
\left. d\mu^\st{L} [\Delta(q,p)] \right|_{q\text{ fixed}}= d\mu^\st{L} (p) \,, \qquad \left. d\mu^\st{R} [\Delta(p,q)]  \right|_{q\text{ fixed}} = d\mu^\st{R} (p) \,,
\end{equation}
it is instructive to prove explicitly these identities:
\begin{equation}
\begin{aligned}
\left. d\mu^\st{L} [\Delta(q,p)]  \right|_{q\text{ fixed}} &=   \left|  \det \partial_{k_\mu}\Delta_\nu  [\Delta (q,p),k]  \right|_{k=o}^{-1} \, \left|  \det \partial_{p_\mu}\Delta_\nu (q,p)  \right|\, d^4 p
  \\
  &=   \left|  \det \partial_{k_\mu}\Delta_\nu  [q,\Delta (p,k)]  \right|_{k=o}^{-1} \, \left|  \det \partial_{p_\mu}\Delta_\nu (q,p) \right|\, d^4 p 
  \\
  &=  \cancel{\left|  \det \partial_{r_\mu} \Delta_\nu  [q,r]  \right|_{r=p}^{-1}}\,
  \left|  \det \partial_{k_\mu}\Delta_\nu (p,k) \right|_{k=o}^{-1} \,\cancel{\left|  \det \partial_{p_\mu} \Delta_\nu (q,p) \right|}\, d^4 p
   \\
  &=  \,
  \left|  \det \partial_{k_\mu} \Delta_\nu (p,k) \right|_{k=o}^{-1}\, d^4 p  
  = d\mu^\st{L}(p) \,,
\end{aligned}
\end{equation}
and the proof for the right-invariant measure goes through analogously.
The two invariant measures are related by the following simple formula:
\begin{equation}\label{LeftRightHaarMeasureRelation}
d\mu^\st{L} (p) =    d\mu^\st{R} [S(p)]  \,, \qquad 
d\mu^\st{R} (p) =    d\mu^\st{L} [S(p)]  \,,
\end{equation}
to prove it, we can use relation~\eqref{AntipodeCoproductAntipode}:
\begin{equation}
\begin{aligned}
(X^\mu_\st{L})_\nu [S(p)] =& \left. \partial_{q_\mu}  \Delta_\nu (S(p),q) \right|_{q=o} 
= \left. \partial_{q_\mu}  S_\nu[\Delta(S(q),p)]\right|_{q=o} \,  
\\
=&  \left. \partial_{k_\rho}  S_\nu(k) \, \partial_{q_\mu} \Delta_\rho(S(q),p) \right|_{k=\Delta (S(q),p)}^{q=o} 
= \left. \partial_{k_\rho} S_\nu(k) \,  \partial_{r_\lambda} \Delta_\rho(r,p)  \, \partial_{q_\mu}  S_\lambda(q) \right|_{k=p}^{q=r=o} 
\\
=&  \left. \partial_{k_\rho} S_\nu(k) \,  (X_\st{R}^\lambda)_\rho (p) \, \partial_{q_\mu} S_\lambda(q)  \right|_{q=o,k=p} 
= -  \partial_{p_\rho} S_\nu(p) \, (X_\st{R}^\mu)_\rho (p)  \,,
\end{aligned}
\end{equation}
where I used $\left. \partial_\mu S_\nu(q)  \right|_{q=o} = - \delta^\mu{}_\nu$. This implies:
\begin{equation}
\left| \det  (X^\mu_\st{L})_\nu [S(p)] \right|^{-1} 
 =  \left| \det \partial_{p_\sigma} S_\rho(p)   \right|^{-1} \left| \det  (X^\mu_\st{R})_\nu (p) \right|^{-1}  \,,
\end{equation}
and since
\begin{equation}
d^4 S(p) = \left| \det \partial_{p_\sigma} S_\rho(p)   \right| d^4 p \,,
\end{equation}
Eq.~\eqref{LeftRightHaarMeasureRelation} follows.
I can now use the left- and right-invariant Haar measures to define the quantum Fourier transform in terms of two generalized Weyl systems in the sense of~\cite{Agostini:2002de}, as
\begin{equation}\label{LeftRightFourierTransformDefinition}
f(\x) = \int d \mu^\st{L} (p) \, \tilde{f}_\st{L} (p) \, \E[p] =\int d \mu^\st{R} (p) \, \tilde{f}_\st{R} (p) \, \E[p] \,, \qquad f(\x) \in C_\ell[\mathbbm{R}^{3,1}]\,,
\end{equation}
where $\tilde{f}_\st{L}$ and $\tilde{f}_\st{R}$ are two (commutative) Schwarz functions, related to each other by
\begin{equation}\label{LeftRightFourierTransformRelation}
d \mu^\st{L} (p) \, \tilde{f}_\st{L} (p) = d \mu^\st{R} (p) \, \tilde{f}_\st{R} (p) 
~~ \Rightarrow ~~
\left| \det  (X^\mu_\st{L})_\nu (p) \right|^{-1} \, \tilde{f}_\st{L}(p)
=
\left| \det  (X^\mu_\st{R})_\nu (p) \right|^{-1} \, \tilde{f}_\st{R}(p) \,.
\end{equation}
Notice that we are associating a noncommutative function $f(\x)$ to a commutative one, $\tilde{f}_\st{L}$ or $\tilde{f}_\st{R}$ (once one of the two is given, the other is completely specified through Eq.~\eqref{LeftRightFourierTransformRelation}), \textit{for all choices of ordering}. The Fourier transforms in different orderings are related via a change of coordinates in momentum space, according to Eq.~\eqref{OrderingChange}, so specifying $\tilde{f}_\st{L}$ (or $\tilde{f}_\st{R}$) in one ordering fixes it in all orderings. The ordering choice amounts, up to this point, to a choice of description, and a reasonable request could be that physics should be invariant under changes of this choice, which corresponds to a sort of principle of general covariance in momentum space, seeing as though changes of ordering coincide with general coordinate transformations in momentum space. The field theories I propose in the present paper can, in principle, be made compatible with such a covariance principle, as long as the physical observables that the theory admits (for example, correlation functions) are independent of the ordering choice. This is a nontrivial request, which ultimately should be checked in the quantum version of these field theories. For now, I will limit myself to observing that it is easy to write down T-Poincar\'e-invariant N-point functions for a free theory, by combining ordered plane waves, and those are independent of the ordering choice. The question remains whether a QFT defined on some T-Minkowski spacetime will be described by such N-point functions.

If the Lie group $\mathcal{G}_\ell$ is \textit{unimodular,} then the left- and right-invariant Haar measures coincide, and so do the left- and right- Fourier transforms $\tilde{f}_\st{L}$ and $\tilde{f}_\st{R}$.

Using $\E^*[p]=\E[S(p)]$ and Eq.~\eqref{LeftRightFourierTransformRelation}, we can prove
\begin{equation}
f^*(\x)  
 = \int d \mu^\st{L} (p) \, \overline{\tilde{f}_\st{L}(p)} \, \E[S(p)] =  
 \int d \mu^\st{L} [S(q)] \, \overline{\tilde{f}_\st{L}[S(q)]}  \, \E[q] 
 =  
 \int d \mu^\st{R}(q) \, \overline{\tilde{f}_\st{L}[S(q)]}   \, \E[q]  \,,
\end{equation}
and, similarly, $f^*(\x)  = 
 \int d \mu^\st{L}(q) \, \overline{\tilde{f}_\st{R}[S(q)]}   \, \E[q]$. This gives us the Fourier transform of a conjugate  function $f^*(\x)$, which exchanges the role of the left- and right-Fourier coefficients:
\begin{equation}\label{FourierTransformStarredFunction}
\widetilde{f^*}_\st{L}(p) = \overline{\tilde{f}_\st{R}[S(p)]}  
\,, 
\qquad
\widetilde{f^*}_\st{R}(p) = \overline{\tilde{f}_\st{L}[S(p)]}  \,.
\end{equation}
For a function to be real, the following condition needs to be satisfied:
\begin{equation}\label{FunctionRealityConditions}
\tilde{f}_\st{L}(p) = \overline{\tilde{f}_\st{R}[S(p)]}  
\,,
\end{equation}
which, since $S[S(p)]=p$, can also be cast in the form $\tilde{f}_\st{R}(p)= \overline{\tilde{f}_\st{L}[S(p)]} $.

\subsection{Noncommutative integral and convolution}

We can now introduce a notion of \textit{definite integral} (over all of spacetime) on noncommutative functions as a linear functional on $C_\ell[\mathbbm{R}^{3,1}]$:
\begin{equation}
 \int  (~.~) \, d^4\x : C_\ell[\mathbbm{R}^{3,1}] \to \mathbbm{C} \,,
\end{equation}
defined on the ordered exponentials as
\begin{equation}\label{NCintegralDefinition1}
\int  \, \E[p] \, d^4\x = \delta^{(4)}(p) \,,
\end{equation}
this implies
\begin{equation}\label{NCintegralDefinition2}
\int  \, f(\x) \, d^4\x =   \tilde{f}_\st{L}(o) =   \tilde{f}_\st{R}(o) \,, 
\end{equation}
which is possible because $(X^\mu_\st{L})_\nu (o) = (X^\mu_\st{L})_\nu (o) = \delta^\mu{}_\nu$, and therefore~\eqref{LeftRightFourierTransformRelation} implies $\tilde{f}_\st{L}(o) =   \tilde{f}_\st{R}(o)$. Notice also that the definition~\eqref{NCintegralDefinition1}, despite being written in terms of a particular basis of ordered exponentials, does not depend on the ordering choice. If we have a second ordering choice, in general the exponential functions ordered in the two ways will be related by Eq.~\eqref{OrderingChange}, repeated here for convenience:
\begin{equation}\label{OrderingChange_repeat}
\E_1[p] = e^{i \, \theta^{\mu\nu} \, \varphi^{12}_{\mu\nu}(q)} \, \E_2[q]  \,,
\end{equation}
where $q_\mu = q_\mu(p)$ is a function of the original momentum $p$.
Then the noncommutative integral, calculated on $\E_2[q]$, gives 
\begin{equation}
\begin{aligned}
\int  \, \E_2[p] \, d^4\x =&  \int  \,  e^{i \, \theta^{\mu\nu} \, \varphi^{12}_{\mu\nu}(p)} \, \E_1[p(q)] \, d^4\x       
= e^{i \, \theta^{\mu\nu} \, \varphi^{12}_{\mu\nu}(o)} \, \left|\det \frac{\partial p_\mu}{\partial q_\nu} \right|^{-1}_{q=o} \, \delta^{(4)}(q) \,,
\end{aligned}
\end{equation}
but, since $\E_1[o] = \E_2[o] = \1$, we know that $
\varphi^{12}_{\mu\nu}(o) = 0$, and $\left. \det  (\partial p_\mu/\partial q_\nu ) \right|_{q=o} = \delta^\nu{}_\mu $, so the integral is the same on all ordered exponentials
\begin{equation}
\int  \, \E_2[p] \, d^4\x =  \int  \, \E_1[p] \, d^4\x = \delta^{(4)}(p) \,.
\end{equation}
The integral is an involutive map, in fact, using Eq.~\eqref{FourierTransformStarredFunction} and the fact that $S(o) =o$,
\begin{equation}\label{IntegralInvolutive}
\int f^*(\x) \, d^4 \x = 
\widetilde{f^*}_\st{L}(o) = \widetilde{f^*}_\st{R}(o) = \overline{\tilde{f}_\st{L}(o)}  = \overline{\tilde{f}_\st{R}(o)}  = \overline{\left( \int f(\x) \, d^4 \x  \right)} \,.
\end{equation}
The integral allows us to define the inverse of the left- and right-Fourier transform as
\begin{equation}
\tilde{f}_\st{L} (p) = {\frac 1 {(2\pi)^4}} \, \int \, \E^*[p] \, f(\x) \,  d^4\x \,, \qquad
\tilde{f}_\st{R} (p) = {\frac 1 {(2\pi)^4}} \, \int  \, f(\x)  \, \E^*[p]\,  d^4\x \,, 
\end{equation}
the reason why these are the inverse of~\eqref{LeftRightFourierTransformDefinition} is by no means trivial. For example:
\begin{equation}
\begin{aligned}
& \int \frac{ \E^*[p] \, f(\x)}{(2\pi)^4}  d^4\x  =   \int \int {\frac{\tilde{f}_\st{L}(q)\,   \E^*[p] \,   \E[q]}{(2\pi)^4}}  d^4\x \, d\mu^\st{L}[q]
=
  \int {\frac{\tilde{f}_\st{L}(q)}{(2\pi)^4}}\, \int  \E[\Delta[S(p),q]] \,  d^4\x \, d\mu^\st{L}[q]
=
\\
&\qquad =  \int {\frac{\tilde{f}_\st{L}(q)}{(2\pi)^4}} \, \delta^{(4)}(\Delta[S(p),q]) \, d\mu^\st{L}[q]
=
 \int {\frac{\tilde{f}_\st{L}(q)}{(2\pi)^4}} \, \left| \det \partial_{q_\nu}  \Delta_\mu[S(p),q] \right|_{q=p}^{-1} \, \delta^{(4)}(q-p) \, d\mu^\st{L}[q]
=
\\
&\qquad =  
\tilde{f}_\st{L}(p) \, \left| \det \partial_{q_\nu}\Delta_\mu[S(p),q] \right|^{-1}_{q=p} \, \left|  \det \partial_{k_\mu} \Delta_\nu (p,k) \right|_{k=o}^{-1} = 
\\
&\qquad =  
\tilde{f}_\st{L}(p) \, \left| \det \partial_{k_\rho} \Delta_\mu[S(p),\Delta (p,k)]  \right|_{k=o}^{-1} 
= 
\tilde{f}_\st{L}(p) \, \left| \det \partial_{k_\rho}  \Delta_\mu[\Delta (S(p),p),k)] \right|_{k=o}^{-1} 
\\
&\qquad =  
\tilde{f}_\st{L}(p)  \,,
\end{aligned}
\end{equation}
the proof goes through similarly for the right-Fourier transform.

The integral, and the inverse Fourier transform, allow me to introduce a deformed notion of convolution. For the left-ordered Fourier transforms, this takes the form
\begin{equation}
\begin{aligned}
\widetilde{(f \, g)}_\st{L} (p) =& \frac{1}{(2\pi)^4}
\int \E^*[p] \,  f(\x) \, g(\x) \, d^4\x
\\
=&
 \frac{1}{(2\pi)^4}
\int \int  \tilde{f}_\st{L}(q) \, \tilde{f}_\st{L}(k) \left( \int \E^*[p] \,  \E[q] \, \E[k] \, d^4\x \right) d\mu^\st{L}(q)\, d\mu^\st{L}(k)
\\
=&
 \frac{1}{(2\pi)^4}
\int \int \tilde{f}_\st{L}(q) \, \tilde{f}_\st{L}(k)  \, e^{i \,\theta^{\mu\nu} \left( \Phi_{\mu\nu} (q,k) +\Phi_{\mu\nu} [S[p],\Delta(q,k)] \right)} \, \delta^{(4)} \left(\Delta[S(p),q,k]\right)  \, d\mu^\st{L}(q)\, d\mu^\st{L}(k)
\\
=&
 \frac{1}{(2\pi)^4}
\int  \tilde{f}_\st{L}(q) \, \tilde{f}_\st{L}(\Delta[S(q),p])  \, e^{i \,\theta^{\mu\nu} \left( \Phi_{\mu\nu} (q,\Delta[S(q),p]) +\Phi_{\mu\nu} [S[p],p] \right)} \, \delta^{(4)} \left(k - \Delta[S(q),p] \right) 
\\
&\qquad \qquad  
\left|  \det \partial_{k_\mu}   \Delta_\nu (S(p),q,k)  \right|_{k=\Delta[S(q),p]}^{-1}
\left|  \det \partial_{r_\mu}   \Delta_\nu (S(q),p,r)  \right|_{r=o}^{-1}
d\mu^\st{L}(q)  \, d^4k\\
=&
 \frac{1}{(2\pi)^4}
\int  \tilde{f}_\st{L}(q) \, \tilde{f}_\st{L}(\Delta[S(q),p])  \, 
e^{i \,\theta^{\mu\nu} \left( \Phi_{\mu\nu} (\Delta[q,S(q)],p)
-\Phi_{\mu\nu} (S[q],p) +\Phi_{\mu\nu} (q,S[q])
\right)}
\, \delta^{(4)} \left(k - \Delta[S(q),p] \right) 
\\
&\qquad \qquad  
\left|  \det \partial_{r_\mu}   \Delta_\nu (S(p),q,\Delta[S(q),p,r])   \right|_{r=o}^{-1}
d\mu^\st{L}(q)  \, d^4k
\\
=&\frac{1}{(2\pi)^4}
\int  \tilde{f}_\st{L}(q) \, \tilde{f}_\st{L}(\Delta[S(q),p])  \, e^{-i \,\theta^{\mu\nu} \Phi_{\mu\nu} (S[q],p)} \, d\mu^\st{L}(q)
\end{aligned}
\end{equation} 
and, in the case of right-ordered Fourier transforms:
\begin{equation}
\begin{aligned}
\widetilde{(f \, g)}_\st{R} (p) =& \frac{1}{(2\pi)^4}
\int f(\x) \, g(\x) \, \E^*[p] \, d^4\x
\\
=&
 \frac{1}{(2\pi)^4}
\int \int \tilde{f}_\st{R}(q) \, \tilde{f}_\st{R}(k) \left(  \int \E[q] \, \E[k] \, \E^*[p] \, d^4\x \right) d\mu^\st{R}(q)\, d\mu^\st{R}(k)
\\
=&
 \frac{1}{(2\pi)^4}
\int \int \tilde{f}_\st{R}(q) \, \tilde{f}_\st{R}(k)  \, e^{i \,\theta^{\mu\nu} \left( \Phi_{\mu\nu} (q,k) +\Phi_{\mu\nu} [\Delta(q,k),S(p)] \right)} \, \delta^{(4)} \left(\Delta[q,k,S(p)]\right)  \, d\mu^\st{R}(q)\, d\mu^\st{R}(k)
\\
=&
 \frac{1}{(2\pi)^4}
\int  \tilde{f}_\st{R}(\Delta[p,S(k)] ) \, \tilde{f}_\st{R}(k)  \,  e^{i \,\theta^{\mu\nu} \left( \Phi_{\mu\nu} (\Delta[p,S(k)],k) +\Phi_{\mu\nu} [\Delta(\Delta[p,S(k)],k),S(p)] \right)}\, \delta^{(4)} \left(q - \Delta[p,S(k)] \right) 
\\
&\qquad \qquad  
\left|  \det \partial_{q_\mu}   \Delta_\nu (q,k,S(p))  \right|_{q=\Delta[p,S(k)] }^{-1}
\left|  \det \partial_{r_\mu}   \Delta_\nu (r,\Delta[p,S(k)])  \right|_{r=o}^{-1}
d\mu^\st{R}(k)  \, d^4q
\\
=&
 \frac{1}{(2\pi)^4}
\int  \tilde{f}_\st{R}(\Delta[p,S(k)] ) \, \tilde{f}_\st{R}(k)  \, e^{i \,\theta^{\mu\nu} \left( \Phi_{\mu\nu} (\Delta[p,S(k)],k) +\Phi_{\mu\nu} [p,S(p)] \right)} \, \delta^{(4)} \left(q - \Delta[p,S(k)] \right) 
\\
&\qquad \qquad  
\left|  \det \partial_{r_\mu}   \Delta_\nu (\Delta[r,p,S(k)],k,S(p))  \right|_{r=o}^{-1}
d\mu^\st{R}(k)  \, d^4q
\\
=&
 \frac{1}{(2\pi)^4}
\int  \tilde{f}_\st{R}(\Delta[p,S(k)] ) \, \tilde{f}_\st{R}(k)  \, e^{- i \,\theta^{\mu\nu} \,\Phi_{\mu\nu} [p, S(k)]}\,
d\mu^\st{R}(k)  \,.
\end{aligned}
\end{equation}

The integral defined in~\eqref{NCintegralDefinition1} and~\eqref{NCintegralDefinition2} \textit{is not, in general, cyclic.} In fact
\begin{equation}\label{IntegralTwoFunctionsFourierTransform}
\begin{aligned}
 &\int \, f(\x) \, g(\x) \, d^4\x  =   \int \int \, \tilde{f}_\st{L}(p) \, \tilde{g}_\st{L}(q)  \int e^{i \, \theta^{\mu\nu} \, \Phi_{\mu\nu}(p,q)}  \, \E[\Delta(p,q)] \, d^4\x   \, d \mu^\st{L}(p)\, d \mu^\st{L}(q) 
\\
&=   \int   \, \tilde{f}_\st{L}(p) \, \tilde{g}_\st{L}(q) \, e^{i \, \theta^{\mu\nu} \, \Phi_{\mu\nu}[p,S(p)]}\,\delta^{(4)}[\Delta(p,q)] \,   d \mu^\st{L}(p)\, \left|  \det \partial_{k_\mu}   \Delta_\nu (q,k)  \right|_{k=o}^{-1} \, d^4 q
\\
&=  (2\pi)^4 \,   \int \, \tilde{f}_\st{L}(p) \, \tilde{g}_\st{L}[S(p)] \left|\det \partial_{q_\nu}  \Delta_\mu(p,q) \right|^{-1}_{q=S(p)} \, \left|  \det \partial_{k_\mu}   \Delta_\nu [S(p),k]  \right|_{k=o}^{-1} \,  d \mu^\st{L}(p)
\\
&=  (2\pi)^4 \,   \int \, \tilde{f}_\st{L}(p) \, \tilde{g}_\st{L}[S(p)] \left|\det \left(\left.\partial_{q_\rho}   \Delta_\mu(p,q) \right|_{q=\Delta [S(p),k]} \,\partial_{k_\nu}   \Delta_\rho [S(p),k]  \right) \right|_{k=o}^{-1} \,  d \mu^\st{L}(p)
\\
&=  (2\pi)^4 \,   \int \, \tilde{f}_\st{L}(p) \, \tilde{g}_\st{L}[S(p)] \left|\det  \partial_{k_\nu}  \Delta_\mu(p,\Delta [S(p),k])    \right|_{k=o}^{-1} \,  d \mu^\st{L}(p)
\\
&=  (2\pi)^4 \,   \int \, \tilde{f}_\st{L}(p) \, \tilde{g}_\st{L}[S(p)] \left|\det   \partial_{k_\nu}  \Delta_\mu(\Delta[p,S(p)],k) \right|_{k=o}^{-1} \,  d \mu^\st{L}(p) 
\\
&=  (2\pi)^4 \,   \int \, \tilde{f}_\st{L}(p) \, \tilde{g}_\st{L}[S(p)] \,  d \mu^\st{L}(p) \,,
\end{aligned} 
\end{equation}
changing variables $p = S(k)$,  
\begin{equation}
 \int \, \tilde{f}_\st{L}(p) \, \tilde{g}_\st{L}[S(p)] \,  d \mu^\st{L}(p)
= \int \, \tilde{f}_\st{L}[S(p)] \, \tilde{g}_\st{L}(p) \,  d \mu^\st{L}[S(p)]
=  \int \, \tilde{g}_\st{L}(p)\, \tilde{f}_\st{L}[S(p)]  \,  d \mu^\st{R}(p)\,,
\end{equation}
which is equal to  $(2\pi)^{-4} \,  \int \, g(\x) \, f(\x) \, d^4\x $ only if $d \mu^\st{R}(p) = d \mu^\st{L}(p)$, \textit{i.e.}, if the left- and right-invariant measures coincide, which means that the Lie group $\mathcal{G}_\ell$ is unimodular. The same proof goes through similarly if we choose to represent the noncommutative functions $f(\x)$ and $g(\x)$ through their right-Fourier transforms, in this case the Fourier representation of the integral is
\begin{equation}
\int \, f(\x) \, g(\x) \, d^4\x  = (2\pi)^4 \,   \int \, \tilde{f}_\st{R}[S(p)]  \, \tilde{g}_\st{R}(p) \,  d \mu^\st{R}(p) \,.
\end{equation}
Notice that we now have an explicit expression for the ``twisted-cyclic'' property of the integral in the case of non-unimodular $\mathcal{G}_\ell$ groups:
\begin{equation} \label{Eq:TwistedCiclicity}
\int f(\x) g(\x) d^4\x = \int \left[ \mathscr{T} \triangleright g(\x) \right] f(\x) d^4\x
= \int g(\x) \left[ \mathscr{T}^{-1} \triangleright f(\x) \right] d^4\x \,,
\end{equation}
where $ \mathscr{T} : C_\ell[\mathbbm{R}^{3,1}] \to C_\ell[\mathbbm{R}^{3,1}]$, is   a linear operator
defined through the following integral kernel:
\begin{equation}
\mathscr{T} \triangleright f(\x)  = \int \left[ \mathscr{T}(p)\, \tilde{f}_\st{L}(p) \right] d\mu^\st{L}(p)  \,,
\end{equation}
where
\begin{equation}\label{Eq:acyclicity_momentum_expression}
\mathscr{T}(p) = \frac{d\mu^\st{R}(p)}{d\mu^\st{L}(p)} =  \frac{\left|  \det \frac{\partial \Delta_\nu (p,r)}{\partial r_\mu} \right|_{r=o}}{\left|  \det \frac{\partial \Delta_\nu (s,p)}{\partial s_\mu} \right|_{s=o}}  \,,
\end{equation}
which, of course, reduces to the identity operator in unimodular cases. The eigenvalue of $\mathscr{T}$ on a plane wave, $\mathscr{T}(p)$ is the so-called \textit{modular function,} a well-known object in harmonic analysis~\cite{deitmar2014_Harmonic_Analysis_book}, see \textit{e.g.}~\cite{Hersent:2020lsr} for a discussion of the modular function within the context of integrations on Lie-algebra-type noncommutative spacetimes.
Notice that this definition allows us to rewrite the relation~\eqref{LeftRightFourierTransformRelation} between the  left and right Fourier transform as
\begin{equation}\label{LeftRightFourierTransformRelation2}
\tilde{\phi}_\st{L} (p)  = \mathscr{T}(p) \, \tilde{\phi}_\st{R} (p)  \,,
\end{equation}
The modular function satisfies the following identities
\begin{equation}
\mathscr{T}[\Delta(p,q)] = \mathscr{T}(p) \, \mathscr{T} (q) \,, \qquad
\mathscr{T}[S(p)] = \mathscr{T}(p)^{-1} \,, \qquad \mathscr{T}[o] = 1 \,,
\end{equation}
which makes $\mathscr{T}$ a group-like element of the Hopf subalgebra of momenta of the T-Poincar\'e algebra (see Sec.~\ref{Sec:T-Poincare_algebra}). The properties above can be deduced directly from Eq.~\eqref{Eq:TwistedCiclicity}. Replacing $g(\x)$ with the identity, for example, yields:
\begin{equation}
\int  f(\x) \, \1 \, d^4\x = \int (\mathscr{T} \triangleright \1) \, f(\x) \, d^4\x ~~ \Rightarrow ~~ \mathscr{T} \triangleright \1 = \1 ~~ \Rightarrow ~~ \mathscr{T}(o) = 1 \,.
\end{equation}
Applying~\eqref{Eq:TwistedCiclicity} to three functions, and exploiting the associativity of the product between functions:
\begin{equation}
\begin{aligned}
\int f(\x)  \, g(\x) \, h(\x) \, d^4\x =&  
\int  \left\{ \mathscr{T} \triangleright [ g(\x) \, h(\x) ] \right\} f(\x)   \, d^4\x 
=
\int \left[ \mathscr{T} \triangleright h(\x) \right] \,f(\x)  \, g(\x) d^4\x \\
=&
\int \left[ \mathscr{T} \triangleright g(\x) \right] \left[ \mathscr{T} \triangleright h(\x) \right] \,f(\x)  d^4\x 
\,,
\end{aligned}
\end{equation}
which implies that
\begin{equation}
\mathscr{T} \triangleright [ g(\x) \, h(\x) ] = [\mathscr{T} \triangleright g(\x) ] [ \mathscr{T} \triangleright h(\x) ] \,.
\end{equation}
One can also prove these properties directly from the definition~\eqref{Eq:acyclicity_momentum_expression}.

The associativity of the coproduct $\Delta(p,q)$ can be exploited to write integrals of multiple functions in a simple way. For example:
\begin{equation}
\begin{aligned}
 \int \, f(\x) \, g(\x) \, h(\x) d^4\x  &=   \int \int\int \, \tilde{f}_\st{L}(p) \, \tilde{g}_\st{L}(q) \, \tilde{h}_\st{L}(k) \int e^{i \, \theta^{\mu\nu} \, \Phi_{\mu\nu}(p,q) + i \, \theta^{\mu\nu} \, \Phi_{\mu\nu}[\Delta(p,q),k]}  \, \cdot
\\
& \qquad \qquad \qquad \cdot \left( \int \E[\Delta(p,q,k)] \, d^4\x   \right) \, d \mu^\st{L}(p)\, d \mu^\st{L}(q)\, d \mu^\st{L}(k) 
\\
&=  \int \int \int\, \tilde{f}_\st{L}(p) \, \tilde{g}_\st{L}(q) \, \tilde{h}_\st{L}(k) \, \int e^{i \, \theta^{\mu\nu} \, \Phi_{\mu\nu}(p,q) + i \, \theta^{\mu\nu} \, \Phi_{\mu\nu}[\Delta(p,q),k]} \cdot
\\
& \qquad \qquad \qquad \cdot \delta^{(4)}[\Delta(p,q,k)] \,   d \mu^\st{L}(p)\,   d \mu^\st{L}(q)\, \left|  \det \partial_{r_\mu}\Delta_\nu (k,r) \right|_{r=o}^{-1} \, d^4 k
\\
&=  \int \int \, \tilde{f}_\st{L}(p) \, \tilde{g}_\st{L}(q) \, \tilde{h}_\st{L}(S[\Delta(p,q)]) \, \left| \det \partial_{k_\nu} \Delta_\mu[\Delta(p,q),k) \right|_{k=S[\Delta(p,q)]}  \cdot
\\
& \qquad \qquad \qquad \cdot  \left|  \det \partial_{r_\mu} \Delta_\nu (S[\Delta(p,q)],r) \right|_{r=o}^{-1} \,  e^{i \, \theta^{\mu\nu} \, \Phi_{\mu\nu}(p,q)} \,  d \mu^\st{L}(p)\,   d \mu^\st{L}(q)
\\
&=  \int \int \, \tilde{f}_\st{L}(p) \, \tilde{g}_\st{L}(q) \, \tilde{h}_\st{L}(S[\Delta(p,q)]) \,  e^{i \, \theta^{\mu\nu} \, \Phi_{\mu\nu}(p,q)} \cdot
\\
& \qquad \qquad \qquad \cdot  \left| \det \partial_{r_\nu}  \Delta_\mu[\Delta(p,q),\Delta (S[\Delta(p,q)],r))\right|_{r=o}^{-1} \,   d \mu^\st{L}(p)\,   d \mu^\st{L}(q)
\\
&=   \int  \int e^{i \, \theta^{\mu\nu} \, \Phi_{\mu\nu}(p,q)} \, \tilde{f}_\st{L}(p) \, \tilde{g}_\st{L}(q) \, \tilde{h}_\st{L}(S[\Delta(p,q)]) ~ d \mu^\st{L}(p)\,   d \mu^\st{L}(q) \,,
\end{aligned} 
\end{equation}
where I used the notation $\Delta(p,q,k)= \Delta[\Delta(p,q),k] = \Delta[p,\Delta(q,k)]$. The integral of a quadruple product can be written:
\begin{equation}
\begin{aligned}
 &\int \, f(\x) \, g(\x) \, h(\x) \, j(\x) \, d^4\x  = 
\\
&=   \int  \int \int e^{i \, \theta^{\mu\nu} \left( \Phi_{\mu\nu}(p,q) +  \Phi_{\mu\nu}[\Delta(p,q),k] \right)}  \, \tilde{f}_\st{L}(p) \, \tilde{g}_\st{L}(q) \, \tilde{h}_\st{L}(k) \, \tilde{j} (S[\Delta(p,q,k)]) ~ d \mu^\st{L}(p)\,   d \mu^\st{L}(q)\,   d \mu^\st{L}(k) \,,
\end{aligned} 
\end{equation}
and so on.

\subsection{Lorentz transforms of momenta and Poincar\'e-invariance of the integral}

In Appendix~\ref{Appendix:Lorentz_transform_momenta} I prove that, under the coaction~\eqref{Coaction_spacetime_coordinates} of the  quantum Poincar\'e group, the ordered exponentials transform as follows
\begin{equation}\label{PoincareTransform_of_Exponential}
\E'[p] = ~ : \, e^{i \, p_\mu \, x'^\mu} \, : ~ = e^{i \, \theta^{\mu\nu} \, \sigma_{\mu\nu} (p,\La) \otimes \1} ~: \, e^{i \, \lambda_\mu (p,\La) \otimes \x^\mu} \, :  \, : \, e^{i \, p_\mu \, \A^\mu  \otimes \1} \,:\,,
\end{equation} 
where I chose to order the translation coordinates $\A^\mu$ to the right of the Lorentz matrix components $\La^\mu{}_\nu$. The coordinates $\A^\mu$  in the  exponential $: \, e^{i \, p_\mu \, \A^\mu  \otimes \1} \,:$ end up ordered among each other in the same way as the coordinates $\x^\mu$ (and this is the meaning of the ordering symbol $: \, . \, :$  when applied on $\A^\mu$ coordinates). $\lambda_\mu(p,\La)$ and $\sigma_{\mu\nu} (p,\La)$ are two (not necessarily linear) functions of $\La^\mu{}_\nu$ and the momentum $p_\mu$. $\lambda_\mu(p,\La)$  describes the transformation law of momenta under Lorentz transformations. In the commutative limit, the functions $\lambda_\mu$  reduce to
\begin{equation}
\lambda_\mu(p,\La) \xrightarrow[\ell \to 0]{} p_\nu \La^\nu{}_\mu \,,
\end{equation}
which is the usual linear Lorentz transformation of the momentum $p_\mu$. Given an ordering choice, $\lambda_\mu(p,\La)$ can be written, in general, as a Taylor series in $\La^\mu{}_\nu$, $p_\mu$ and the structure constants $f^{\mu\nu}{}_\rho$, but the full analytic expression of $\lambda_\mu(p,\La)$ can only be calculated, once a specific model is chosen, as the solution to some differential equations (see Appendix~\ref{Appendix:Lorentz_transform_momenta}). We can, however, list a series of properties that $\lambda_\mu(p,\La)$ needs to have, just by consistency with the homomorphism property of the coaction $( \, \cdot \,)'$ and the Hopf-algebra structure of $C_\ell[ISO(3,1)]$:
\begin{itemize}
\item $\lambda_\mu(p,\La)$ is a (in general nonlinear) representation of the Lorentz group:
\begin{equation}
\lambda \left[ \lambda(p,\La),\La' \right] = \lambda_\mu(p,\La^\nu{}_\rho \La'^\rho{}_\mu) \,,
\end{equation}
or, written in terms of the coproduct of $C_\ell[ISO(3,1)]$, $\lambda(\lambda(p,\La\otimes \1),\1 \otimes \La) = \lambda (p, \Delta[\La])$.
 The proof goes as follows: applying twice the Poincar\'e-transform homomorphism
$\x''^\mu = \La^\mu{}_\nu \otimes \La^\nu{}_\rho \otimes \x^\rho + \La^\mu{}_\nu \otimes \A^\nu \otimes \1 +   \A^\mu \otimes \1 \otimes \1 = \Delta[\La^\mu{}_\nu] \otimes \x^\nu + \Delta[\A^\mu] \otimes \1
$ to the exponential:
\begin{equation} 
\begin{aligned}
\E''[p] =&  e^{i \, \theta^{\mu\nu} \, \sigma_{\mu\nu} (p,\La) \otimes \1 \otimes \1} ~: \, e^{i \, \lambda_\mu (p,\La) \otimes x'^\mu} \, :  \, : \, e^{i \, p_\mu \, \A^\mu  \otimes \1 \otimes \1} \,: \\
=&
e^{i \, \theta^{\mu\nu} \, \sigma_{\mu\nu} (p,\La) \otimes \1 \otimes \1 + i \, \theta^{\mu\nu} \, \sigma_{\mu\nu} (\lambda(p,\La\otimes \1),\1 \otimes \La)  \otimes \1} ~: \, e^{i \, \lambda_\mu (\lambda(p,\La \otimes \1) , \1 \otimes \La) \otimes \x^\mu} \, :  
\\
& : \, e^{i \, \lambda_\mu(p,\La \otimes \1) \otimes \A^\mu  \otimes \1  } \, :  \,: \, e^{i \, p_\mu \, \A^\mu  \otimes \1 \otimes \1} : 
\\
=&
(\Delta \otimes \text{\it id} ) \left( E'[p] \right) = e^{i \, \theta^{\mu\nu} \, \sigma_{\mu\nu} (p,\Delta[\La]) \otimes \1} ~: \, e^{i \, \lambda_\mu (p,\Delta[\La]) \otimes \x^\mu} \, :  \, : \, e^{i \, p_\mu \, \Delta[\A^\mu]  \otimes \1} \,: \,,
\end{aligned}
\end{equation} 
and isolating the $\x^\mu$-dependent terms, we deduce that
\begin{equation}
\lambda_\mu [p,\Delta(\La)] = \lambda_\mu \left[\lambda(p,\La \otimes \1) , \1 \otimes \La \right]\,,
\end{equation}
and
\begin{equation}
\begin{aligned}
: \, e^{i \, p_\mu \, \Delta[\A^\mu]  \otimes \1} \,:
=&
e^{i \, \theta^{\mu\nu} \left[ \sigma_{\mu\nu} (p,\La) \otimes \1 +  \sigma_{\mu\nu} (\lambda(p,\La\otimes \1),\1 \otimes \La)  -  \sigma_{\mu\nu} (p,\Delta[\La]) \right]  \otimes \1 } 
\\
&: \, e^{i \, \lambda_\mu(p,\La \otimes \1) \otimes \A^\mu  \otimes \1  } \, :  \,: \, e^{i \, p_\mu \, \A^\mu  \otimes \1 \otimes \1} : 
 \,,
 \end{aligned}
\end{equation}
which is compatible with the commutation relations between $\A^\mu$ and $\La^\mu{}_\nu$, Eq.~\eqref{T-PoincareGroup_commutators}.
\item $\lambda^{-1}(p,\La) =\lambda(p,\La^{-1})  $. This is a straightforward consequence of the representation property $\lambda \left[ \lambda(p,\La),\La' \right] = \lambda_\mu \left[ p,\La^\nu{}_\rho \La'^\rho{}_\mu \right]$.

\item  $\lambda_\mu(p,\delta) = p_\mu$, and $\sigma_{\mu\nu}(p,\delta) = 0$, where with $\delta$ I mean the identity of the Lorentz group. This  can be seen by acting on both sides of the identity~\eqref{PoincareTransform_of_Exponential} with the map $\varepsilon \otimes \text{\it id}$, where $\varepsilon$ is the counit of the T-Poincar\'e Hopf algebra, $\varepsilon(\La^\mu{}_\nu) =  \delta^\mu{}_\nu$, $\varepsilon(\A^\mu) = 0$. Then one gets
\begin{equation}
:e^{i \, p_\mu \, \x^\mu} := e^{i \, \theta^{\mu\nu} \, \sigma_{\mu\nu} (p,\delta) } : e^{i\, \lambda_\mu (p,\delta)\, \x^\mu}   : \,.
\end{equation}

\item  $\lambda_\mu(p,\La) = 0  ~~ \Leftrightarrow ~~ p = o$. The $\Leftarrow$ implication can be proved by taking  the limit $p \to o$ in both sides of~\eqref{PoincareTransform_of_Exponential}.
Suppose now that there exists a $p$, call it \textit{e.g.} ${\bar p}$, such that $\lambda_\mu({\bar p},\La) = 0 $. Then the identity~\eqref{PoincareTransform_of_Exponential}, for this value of $p_\mu$, turns into
\begin{equation}
: \, e^{i \, {\bar p}_\mu \, \La^\mu{}_\nu \otimes  \x^\nu + i \, {\bar p}_\mu \, \A^\mu  \otimes \1} \, : ~ = e^{i \, \theta^{\mu\nu} \, \sigma_{\mu\nu} ({\bar p},\La) \otimes \1} ~   \, : \, e^{i \, {\bar p}_\mu \, \A^\mu  \otimes \1} \,: \,,
\end{equation}
we can apply the map $\varepsilon \otimes \text{\it id}$ to both sides above, and get:
\begin{equation}
: e^{i \, {\bar p}_\mu \,  \x^\mu } = \1 \,,
\end{equation}
which of course implies $\bar p = o$.

\end{itemize}

Consider now a noncommutative scalar function $f(\x')$ which has been acted upon by the coaction~\eqref{Coaction_spacetime_coordinates} of the T-Poincar\'e group:
\begin{equation}
f(\x') = \int d\mu^\st{L}(p) \, \tilde{f}_\st{L}(p) \, \E'[p]
= \int d\mu^\st{L}(p) \, \tilde{f}_\st{L}(p) \, : \, e^{i \, \lambda_\mu (p,\La) \otimes \x^\mu} \, :  \, : \, e^{i \, p_\mu \, \A^\mu  \otimes \1} \,: 
 \,,
\end{equation}
the integral of the above function, seen as an operator acting on the right-hand side of the tensor product,\footnote{Although a notation like $\text{\it{id}} \otimes \int d^4\x$ might be more precise in this case, I think it would be unnecessarily clumsy.}
\begin{equation}
\begin{aligned}
\int \, f(\x') \, d^4\x =&  \int d\mu^\st{L}(p) \, \tilde{f}_\st{L}(p) \, \int : \, e^{i \, \lambda_\mu (p,\La) \otimes \x^\mu} \, : \, d ^4 \x  \, : \, e^{i \, p_\mu \, \A^\mu  \otimes \1} \,: 
\\
=&  \int d\mu^\st{L}(p) \, \tilde{f}_\st{L}(p) \,  \delta^{(4)}\left[ \lambda_\mu (p,\La)\right] \, : \, e^{i \, p_\mu \, \A^\mu} \,: 
 \,,
\end{aligned}
\end{equation}
and since $\lambda_\mu(p,\La) = o$ iff $p=o$,
\begin{equation}
\begin{aligned}
\int \, f(\x') \, d^4\x =&  \int  \frac{d\mu^\st{L}(p) \, \tilde{f}_\st{L}(p)\delta^{(4)}(p)}{\left| \det  \partial_{p_\nu} \lambda_\mu(p,\La)  \right|} \, : \, e^{i \, p_\mu \, \A^\mu } \,: 
~ =  \1 \, \tilde{f}_\st{L}(o) \,\left| \det \partial_{p_\nu} \lambda_\mu(p,\La) \right|^{-1}_{p=o} \!\!\! =  \tilde{f}_\st{L}(o) \, \1 
 \,,
\end{aligned}
\end{equation}
where the remaining $\1$ belongs to the algebra $C_\ell[ISO(3,1)]$. The fact that the Jacobian determinant $\partial_\nu \lambda_\mu(p,\La)$ goes to one when $p \to o$ can be seen by considering the Taylor expansion of $\lambda_\mu(p,\La)$, which begins with an undeformed lowest-order $p_\nu \,\La^\nu{}_\mu$, and, for dimensional reasons, all the other orders are at least quadratic in $p_\mu$. The determinant of the lowest order is the determinant of $\La^\nu{}_\mu$, which is one. A similar proof goes through also for integrals of  multiple functions, for example
\begin{equation}
\begin{aligned}
 &\int \, f(\x') \, g(\x') \, d^4 \x   
= (2\pi)^4 \1  \int \, \tilde{f}_\st{L}(p) \, \tilde{g}_\st{L}[S(p)] \, \left|\det  \partial_{k_\nu} \lambda_\mu(k,\La)  \right|^{-1}_{k=o} \!\!\!  d \mu^\st{L}(p)   
= \1 \int \, f(\x) \, g(\x) \, d^4 \x  \,,
\end{aligned} 
\end{equation}
and, in general:
\begin{equation}
\int \, f_1(\x') \, f_2(\x') \, \dots \, f_N(\x') \, d^4 \x   = \1 ~ \int \, f_1(\x) \, f_2(\x) \, \dots \, f_N(\x) \, d^4 \x \,. 
\end{equation}
An important \textit{caveat:} all the derivations carried out in the present paper refer to \textit{local} properties of momentum space, that strictly hold only in a neighbourhood of the origin $o$. In certain specific examples of T-Minkowski noncommutative spacetimes, global issues can arise. For example, in the lightlike $\kappa$-Minkowski model studied in~\cite{Lizzi:2021rlb,DiLuca:2022idu,Fabiano:2023xke}, the Lorentz transform of momenta $\lambda_\mu(p,\La)$ has some singularities at certain finite values of the momenta and Lorentz matrix. This was already known in the closely related case of \textit{timelike} $\kappa$-Minkowski~\cite{Majid:2006xn}, and in~\cite{DiLuca:2022idu} this fact is interpreted as a \textit{coordinate singularity,} by supplementing momentum space with a new sector. In fact, the basis of ordered exponentials used up to that point to expand functions in Fourier series is not closed under Lorentz transforms: a new kind of exponentials arise as a result of certain Lorentz transforms, which make the function $\lambda_\mu(p,\La)$ complex. These exponentials can be interpreted as belonging to a connected component of the group $\mathcal{G}_\ell$ that is disconnected from the identity, so one cannot reach these waves via the coproduct, \textit{i.e.} by multiplying ordinary waves with each other. However, they can be reached with a Lorentz transform, and they should be included in the Fourier expansion of fields, in order to restore the invariance of the theory under Lorentz transformations. A similar observation was made in the timelike $\kappa$-Minkowski model in~\cite{Arzano:2020jro}.

\subsection{Backreaction and Lorentz invariance of the Fourier transform}

It is convenient to introduce an additional map describing the non-commutative interactions between Lorentz matrices and translations. The commutator between $\La^\mu{}_\nu$ and $\A^\mu$ will be  reinterpreted as a (right-) action of the momenta on the Lorentz group.  In fact, the translations acting as vector fields on the Lorentz group allow us to write the  adjoint action of an ordered exponential of translation coordinates on a Lorentz matrix as
\begin{equation}\label{Eq:Def_BackReaction}
: \, e^{i \, q_\mu \, \A^\mu} \, :\, \La^\mu{}_\nu   = (\La \triangleleft q)^\mu{}_\nu  \, : \, e^{i \, q_\mu \, \A^\mu} \, : \,,
\end{equation}
where $\triangleleft  : SO(3,1) \otimes \mathcal{A}^* \to SO(3,1)$ is the anticipated right-action of momentum space on the Lorentz group (called ``backreaction'' in~\cite{Majid:2006xn,Gubitosi:2011hgc}). $(\La \triangleleft q)^\mu{}_\nu $ is the solution of the of a set of differential equations.
The form of these equations depend on the ordering choice codified in $: \, . \, :$ . The exponential of a single linear combination of generators $i\, k_\mu \, \A^\mu$ (\textit{i.e.} a \textit{Weyl-ordered} exponential), defined as the formal power series:
\begin{equation}\label{Eq:Weyl-ordered_exponential_Amu}
e^{i \, k_\mu \, \A^\mu} = \sum_{n=0}^\infty \frac{1}{n!} \, \left( i \, k_\mu \, \A^\mu \right)^n \,,
\end{equation}
acts on the Lorentz group as the exponential map of the vector field
\begin{equation}\label{Eq:VectorFieldExponentialMap}
-k_\nu \,  f^{\alpha\beta}{}_\gamma
\left(  \La^{\mu}{}_\alpha \, \La^{\nu}{}_\beta \, \delta^\gamma{}_\rho  - \delta^{\mu}{}_\alpha \, \delta^{\nu}{}_\beta \, \La^\gamma{}_\rho \right) 
\frac{\partial }{ \partial \La^{\mu}{}_\rho } \,, 
\end{equation} 
so that Eq.~\eqref{Eq:Def_BackReaction} can be written, in this case,
\begin{equation}\label{Eq:ExponentialMapActionOnLambda}
e^{i k_\mu \, a^\mu} \, \La^\mu{}_\nu = \Upsilon^\mu{}_\nu(k,\La) \, e^{i k_\mu \, a^\mu} \,,
\end{equation}
where the function $\Upsilon : SO(3,1) \times \mathbbm{R}^4 \to SO(3,1)$ is the exponential map associated to~\eqref{Eq:VectorFieldExponentialMap}. 
$\Upsilon^\mu{}_\nu$ is defined as the solution of the following differential equation:
\begin{equation}\label{Eq:ExponentialMapDiffEq}
\frac{d\Sigma^\mu{}_\nu  }{d  t} = 
i \, k_\rho \, f^{\alpha\beta}{}_\gamma
\left[ \Sigma^{\mu}{}_\alpha\, \Sigma^{\rho}{}_\beta \, \delta^\gamma{}_\nu  - \delta^{\mu}{}_\alpha \, \delta^{\rho}{}_\beta \, \Sigma^\gamma{}_\nu \right] \,,
\end{equation}
with the initial condition
\begin{equation}\label{Eq:ExponentialMapDiffEqInitialCondition}
\Sigma^\mu{}_\nu \big{|}_{t=0}  = \La^\mu{}_\nu  \,.
\end{equation}
The exponential map is obtained as
\begin{equation}
\Upsilon^\mu{}_\nu (k,\La) =  \Sigma^\mu{}_\nu(k,\La,t) \big{|}_{t=1} \,.
\end{equation}
In general, one can write an arbitrarily-ordered exponential as a product of exponentials of the form~\eqref{Eq:Weyl-ordered_exponential_Amu}, as follows:
\begin{equation}
: \, e^{i \, q_\mu \, \A^\mu} \, : = e^{i \, q^{(1)}_\mu \, \A^\mu} \, e^{i \, q^{(2)}_\mu \, \A^\mu} \, \dots \, e^{i \, q^{(n)}_\mu \, \A^\mu} \,,
\end{equation}
then the adjoint action~\eqref{Eq:Def_BackReaction} is the result of applying consecutively $n$ exponential maps: 
\begin{equation}
(\La \triangleleft q)^\mu{}_\nu  = \Upsilon^\mu{}_\nu \left[ q^{(n)} , \Upsilon \left[ q^{(n-1)} , \dots , \Upsilon [ q^{(1)} , \La ] \right] \dots   \right] \,.
\end{equation}
In the particular case of Weyl ordering,
\begin{equation}
: \, e^{i \, q_\mu \, \A^\mu} \, : ~ =  e^{i \, q_\mu \, \A^\mu} \,, \qquad \Rightarrow \qquad q^{(1)}_\mu = q_\mu \,, n = 1\,,
\end{equation} 
the backreaction coincides with the exponential map:
\begin{equation}
(\La \triangleleft q)^\mu{}_\nu  = \Upsilon^\mu{}_\nu \big{[} q  , \La \big{]} \,. 
\end{equation}
The backreaction map is a                                                     homomorphism for the commutative product between                                                          Lorentz group functions, so for any function $F$ of Lorentz group coordinates:
\begin{equation}
: \, e^{i\, q_\mu \, \A^\mu} \,: \, F(\La) = F(\La \triangleleft q) \, : \, e^{i\, q_\mu \, \A^\mu} \,: \,,
\end{equation}
this can be proven by assuming that $F$ can be expanded as a formal power series in
$\La^\mu{}_\nu$:
\begin{equation}
F(\La) = \sum_{n=0}^\infty F^{\nu_1\nu_2\dots\nu_n}_{\mu_1\mu_2\dots\mu_n}  \,   \La^{\mu_1}{}_{\nu_1} \,  \La^{\mu_2}{}_{\nu_2} \dots  \La^{\mu_n}{}_{\nu_n} \,,
\end{equation}
and observing that
\begin{equation}
\begin{aligned}
: \, e^{i\, q_\mu \, \A^\mu} \,: \, \La^{\mu_1}{}_{\nu_1} \,  \La^{\mu_2}{}_{\nu_2} \dots  \La^{\mu_n}{}_{\nu_n} &= (\La \triangleleft q)^{\mu_1}{}_{\nu_1}  \, : \, e^{i\, q_\mu \, \A^\mu} \,: \,   \La^{\mu_2}{}_{\nu_2} \dots  \La^{\mu_n}{}_{\nu_n} \\
&= (\La \triangleleft q)^{\mu_1}{}_{\nu_1} \,  (\La \triangleleft q)^{\mu_2}{}_{\nu_2} \dots  (\La \triangleleft q)^{\mu_n}{}_{\nu_n}\, : \, e^{i\, q_\mu \, \A^\mu} \,: \,.
\end{aligned}
\end{equation}
Acting twice on $\La^\mu{}_\nu$ with different momenta $p_\mu$, $q_\mu$ gives:
\begin{equation}
: \, e^{i \, p_\mu \, \A^\mu} \, :\,: \, e^{i \, q_\mu \, \A^\mu} \, : \, \La^\mu{}_\nu   = [(\La \triangleleft q )\triangleleft p]^\mu{}_\nu  \, : \, e^{i \, p_\mu \, \A^\mu} \, : \, : \, e^{i \, q_\mu \, \A^\mu} \, :  \,,
\end{equation}
but the commutation relations~\eqref{T-PoincareGroup_commutators} imply that the
product of the two exponentials can be written as a part that commutes with $\La^\mu{}_\nu$, multiplied by $: \,\exp [ i \, \Delta_\mu(p,q) \, \A^\mu] \, :$. Therefore $\triangleleft$ is a homomorphism with respect to the group multiplication $\Delta(p,q)$, describing the Lie group $\mathcal{G}_\ell$:
\begin{equation}
(\La \triangleleft q) \triangleleft p  = \La \triangleleft \Delta(p,q) \,.
\end{equation}
Compatibility with the counit also implies that the coaction $\triangleleft$ leaves the identity matrix invariant, and, of course, the ``$o$'' momentum does not change the Lorentz matrix it acts upon:
\begin{equation}
\delta^\mu{}_\nu \triangleleft p = \delta^\mu{}_\nu \,, \qquad (\La \triangleleft o)^\mu{}_\nu = \La^\mu{}_\nu \,.
\end{equation}

Armed with the backreaction of momenta on Lorentz transformation, we can  now consider the multiplication rule of exponentials~\eqref{ProductBetweenExponentials}, $\E[p] \, \E[q] =  e^{i \, \theta^{\mu\nu} \, \Phi_{\mu\nu}(p,q)} \, \E[\Delta(p,q)] $, and Poincar\'e-transform both sides with the coaction~\eqref{Coaction_spacetime_coordinates} (which, I recall, is a homomorphism for the product of $C_\ell[\mathbbm{R}^{3,1}]$):\footnote{Notice that, in Eq.~\eqref{Eq:PoincareTransfProductExp} I applied the definition of backreaction~\eqref{Eq:Def_BackReaction} on an ordered exponential of the coordinates, $:   \exp[i \, \lambda_\mu (q,\La) \otimes \x^\mu] :$, whose wavevector depends on the Lorentz matrices.  The presence of the noncommutative coordinates $\x^\mu$ might 
cast doubts on the validity of formula~\eqref{Eq:Def_BackReaction} in this case, however we can explicitly prove the equation this way:
 $: \, e^{i \, p_\mu \, \A^\mu  \otimes \1} \,:
 ~: \, \exp[i \, \lambda_\mu (q,\La) \otimes \x^\mu] \, : =  \sum_{n=0}^\infty \frac{i^n}{n!}: \, e^{i \, p_\mu \, \A^\mu  \otimes \1} \,: \,  \lambda_{\mu_1} (q,\La)\dots \lambda_{\mu_n}(q,\La) \otimes :\x^{\mu_1} \dots\x^{\mu_n}: =  \sum_{n=0}^\infty \frac{i^n}{n!} \lambda_{\mu_1} (q \triangleleft p,\La)\dots \lambda_{\mu_n}(q\triangleleft p,\La) \, : \, e^{i \, p_\mu \, \A^\mu  \otimes \1} \,: \otimes :\x^{\mu_1} \dots \x^{\mu_n}:= : \, \exp[i \, \lambda_\mu (q\triangleleft p,\La) \otimes \x^\mu] \, :  ~: \, e^{i \, p_\mu \, \A^\mu  \otimes \1} \,:$.}
 \begin{equation}
\begin{aligned}\label{Eq:PoincareTransfProductExp}
&\E'[p] \, \E'[q] =  e^{i \, \theta^{\mu\nu} \, [ \sigma_{\mu\nu} (p,\La) + \sigma_{\mu\nu} (q,\La)] \otimes \1}  ~: \, e^{i \, \lambda_\mu (p,\La) \otimes \x^\mu} \, :  \, : \, e^{i \, p_\mu \, \A^\mu  \otimes \1} \,:
 ~ : \, e^{i \, \lambda_\mu (q,\La) \otimes \x^\mu} \, :  ~ : \, e^{i \, q_\mu \, \A^\mu  \otimes \1} \,:
\\
&= e^{i \, \theta^{\mu\nu} \, [ \sigma_{\mu\nu} (p,\La) + \sigma_{\mu\nu} (q,\La)] \otimes \1} 
 ~ : \, e^{i \, \lambda_\mu (p,\La) \otimes \x^\mu} \, :   ~ : \, e^{i \, \lambda_\mu (q,\La  \triangleleft p) \otimes \x^\mu} \, :  ~  : \, e^{i \, p_\mu \, \A^\mu  \otimes \1} \,: ~ : \, e^{i \, q_\mu \, \A^\mu  \otimes \1} \,:
 \\
&= e^{i \, \theta^{\mu\nu} \, \left\{ \sigma_{\mu\nu} (p,\La) + \sigma_{\mu\nu} (q,\La) + \Phi_{\mu\nu}[\lambda_\mu (p,\La),\lambda_\mu (q,\La  \triangleleft p)] \right\} \otimes \1  } 
 \\
& ~~ : \, e^{i \, \Delta[\lambda_\mu (p,\La),\lambda_\mu (q,\La  \triangleleft p)] \otimes \x^\mu} \, :  ~  : \, e^{i \, p_\mu \, \A^\mu  \otimes \1} \,: ~ : \, e^{i \, q_\mu \, \A^\mu  \otimes \1} \,:
\\
&=
\E'[\Delta(p,q)] =    e^{i \, \theta^{\mu\nu} \, \sigma_{\mu\nu} \left[ \Delta(p,q),\La \right] \otimes \1} ~: \, e^{i \, \lambda_\mu \left[\Delta(p,q),\La \right] \otimes \x^\mu} \, :  \, : \, e^{i \, \Delta_\mu(p,q)\, \A^\mu  \otimes \1} \,:
\,.
\end{aligned}
\end{equation}
Identifying, in the expression above, all the terms of the type $ \hat{T} \otimes \1$ and those of the type $ \hat{T} \otimes \x^\mu $ ($\hat{T} \in C_\ell[ISO(3,1)]$), we conclude that
\begin{equation}
\begin{aligned}\label{Eq:Coproduct_of_a_mu}
: \, e^{i \, p_\mu \, \A^\mu  \otimes \1} \,: ~ : \, e^{i \, q_\mu \, \A^\mu  \otimes \1} \,:
~ = ~ &
  e^{i \, \theta^{\mu\nu} \, \left\{  \sigma_{\mu\nu} \left[ \Delta(p,q),\La \right] - \sigma_{\mu\nu} (p,\La) - \sigma_{\mu\nu} (q,\La) - \Phi_{\mu\nu}[\lambda_\mu (p,\La),\lambda_\mu (q,\La  \triangleleft p)] \right\} \otimes \1  } \\
&
: \, e^{i \, \Delta_\mu(p,q)\, \A^\mu  \otimes \1} \,:
\,,
\end{aligned}
\end{equation}
and
\begin{equation}\label{MomentumLorentzTransf_Coproduct}
\lambda \left[\Delta(p,q),\La \right] =  \Delta[\lambda (p,\La),\lambda (q,\La  \triangleleft p)] \,.
\end{equation}
Eq.~\eqref{Eq:Coproduct_of_a_mu} is a multiplication law for exponentials of the translations [which, I recall, do not close a copy of the algebra  $C_\ell[\mathbbm{R}^{3,1}]$ unless $\theta^{\mu\nu} =0$, viz.~\eqref{T-PoincareGroup_commutators}]. Eq.~\eqref{MomentumLorentzTransf_Coproduct} is a sort of twisted homomorphism rule between the Lorentz transform of momenta and the coproduct of momenta.\footnote{Notice that, in \eqref{MomentumLorentzTransf_Coproduct}, the backreaction term $\La \triangleleft p$ appears on the right-hand side because I have chosen to order the translation coordinates $\A^\mu$ to the right of the Lorentz matrices.} Setting $p=S(q)$ in Eq.~\eqref{MomentumLorentzTransf_Coproduct}, we deduce a new identity, relating the Lorentz transform, the antipode and the backreaction:
\begin{equation}\label{MomentumLorentzTransf_Antipode_Backreaction}
\lambda(S(q),\La) = S[\lambda(q,\La \triangleleft S(q))] \,.
\end{equation}
 
Eq.~\eqref{MomentumLorentzTransf_Coproduct} allows us to derive the action of the momentum-space Lorentz transform $\lambda(p,\La)$ onto the left-invariant measure. Deriving~\eqref{MomentumLorentzTransf_Coproduct} w.r.t. $q$, and setting   $q=o$: 
 \begin{equation}
\left. \partial_{k_\nu} \lambda_\mu (k,\La) \right|_{k=\Delta(p,q=o)} ~ \left.\partial_{q_\rho}\Delta_\nu(p,q)  \right|_{q=o} =
\left.\partial_{k_\nu}\Delta_\mu(\lambda ( p, \La),k) \right|_{k=\lambda ( q, \La \triangleleft p )} 
~\left. \partial_{q_\rho} \lambda_\nu ( q, \La \triangleleft p )  \right|_{q=o}\,,
\end{equation} 
which leads to
 \begin{equation}
(X_\st{L}^\nu)_\mu [\lambda ( p, \La)]
~\left. \partial_{q_\rho} \lambda_\nu ( q, \La \triangleleft p )  \right|_{q=o}
=
  \partial_{p_\nu} \lambda_\mu (p,\La) ~ (X_\st{L}^\rho)_\nu (p) 
\,,
\end{equation} 
now recall that $\lambda_\mu ( q, \La )= q_\nu \, \La^{\nu}_\mu+ \mathcal{O}(q^2)$, which implies that $\left. \partial_{q_\rho} \lambda_\nu ( q, \La )  \right|_{q=o} = \La^\rho{}_\nu$, so 
 \begin{equation}
(X_\st{L}^\nu)_\mu [\lambda ( p, \La)]
~  (\La \triangleleft p ) ^\rho{}_\nu 
=
  \partial_{p_\nu} \lambda_\mu (p,\La) ~ (X_\st{L}^\rho)_\nu (p) 
\,,
\end{equation} 
therefore, the Lorentz transform of the left-invariant vector fields is
 \begin{equation}
(X_\st{L}^\sigma)_\mu [\lambda ( p, \La)]
=
 [(\La \triangleleft p )^{-1}]^\sigma{}_\rho  \partial_{p_\nu} \lambda_\mu (p,\La) ~ (X_\st{L}^\rho)_\nu (p) 
\,,
\end{equation}
and, taking the determinant of both sides:
\begin{equation}
\begin{aligned}
\det (X_\st{L}^\mu)_\nu [\lambda ( p, \La)] &= \det [(\La \triangleleft p)^{-1}] \, \det [\partial_{p_\mu} \lambda_\nu (p,\La)] \, \det [(X_\st{L}^\mu)_\nu (p) ] 
\\
&=  \det [\partial_{p_\mu} \lambda_\nu (p,\La)] \, \det [(X_\st{L}^\mu)_\nu (p) ] \,,
\end{aligned}
\end{equation}
where I used the fact that $\La \triangleleft p$ is still a Lorentz matrix, and therefore has unit determinant.
This implies that the left-invariant measure is Lorentz-invariant:
\begin{equation}
d\mu^\st{L}[\lambda(p,\La)] = \left| \det  (X^\mu_\st{L})_\nu [\lambda(p,\La)] \right|^{-1} \, d^4 p  \, \left| \det  \partial_{p_\nu} \lambda_\mu(p,\La)  \right| =   d\mu^\st{L}(p)\,.
\end{equation}
Recalling that $d\mu^\st{R}(p) = d\mu^\st{L}[S(p)]$ [Eq.~\eqref{LeftRightHaarMeasureRelation}], we immediately deduce the Lorentz transformation rule of the right-invariant measure
\begin{equation}
d\mu^\st{R}\left(\lambda[p,\La\triangleleft S(p)] \right) =    d\mu^\st{R}(p)\,,
\end{equation}
where I used the relation~\eqref{MomentumLorentzTransf_Antipode_Backreaction}.

\section{T-Poincar\'e algebra}\label{Sec:T-Poincare_algebra}

The transformation rule~\eqref{PoincareTransform_of_Exponential} of an ordered exponential under finite T-Poincar\'e transformations can be expanded as a power series in the group parameters $\A^\mu$ and $\La^\mu{}_\nu$, around $\A^\mu =0$, $\La^\mu{}_\nu = \delta^\mu{}_\nu$. Up to first order the expansion looks like: 
\begin{equation}
\begin{aligned}
\E'[p] =&  e^{i \, \theta^{\mu\nu} \, \sigma_{\mu\nu} (p,\La) \otimes \1} ~: \, e^{i \, \lambda_\mu (p,\La) \otimes \x^\mu} \, :  \, : \, e^{i \, p_\mu \, \A^\mu  \otimes \1} \,:\\
=&
\1 \otimes \E[p] + i \, p_\mu \, \A^\mu \otimes \E[p]
+
i \, \Sigma_{\rho\sigma}(p) \, \eta^{\lambda[\sigma} (\La^{\rho]}{}_\lambda - \delta^{\rho]}{}_\lambda \,\1 )\otimes \E[p]
\\
&+
i \, \Xi^\mu_{\rho\sigma}   (p) \, \eta_{\mu\nu} \,  \eta^{\lambda[\sigma} (\La^{\rho]}{}_\lambda  - \delta^{\rho]}{}_\lambda \,\1)\otimes \, : \,  \frac{\partial}{\partial p_\nu} e^{i \,p_\mu \, \x^\mu} \, :  + \mathcal{O}\left[\A^2 , (\La - \delta)^2 ,  (\La - \delta)\A \right]
\,,
\end{aligned}
\end{equation}
where
\begin{equation}
\Sigma_{\rho\sigma}(p) =  \eta_{\lambda[\sigma} \,\theta^{\mu\nu} \, \left. \frac{\partial \sigma_{\mu\nu} (p,\La)}{\partial \La^{\rho]}{}_\lambda} \right|_{\La = \delta}  \,, \qquad
\Xi^\mu_{\rho\sigma}  (p) =  \eta^{\mu\nu}  \eta_{\lambda[\sigma} \left. \frac{\partial \lambda_\nu (p,\La)  }{\partial \La^{\rho]}{}_\lambda}\right|_{\La = \delta} \,.
\end{equation}
Notice now that the Lorentz matrices (who close an Abelian subalgebra of $C_\ell[ISO(3,1)]$) can always be written as the matrix exponential of a matrix $\hat{\omega}^{\mu\rho}\eta_{\rho\nu}$, where $\hat{\omega}^{\mu\nu} = - \hat{\omega}^{\nu\mu}$:
\begin{equation}
\La^\mu{}_\nu = \left( e^{\eta \, \hat \omega} \right)^\mu{}_\nu \,, \qquad \La^\mu{}_\nu = \1 \, \delta^\mu{}_\nu + \hat{\omega}^{\mu\rho}\eta_{\rho\nu} + \mathcal{O} \left(\hat{\omega}^2\right)\,,
\end{equation}
\begin{equation}
\begin{aligned}
\E'[p] =& \1 \otimes \E[p] + i \, \A^\mu \otimes \, p_\mu  \E[p] 
+ i \, \hat{\omega}^{\rho\sigma} \otimes \left( \Sigma_{\rho\sigma}(p) + \Xi^\mu_{\rho\sigma}(p)  \, \eta_{\mu\nu}  \frac{\partial }{\partial p_\nu} \right)\E[p] 
  + \mathcal{O}\left[\A^2 , \hat{\omega}^2 , \A \hat{\omega}\right] \,.
\end{aligned}
\end{equation}
This can be recasted as a \textit{left action} on plane waves by linear operators $P^\ell_\mu$, $M^\ell_{\mu\nu}$
\begin{equation}\label{Eq:InfTransf_Exponential_P_M}
\begin{aligned}
\E'[p] =&  
\1 \otimes \E[p] +   i \,  \A^\mu \otimes \left( P^\ell_\mu \triangleright \E[p] \right) + i \, \hat{\omega}^{\rho\sigma} \otimes  \left( M^\ell_{\rho\sigma} \triangleright \E[p] \right)  + \dots
\,,
\end{aligned}
\end{equation} 
where
\begin{equation}
P^\ell_\mu \triangleright \E[p] =   p_\mu \triangleright \E[p] \,,
\qquad
M^\ell_{\rho\sigma} \triangleright \E[p] =   
\Sigma_{\rho\sigma}(p) \, \E[p] 
+   \Xi^\mu_{\rho\sigma}(p) \, \eta_{\mu\nu} \, \frac{\partial \E[p]}{\partial p_\nu}  \,.
\end{equation}
The formula~\eqref{Eq:InfTransf_Exponential_P_M} allows me to write the infinitesimal (or, rather, first-order) action of the T-Poincar\'e group on arbitrary functions $f(\x) \in C_\ell[\mathbbm{R}^{3,1}]$, by means of the Fourier transform:
\begin{equation}
f(\x') = 
\1 \otimes f(\x) +  \A^\mu \otimes \left[ P^\ell_\mu \triangleright f(\x) \right]
 +  \hat{\omega}^{\rho\sigma} \otimes  \left[ M^\ell_{\rho\sigma} \triangleright f(\x) \right]  + \dots
\,,
\end{equation}
where
\begin{equation}\label{Eq:Action_P_M}
\begin{aligned}
&P^\ell_\mu \triangleright f(\x) = \int d  \mu^\st{L}(p) \, p_\mu \tilde{f}_\st{L}(p) \, \E[p] \,,
\\
&M^\ell_{\rho\sigma} \triangleright f(\x) =
  \int d  \mu^\st{L}(p) \,  \left( \Sigma_{\rho\sigma}(p) \, \tilde{f}_\st{L}(p)\E[p]  + \Xi^\mu_{\rho\sigma} (p) \, \eta_{\mu\nu} \,  \tilde{f}_\st{L}(p) \, \frac{\partial \E[p]}{\partial p_\nu} \right) \,  \,,
\end{aligned}
\end{equation}
where the derivative of $\E[p]$ in the last line can be integrated by part and ends up acting on the integration measure mutltiplied by $\Xi^\mu_{\rho\sigma} (p) \, \eta_{\mu\nu} \,  \tilde{f}_\st{L}(p)$. A similar expression holds for the right-Fourier transforms.

The linear operators $P^\ell_\mu , M^\ell_{\mu\nu} : C_\ell[\mathbbm{R}^{3,1}] \to C_\ell[\mathbbm{R}^{3,1}] $ close a Hopf algebra $U_\ell[iso(3,1)]$, which is a deformation of the universal enveloping algebra $U[iso(3,1)]$ of the Poincar\'e algebra. This Hopf algebra is the dual  $C^*_\ell[ISO(3,1)]$ of the T-Poincar\'e algebra (\textit{i.e.} the algebra of linear functionals on $C_\ell[ISO(3,1)]$), which is an Hopf algebra itself by virtue of the self-duality property of Hopf algebras. The Hopf algebra maps of $U_\ell[iso(3,1)]$ can all be deduced from the expressions above. In particular, from the form of~\eqref{Eq:Action_P_M}, we can immediately deduce that the translation generators commute, 
\begin{equation}
[P^\ell_\mu , P^\ell_\nu ] = 0 \,,
\end{equation}
and that the commutator between boost and translation generators is a (nonlinear) function of $P_\mu$:
\begin{equation}
[P^\ell_\mu , M^\ell_{\nu\rho} ] = u_{\mu\nu\rho}(P^\ell) \,,
\end{equation}
Moreover, the commutator between boosts is the sum of a term that is linear in the $\partial/\partial p_\mu$ derivatives, and a function of the momenta alone. This means that this commutator can be written as a linear combination of boost generators (whose coefficients are functions of $P_\mu$) plus a function of the translation generators:
\begin{equation}
[M^\ell_{\mu\nu} , M^\ell_{\rho\sigma} ] = v_{\mu\nu\rho\sigma}(P^\ell) +  w^{\lambda\kappa}_{\mu\nu\rho\sigma}(P^\ell) \, M^\ell_{\lambda\kappa} \,.
\end{equation}

The coproducts and antipodes of the translation generators follow from the definition, by applying it to a product of two functions:
\begin{equation}
\Delta(P^\ell_\mu) = \Delta_\mu(P^\ell\otimes \1 , \1 \otimes P^\ell) \,, \qquad
S(P^\ell_\mu) = S(P^\ell)_\mu \,,
\end{equation}
where $\Delta$ and $S$ are the previously-defined multiplication and inverse maps of the group $\mathcal{G}_\ell$, Eqs. \eqref{Eq:Definition_Coproduct} and \eqref{Eq:Definition_Antipode}.
The coproduct and antipode of $M^\ell_{\mu\nu}$ are more complicated, but they can be determined by acting on a product of functions, and Eqs.~\eqref{Eq:Action_P_M} (in particular, the fact that the action of $M_{\mu\nu}$ involves up to first derivatives in the momenta) imply that both maps are (up to) linear in $M^\ell_{\mu\nu}$, but nonlinear in the momenta:
\begin{equation}
\Delta(M^\ell_{\mu\nu}) =  A_{\mu\nu}^{\rho\sigma}(P^\ell \otimes \1 , \1 \otimes P^\ell)  \,  M^\ell_{\rho\sigma} \otimes \1  +  B_{\mu\nu}^{\rho\sigma}(P^\ell\otimes \1 , \1 \otimes P^\ell) \, \1 \otimes M^\ell_{\rho\sigma} 
+ C^{\rho\sigma}(P^\ell\otimes \1 , \1 \otimes P^\ell)  \,.
\,
\end{equation}
Notice that $U_\ell[iso(3,1)]$ can be deduced from the Drinfel'd twist much more straightforwardly (see Eq.~\eqref{Eq:Twisted_Hopf_algebra}). However what one gets with such a procedure is the same Hopf algebra, but written in a different basis. This basis is nonlinearly related to the \textit{bicrossproduct basis} that one gets from the procedure described in this Section, and the hard part is to find this nonlinear relation.

\newpage

\section{Differential geometry constructions}\label{Sec:DifferentialGeometry}

Following Woronowicz~\cite{Woronowicz:1989}, we can introduce a exterior differential operator $d : C_\ell[\mathbbm{R}^{3,1}] \to \Gamma_\ell^1$ such that:
\begin{equation}\label{Eq:Exterior_differential_def}
    d( \x^\mu) = d\x^\mu \,, \qquad d (\hat{f} \, \hat{g}) = d\hat{f} \, \hat{g} + \hat{f} \, d \hat{g} \,, \qquad d^2 = 0 \,,
\end{equation}
its action on generic functions can be described in terms of two linear operators  $\X_\mu, \, \chi^\mu{}_\nu : C_\ell[\mathbbm{R}^{3,1}] \to C_\ell[\mathbbm{R}^{3,1}]$, which can be seen as elements of the translation subalgebra of $U_\ell[iso(3,1)]$, or, equivalently, as functions on momentum space.  The $\X_\mu$ functions are such that
\begin{equation} \label{ExteriorDifferentialExplicitForm}
d \hat{f} = i \, d \x^\mu \, (\X_\mu \triangleright \hat{f}) \,,
~~
\Delta(\X_\mu) = \X_\mu \otimes \1 + \chi^\nu{}_\mu  \otimes \X_\nu \,,
~~
S(\X_\mu) = - \X_\nu \,(\chi^{-1})^\nu{}_\mu \,, \qquad \epsilon(\X_\mu) = 0 \,,
\end{equation}
while $\chi^\mu{}_\nu$ is a grouplike element of $U_\ell[iso(3,1)]$:
\begin{equation}
\Delta ( \chi^\mu{}_\nu )= \chi^\mu{}_\rho \otimes \chi^\rho{}_\nu \,,
\qquad
S(\chi^\mu{}_\nu) = (\chi^{-1})^\mu{}_\nu \,, \qquad
\epsilon(\chi^\mu{}_\nu) = \delta^\mu{}_\nu \,.
\end{equation}
Applying the exterior differential to a product of two functions, and with the definition~\eqref{ExteriorDifferentialExplicitForm}, we get:
\begin{equation}
\begin{gathered}
d (\hat{f} \, \hat{g} ) =  i \, d\x^\mu \, \X_\mu \triangleright (\hat{f} \, \hat{g} ) =
 i \, d\x^\mu \left[ (\X_\mu \triangleright \hat{f}) \, \hat{g} + i \, (\chi^\nu{}_\mu \triangleright \hat{f}) (\X_\nu \triangleright \hat{g})
\right]
\\
= d\hat{f} \, \hat{g} + \hat{f} \, d \hat{g} = i \, (d\x^\mu \, \X_\mu \triangleright \hat{f} ) \, \hat{g} + i \,  \hat{f} \, (  d\x^\mu \, \X_\mu \triangleright \hat{g})
\end{gathered}
\end{equation}
which implies that $\chi^\mu{}_\nu$ implements the commutator between $d\x^\mu$ and functions:
\begin{equation}\label{ChiDefinition}
\hat{f} \,  d\x^\mu =  d\x^\nu  \, ( \chi^\mu{}_\nu \triangleright \hat{f} )\,.
\end{equation}

Had we chosen to order the differentials on the right-hand side of functions, we would have found the antipodal operators to $\X_\mu$ and $	\chi^\mu{}_\nu$:
\begin{equation}  
d \hat{f} =  - i \, ( S(\X)_\mu \triangleright \hat{f}) \, d \x^\mu \,,
\qquad
d\x^\mu \, \hat{f}  = ( S(\chi)^\mu{}_\nu \triangleright \hat{f} ) \, d\x^\nu \,.
\end{equation}

The form of $\chi^\mu{}_\nu$ can be read right off the commutation relations~\eqref{xdx_comm_rel}. In fact, the spacetime coordinates $\x^\mu$ act linearly, upon commutator, on $d \x^\mu$:
\begin{equation} 
 [ \x^{\mu} , d\x^{\nu} ]   
=
i \,  (K^\mu)^\nu{}_\rho   \,    d \x^\rho   \,.
\end{equation}
where $K^\mu$ are the four Lorentz algebra matrices introduced in~\eqref{K_as_M}. Therefore, an ordered exponential acts via adjoint action on $d \x^\mu$:
\begin{equation} 
\E[p] \, d\x^{\mu} \, \E[p]^*   
=
\left( : e^{-p_\rho \, K^\rho} : \right)^\mu{}_\nu \, d \x^\nu \, 
\end{equation}
which implies:
\begin{equation}
\chi^\mu{}_\nu(p) = \left(: e^{- p_\rho \, K^\rho} : \right)^\mu{}_\nu \,,
\end{equation}
and, since $K^\mu$ are linear combinations of Lorentz generators, $\chi^\mu{}_\nu(p)$ is an $SO(3,1)$ matrix, \textit{i.e.}
\begin{equation}
\eta_{\mu\nu} \, \chi^\mu{}_\rho(p) \, \chi^\nu{}_\sigma(p) = \eta_{\rho\sigma} \,, \qquad
\eta^{\rho\sigma} \, \chi^\mu{}_\rho(p) \, \chi^\nu{}_\sigma(p) = \eta^{\mu\nu} \,.
\end{equation}
the above relation implies that, if we define a Minkowski line element as an element of the tensor product between one-forms,
\begin{equation}
d \hat{s}^2 = \eta_{\mu\nu} \, d\x^\mu \otimes d \x^\nu \,,
\end{equation}
it will commute with any element of $C_\ell[\mathbbm{R}^{3,1}]$:
\begin{equation}
\hat{f} \, d \hat{s}^2   = \eta_{\mu\nu} \, d\x^\rho (  \chi^\nu{}_\rho \triangleright \hat{f} ) \otimes d \x^\nu = \eta_{\mu\nu} \, d\x^\rho \otimes d \x^\sigma ( \chi^\mu{}_\sigma \chi^\nu{}_\rho \triangleright \hat{f}) =   d \hat{s}^2 \, \hat{f}  \,.
\end{equation}

\subsection{Explicit formulas for the differential}

We can calculate explicitly the form of $\X_\mu$ as well. Consider the following map~\cite{Podles:1995qy}:
\begin{equation}\label{MatrixHomomorphismL}
\mathcal{L} : C_\ell[\mathbbm{R}^{3,1}] \to C_\ell[\mathbbm{R}^{3,1}] \otimes GL(5)
\,, \qquad
\mathcal{L}  (\hat{f}) =
\left(
\begin{array}{c|c}
(\chi^{-1})^\mu{}_\nu \triangleright \hat{f} & i \, \X^\mu \triangleright \hat{f} 
\\
\hline
0 & \hat{f} 
\end{array}
\right) \,,
\end{equation}
this a homomorphism for the product of $C_\ell[\mathbbm{R}^{3,1}]$. In fact, the coproducts of $\X^\mu = \eta^{\mu\nu}\X_\nu$ and $\chi^\mu{}_\nu$ translate into a matrix product:
\begin{equation}
\begin{aligned}
\mathcal{L} (\hat{f}\hat{g}) =&
\left(
\begin{array}{c|c}
((\chi^{-1})^\mu{}_\rho \triangleright \hat{f} )((\chi^{-1})^\rho{}_\nu \triangleright \hat{g}) & i \eta^{\mu\alpha}[(\xi_\alpha\triangleright \hat{f})\hat{g}+(\chi^{\gamma}{}_\alpha\triangleright\hat{f})(\xi_\gamma\triangleright\hat{g})]
\\
\hline
0 & \hat{f} \hat{g}
\end{array}
\right) = 
\\
=&
\left(
\begin{array}{c|c}
((\chi^{-1})^\mu{}_\rho \triangleright \hat{f} )((\chi^{-1})^\rho{}_\nu \triangleright \hat{g}) & i (\xi^\mu\triangleright \hat{f})\hat{g}+i((\chi^{-1})^\mu{}_\nu\triangleright \hat{f})(\xi^\nu \triangleright \hat{g})
\\
\hline
0 & \hat{f} \hat{g}
\end{array}
\right) = 
\\
=&
\left(
\begin{array}{c|c}
(\chi^{-1})^\mu{}_\rho \triangleright \hat{f} & i \, \X^\mu \triangleright \hat{f}
\\
\hline
0 & \hat{f} 
\end{array}
\right) 	\cdot
\left(
\begin{array}{c|c}
(\chi^{-1})^\rho{}_\nu \triangleright \hat{g} & i \, \X^\rho \triangleright \hat{g}
\\
\hline
0 & \hat{g} 
\end{array}
\right) = \mathcal{L} (\hat{f}) \cdot \mathcal{L} (\hat{g})\,,
\end{aligned}
\end{equation}
where in the second inequality we have used the fact that $\eta^{\mu\alpha}\chi^{\gamma}{}_{\alpha}\eta_{\gamma\nu}=(\chi^{-1})^\mu{}_\nu$, given that $\chi^\mu{}_\nu$ is an SO(3,1) matrix. 
The definitions~\eqref{ExteriorDifferentialExplicitForm} and~\eqref{ChiDefinition} imply $(\chi^{-1})^\rho{}_\sigma \triangleright \x^\mu = \delta^\rho{}_\sigma \, \x^\mu - i \, (K^\mu)^\rho{}_\sigma $, and $\X^\mu \triangleright \x^\nu = - i \, \eta^{\mu\nu}$, which allows us to calculate the action of $\mathcal{L}$ on a coordinate:
\begin{equation}
\label{lofx}
\mathcal{L}  (\x^\alpha) =
\left(
\begin{array}{c|c}
\delta^\mu{}_\nu \, \x^\alpha - i \, (K^\alpha)^\mu{}_\nu &   \eta^{\mu\alpha}
\\
\hline
0 & \x^\alpha 
\end{array}
\right)
=
\x^\alpha \, I
+\rho'(\hat{t}^\alpha)
\,,
\end{equation}
where $I$ is the $5\times 5$ identity matrix, and $\rho(\hat{t}^\mu)$ are the 5-dimensional representation of $\mathcal{A}_\ell$, introduced in~\eqref{Eq:t_matrices}, satisfying:
\begin{equation}
[\rho(\hat{t}^\mu) , \rho(\hat{t}^\nu)] = i \, c^{\mu\nu}{}_\rho \, \rho(\hat{t}^\rho) \,.
\end{equation}
Using this expression, and the homomorphism property of $\mathcal{L}$, I can calculate the action on an ordered exponential
\begin{equation}\label{LmapOnExponentials}
\mathcal{L}  \left( \E[p] \right) = ~ :\, e^{i \, p_\mu \, \x^\mu} : ~ :\, e^{-i \, p_\mu \, \rho(\hat{t}^\mu)} : ~ = \E[p] ~ :\, e^{-i \, p_\mu \, \rho(\hat{t}^\mu)} : ~ = \left(
\begin{array}{c|c}
(\chi^{-1})^\mu{}_\nu \triangleright \E[p] & i \, \X^\mu \triangleright \E[p] 
\\
\hline
0 & \E[p]
\end{array}
\right) ~ \,,
\end{equation}
where the ordering choice $: \, \cdot \, :$ is the same for $\x^\mu$ and $\rho(\hat{t}^\mu)$. This implies that the $\E[p]$ are eigenfunctions of both $\chi^\mu{}_\nu$ and $\X^\mu$, with eigenvalues given by:
\begin{equation}
\left(
\begin{array}{c|c}
(\chi^{-1})^\mu{}_\nu (p) & i \, \X^\mu  (p)
\\
\hline
0 & 1
\end{array}
\right) = ~  :\, e^{-i \, p_\mu \, \rho(\hat{t}^\mu)} : ~
= : \exp
\left(
\begin{array}{c|c}
  p_\alpha \, (K^\alpha)^\mu{}_\nu & i p^\mu 
\\
\hline
0 & 0
\end{array}
\right) : 
\,.
\end{equation}
This gives us an explicit formula to calculate $\X^\mu (p)$ and $\chi^\mu{}_\nu(p)$ in a given coordinate system on momentum space, associated to a particular ordering choice.

\subsection{Behaviour under involution of the differential calculus}

The elements of $U_\ell[iso(3,1)]$ act on $C_\ell[\mathbbm{R}^{3,1}]$ as linear operators, however this action does not commute with the involution. This can be seen by acting on plane waves, \textit{e.g.}:
\begin{equation}
P^\ell_\mu \triangleright \E[k] = k_\mu \, \E[k] \,, \qquad
(P^\ell_\mu \triangleright \E[k])^* = k_\mu \, \E^*[k] \neq
P^\ell_\mu \triangleright \E^*[k] = S(k_\mu) \, \E^*[k] \,,
\end{equation}
the correct expression is:
\begin{equation}
(P^\ell_\mu \triangleright \E[k])^* = k_\mu \, \E^*[k]  = S(P^\ell_\mu) \triangleright \E^*[k] \,.
\end{equation}
We conclude that the involution is represented on functions on momentum space via the antipode. In the case of  $\X_\mu$ and $\chi^\mu{}_\nu$, which are real functions of the momenta, we have the following 
\begin{equation}\label{Eq:Conjugate_xi_chi}
\begin{aligned}
&[ \X_\mu \triangleright f(\x) ]^* =    S(\X_\mu) \triangleright f^*(\x)   = - (\X_\nu \,(\chi^{-1})^\nu{}_\mu) \triangleright f^*(\x) \,,
\\
&[ \chi^\mu{}_\nu \triangleright f(\x) ]^*   =  S(\chi^\mu{}_\nu) \triangleright f^*(\x)   =  (\chi^{-1})^\mu{}_\nu \triangleright f^*(\x)   \,.
\end{aligned}
\end{equation}
These relations are consistent with 
\begin{equation}
\begin{aligned}
[d f(\x)]^* =& [i \, d \x^\mu \, \X_\mu \triangleright f(\x)]^* = -i \, [\X_\mu \triangleright f(\x)]^* \, (d \x^\mu)^*
= i \,  S(\X_\mu) \triangleright f^*(\x) \, d \x^\mu
= -i \, (\X_\nu \,(\chi^{-1})^\nu{}_\mu) \triangleright f^*(\x) \, d \x^\mu
\\
=& -i \, d\x^\rho \, (\chi^\mu{}_\rho \, \X_\nu \,(\chi^{-1})^\nu{}_\mu)  \triangleright f^*(\x) 
= i \,  d\x^\mu \, \X_\mu \triangleright f^*(\x)  
=  d f^*(\x) \,.
\end{aligned}
\end{equation}

\subsection{Lorentz covariance of the differential calculus}

The covariance of $\X_\mu$ under quantum Poincar\'e transforms follows from the Lorentz-invariance of the expression 
\begin{equation}\label{Eq:dEE*prime}
(d \E[p]) \E[p]^* = i \, \X_\mu(p) \, d\x^\mu \,.
\end{equation}
Appliying the T-Poincar\'e coaction on the left-hand side:
\begin{equation}
(d \E'[p]) \E'[p]^* = i \,  \X_\mu[\lambda(p,\La)] \, d\x^\mu  \,,
\end{equation}
where $d \E'[p]$ is understood in the sense of $(\text{id}\otimes d ) \, \E'[p]$.  Now, applying the coaction on the right-hand side of~\eqref{Eq:dEE*prime}, recalling that $d\x'^\mu = \La^\mu{}_\nu \otimes  d\x^\nu$, one gets $i \, \X_\mu(p) \, \La^\mu{}_\nu \otimes d \x^\mu$. Identifying the two sides, one obtains an identity implying that $\X_\mu$ transforms as a covariant vector under deformed Lorentz transformations:
\begin{equation}\label{Eq:Lorentz_transform_xi}
\X_\mu[\lambda(p,\La)] =  \La^\nu{}_\mu \,\X_\nu(p) \,.
\end{equation}

The transformation rule of the $\chi^\mu{}_\nu$ operator can be deduced from the coproduct of $\X_\mu$, Eq.\eqref{ExteriorDifferentialExplicitForm}, and the identity~\eqref{MomentumLorentzTransf_Coproduct}, relating  the deformed Lorentz transformation of momenta and the momentum composition law, $\lambda[\Delta(p,q),\La] = \Delta[\lambda(p,\La),\lambda(q,\La\triangleleft p)]$. On one hand, Eq.~\eqref{Eq:Lorentz_transform_xi} implies that $\X_\mu$ calculated on the Lorentz transform $\lambda[\Delta(p,q),\La]$ of the composition $\Delta(p,q)$ of two momenta is:
\begin{equation}\label{Eq:Proof_Lorentztransf_chi_1}
\X_\mu[\lambda[\Delta(p,q),\La]] =  \X_\nu[\Delta(p,q)] \, \La^\nu{}_\mu = \left( \X_\nu(p)  + \chi^\rho{}_\nu(p) \, \X_\rho(q) \right) \La^\nu{}_\mu \,,
\end{equation}
on the other hand, using Eq.~\eqref{MomentumLorentzTransf_Coproduct} first, we get
\begin{equation}\label{Eq:Proof_Lorentztransf_chi_2}
\begin{aligned}
\X_\mu(\lambda[\Delta(p,q),\La])  =& \X_\mu(\Delta[\lambda(p,\La),\lambda(q,\La\triangleleft p)]) = \X_\mu[\lambda(p,\La)] + \chi^\nu{}_\mu[\lambda(p,\La)] \, \X_\nu[\lambda(q,\La\triangleleft p)]
\\
=&
 \X_\nu(p) \, \La^\nu{}_\mu  + \chi^\nu{}_\mu[\lambda(p,\La)] \, \X_\rho(q) \, (\La \triangleleft p )^\rho{}_\nu  \,.
\end{aligned}
\end{equation}
Equating Eqs.\eqref{Eq:Proof_Lorentztransf_chi_1} and \eqref{Eq:Proof_Lorentztransf_chi_2} implies
\begin{equation}\label{Eq:Proof_Lorentztransf_chi}
\chi^\mu{}_\nu[\lambda(p,\La)] = [(\La \triangleleft p)^{-1}]^\mu{}_\rho\,  \chi^\rho{}_\sigma(p) \, \La^\sigma{}_\nu \,.
\end{equation}

\subsection{Relation between $\X$ and left-invariant vector fields}

Consider the coproduct rule~\eqref{ExteriorDifferentialExplicitForm} of $\X_\mu$,
\begin{equation}
\X_\mu[\Delta(p,q)] = \X_\mu(p) + \chi^\nu{}_\mu (p) \, \X_\nu (q) \,,
\end{equation}
differentiate both sides w.r.t. $q_\alpha$:
\begin{equation}
\frac{\partial \X_\mu (p)}{\partial p_\beta} 
 \frac{\partial \Delta_\beta(p,q)}{\partial q_\alpha} =   \chi^\nu{}_\mu (p) \,   \frac{\partial \X_\nu (q)}{\partial q_\alpha} \,, 
\end{equation}
if we now set $q=o$, the derivative on the r.h.s. becomes the identity matrix:
\begin{equation}\label{Eq:Derivative_xi_in_zero}
\left. \frac{\partial \X_\nu (q)}{\partial q_\alpha} \right|_{q=o}
= \delta^\alpha{}_\nu \,,
\end{equation}
this follows directly from the definition:
\begin{equation}
\left(
\begin{array}{c|c}
(\chi^{-1})^\mu{}_\nu (p) & i \,  \X^\mu (p)
\\
\hline
0 & 1
\end{array}
\right)
= :e^{i \, p_\lambda \, \rho(\hat{t}^\lambda)}:  = I + i \, p_\lambda \, \rho(\hat{t}^\lambda ) + \mathcal{O}(p^2) \,,
\end{equation}
so
\begin{equation}
\left. \frac{\partial \X^\mu (q)}{\partial q_\alpha} \right|_{q=o}
= \rho(\hat{t}^\lambda )^\mu{}_4  =  \eta^{\lambda\mu} \,,
\end{equation}
and lowering the $\mu$ index one gets Eq.~\eqref{Eq:Derivative_xi_in_zero}. Then:
\begin{equation}
\frac{\partial \X_\mu (p)}{\partial p_\beta} 
\left. \frac{\partial \Delta_\beta(p,q)}{\partial q_\alpha}\right|_{q=o} =   \chi^\alpha{}_\mu (p) \,.
\end{equation}
Taking the determinant of both sides, and recalling that $ \chi^\alpha{}_\mu (p) $ is a Lorentz matrix,
\begin{equation}\label{Eq:Xi_Jacobians_identity}
\det \left( \frac{\partial \X_\mu (p)}{\partial p_\nu} \right) \det \left( 
\left. \frac{\partial \Delta_\rho(p,q)}{\partial q_\sigma}\right|_{q=o} \right)  =  1 \,.
\end{equation}
Now consider the left-invariant measure~\eqref{Eq:Invariant_measures_definitions}:
\begin{equation}
d\mu^\st{L} (p)  =
 \left|  \det \frac{\partial \Delta_\nu (p,q)}{\partial q_\mu} \right|_{q=o}^{-1} \, d^4 p \,,
\end{equation}
we can change variables from $p_\mu$ to $\X_\mu$, by introducing of the inverse of the Jacobian $\partial \X_\mu (p) / \partial p_\nu$
\begin{equation}
d\mu^\st{L} (p)  =
 \left|  \det \frac{\partial \Delta_\nu (p,q)}{\partial q_\mu} \right|_{q=o}^{-1}  \, \left| \det  \frac{\partial \X_\rho (p)}{\partial p_\sigma} \right|^{-1} d^4 \X  = d^4 \X  \,,
\end{equation}
because of Eq.~(\ref{Eq:Xi_Jacobians_identity}). The $\X_\mu$ are a coordinate system on momentum space in which the left-invariant measure reduces to the identity, and Lorentz transformations act linearly. Notice, however, that these results come with an important caveat: they are all \textit{local,} in the sense that they hold in a neighbourhood of the origin of momentum space. Global issues, related to the topology of the group $\mathcal{G}_\ell$ can complicate these issues. In particular, the map $k_\mu \to \X_\mu(k)$ between the bicrossproduct-like coordinates  and the Lorentz-covariant coordinates $\X_\mu$ might not be bijective. For example, in the case of the $\kappa$-lightlike spacetime, the map is not surjective, and the whole
range of bicrossproduct coordinates maps to a region covering only half of the range of coordinates $\X_\mu$. This region is not closed under Lorentz transformations, and this would amount to a breakdown of Lorentz invariance, unless momentum space is extended into a whole new region, by including a second type of plane waves in the Fourier expansion of fields~\cite{Lizzi:2021rlb,DiLuca:2022idu,Fabiano:2023xke}. These global issues can only be discussed separately in each model. In the present paper, I am only interested in the general features of the T-Minkowski models, and will therefore postpone any discussion of global nature to more specialized studies.

\subsection{Exterior algebra and Hodge star}
 
Contrary to the models considered in~\cite{Meier:2023kzt,Meier:2023lku},  in any T-Minkowski spacetime, since the Lorentz matrices commute with each other, a standard/undeformed exterior algebra is Lorentz-covariant:
\begin{equation}\label{WedgeProductDefinition}
    d \x^\mu \wedge d \x^\nu = - d \x^\nu \wedge d \x^\mu \,,
\end{equation}
meaning that
\begin{equation}
  d \x'^\mu \wedge d \x'^\nu = 
\La^\mu{}_\rho \, \La^\nu{}_\sigma \,  d \x^\rho \wedge d \x^\sigma = -
\La^\mu{}_\rho \, \La^\nu{}_\sigma \,  d \x^\sigma \wedge d \x^\rho
  = - d \x'^\nu \wedge d \x'^\mu \,.
\end{equation}
and so on for higher-degree forms.  We can then generalize $\Gamma_\ell^1$ to a whole differential complex $\Gamma^{\wedge}_\ell = C_\ell[\mathbbm{R}^{3,1}] \oplus \Gamma_\ell^1 \oplus \Gamma_\ell^2\oplus \Gamma_\ell^3 \oplus \Gamma_\ell^4$, generated by
\begin{equation}\label{WedgeProductDefinition2}
d \x^\mu \,,
\qquad
d \x^\mu \wedge d \x^\nu \,,
\qquad
d \x^\mu \wedge d \x^\nu \wedge d \x^\rho \,,
\qquad
d^4 \x = d\x^0 \wedge d\x^1 \wedge d\x^2 \wedge d\x^3 \,,
\end{equation} 
with all the possible (non-repeating) choices of indices, modded by the antisymmetry relations~\eqref{WedgeProductDefinition}, and the associativity of the wedge product.  $\Gamma^{\wedge}_\ell $ is spanned by all the products of the generators~~\eqref{WedgeProductDefinition2}  with elements of $C_\ell[\mathbbm{R}^{3,1}]$.  The involution of $C_\ell[\mathbbm{R}^{3,1}]$ extends straightforwardly to $\Gamma^{\wedge}_\ell$:
\begin{equation}
\begin{gathered}
(d \x^\mu)^* = d \x^\mu \,,
\qquad
(d \x^\mu \wedge d \x^\nu)^* = d \x^\nu \wedge d \x^\mu = -d \x^\mu \wedge d \x^\nu \,,
\\
(d \x^\mu \wedge d \x^\nu \wedge d \x^\rho)^* =  
d \x^\rho \wedge d \x^\nu \wedge d \x^\mu  = d \x^\mu \wedge d \x^\nu \wedge d \x^\rho \,,
\qquad
(d^4 \x)^* = d^4 \x \,.
\end{gathered}
\end{equation} 
So, a generic $n$-form can always be written as~\cite{Woronowicz:1989}
\begin{equation}
\omega = d\x^{\mu_1} \wedge d\x^{\mu_2} \wedge  \dots \wedge d\x^{\mu_n} \, \omega_{\mu_1 \mu_2 \dots \mu_n}(\x) \,,
\end{equation}
(or, equivalently, factorizing the $\omega_{\mu_1 \mu_2 \dots \mu_n}(\x)$ to the left. The wedge product should be treated as an associative product on $\Gamma_\ell^{\wedge}$, which reduces to the product of $ C_\ell[\mathbbm{R}^{3,1}]$ when acting on two noncommutative functions,  to the comodule product of $\Gamma^1_\ell$ when acting on a function and a one-form, and to the wedge product~\eqref{WedgeProductDefinition} when acting on pairs of basis one-forms.
Notice that I called the basis 4-form $d^4 \x$, with the same symbol that I used to indicate the noncommutative integral~\eqref{NCintegralDefinition1}. This is not a coincidence: using Eq.~\eqref{xdx_comm_rel} one can calculate the commutator of $d^4 \x$ with a generic function
\begin{equation}
\begin{aligned}
        [x^\mu , d^4 \x ]  &= 
   i\,  \bigg{(}
d \x^\nu \, f^{0\mu }{}_\nu\wedge d \x^1 \wedge d \x^2 \wedge d \x^3
+
d \x^0 \wedge d \x^\nu \, f^{1\mu }{}_\nu \wedge d \x^2 \wedge d \x^3
+
\\
& \qquad+
d \x^0 \wedge d \x^1 \wedge d \x^\nu \, f^{2\mu }{}_\nu \wedge d \x^3
+
d \x^0 \wedge d \x^1 \wedge d \x^2 \wedge d \x^\nu \, f^{3\mu }{}_\nu        \bigg{)}
        \\
     =&   i \, \left(
  f^{0\mu}{}_0 +  f^{1\mu}{}_1 + f^{2\mu}{}_2 + f^{3\mu }{}_3
    \right) d^4 \x 
\end{aligned}
\end{equation}
so if $f^{\nu\mu}{}_\nu = 0$, the maximum-degree form is commutative.
This is guaranteed, as the matrices  $(K^\mu)^ \alpha{}_\beta = f^{\alpha\mu}{}_\beta$ are Lorentz algebra matrices, and therefore traceless.
The algebra of maximum-degree forms is then isomorphic to the algebra of 0-forms:
\begin{equation}
\Gamma^4_\ell = \Gamma^0_\ell = C_\ell[\mathbbm{R}^{3,1}] \,,
\end{equation}
and the integral, so far defined as an involutive linear functional from $ C_\ell[\mathbbm{R}^{3,1}] $ to $\mathbbm{C}$, can be straightforwardly extended to  $\Gamma^4_\ell$. This is the reason behind the choice of using a single symbol $d^4\x$: all the integrals written above can be thought of as acting on 4-forms instead of scalar functions.
 
The differential  can now be promoted to a map on the whole differential complex:
\begin{equation}
d : \Gamma^n_\ell \to \Gamma^{n+1}_\ell \,, \qquad d(\omega \wedge \rho )= d (\omega )\wedge + (-1)^n \omega \wedge d(\rho) \,, ~~ \omega
\in \Gamma^n_\ell \,, ~~ \rho \in \Gamma^m_\ell\,.
\end{equation} 
The fact that ordered exponentials are eigenfunctions of $\X_\mu$  
implies that two consecutive actions of $\X_\mu$ on Fourier-transformable functions commute. Then the nilpotency of the differential follows:
\begin{equation}
d^2 \omega = d  \left[ (i \,\X_\nu \triangleright \omega_{\mu_1\dots\mu_n}) \,d\x^\nu \wedge d\x^{\mu_1} \wedge \dots \wedge d\x^{\mu_n} \right] = - (\X_\rho \X_\nu \triangleright \omega_{\mu_1\dots\mu_n}) \, d \x^\rho \wedge d\x^\nu \wedge d\x^{\mu_1} \wedge \dots \wedge d\x^{\mu_n} \,,
\end{equation}
which is zero because $(\X_\rho \X_\nu \triangleright \omega_{\mu_1\dots\mu_n}) \, d \x^\rho \wedge d\x^\nu = (\X_{[\rho} \X_{\nu]} \triangleright \omega_{\mu_1\dots\mu_n}) \, d \x^\rho \wedge d\x^\nu  =0$.

The Hodge-$\bm{\ast}$ will be defined as an involutive map
\begin{equation}
\bm{\ast} : \Gamma_\ell^n \to \Gamma_\ell^{4-n} \,, \qquad \bm{\ast} \circ \bm{\ast} = (-1)^{n(4-n)} \, \text{\textit{id}}\,,
\end{equation}
which is bilaterally $C_\ell[\mathbbm{R}^{3,1}]$-linear:
\begin{equation}
\bm{\ast} \left[ f(\x) \, \omega \right] = f(\x) \, \bm{\ast} ( \omega  )  \,,
\qquad
\bm{\ast} \left[ \omega  \, f(\x) \right] = \bm{\ast} \left( \omega \right) \, f(\x) \,,
\qquad
\omega \in \Gamma_\ell^n \,,
\end{equation}
and such that the following Hermitian sesquilinear form is a nondegenerate (indefinite) inner product between forms of the same degree:
\begin{equation}\label{InnerProductBetweenForms}
(\omega , \rho ) = \int \omega^* \wedge \bm{\ast} (\rho) \,, \qquad
\overline{(\rho ,\omega)} = (\omega , \rho )  \,, \qquad  \omega,\rho \in \Gamma_\ell^n \,.
\end{equation}
The following rules
\begin{equation}
\begin{gathered}
\bm{\ast} (\1) = d^4 \x \,,
\qquad
\bm{\ast} (d \x^\mu) = {\sfrac 1 {3!}} \, \varepsilon^\mu{}_{\nu\rho\sigma} \, d \x^\nu \wedge d \x^\rho \wedge d\x^\sigma \,,
\qquad
\bm{\ast} (d \x^\mu \wedge  d\x^\nu) =  {\sfrac 1 {2!}} \,  \varepsilon^{\mu\nu}{}_{\rho\sigma} \, d \x^\rho \wedge d \x^\sigma \,, 
\\
\bm{\ast} (d \x^\mu \wedge d \x^\nu \wedge d\x^\rho) = \varepsilon^{\mu \nu\rho}{}_\sigma \, d \x^\sigma \,, 
\qquad
\bm{\ast} (d^4 \x ) = \1 \,,
\end{gathered}
\end{equation}
together with the left- and right-$C_\ell[\mathbbm{R}^{3,1}]$-linearity, completely fix the action of the Hodge-$\bm{\ast}$ on arbitrary forms. First, one can verify the consistency of the left- and right-$C_\ell[\mathbbm{R}^{3,1}]$-linearity, by considering the identity~\eqref{ChiDefinition}, and applying the Hodge-$\bm{\ast}$ to both sides. The left-hand side gives:
\begin{equation}\label{ConsistencyBilateralLinearity1}
 \bm{\ast} [ f(\x) \, d\x^\mu ]
 =\varepsilon^\mu{}_{\nu\rho\sigma} \, f(\x)\, d\x^\nu \wedge d\x^\rho \wedge d\x^\sigma  \,,
\end{equation}
while the right-hand side:
\begin{equation}\label{ConsistencyBilateralLinearity2}
\bm{\ast} [d\x^\nu  \, ( \chi^\mu{}_\nu \triangleright \hat{f} )] =
\varepsilon^\nu{}_{\lambda\rho\sigma} \,  d\x^\lambda \wedge d\x^\rho \wedge d\x^\sigma \, ( \chi^\mu{}_\nu \triangleright \hat{f} )
\,,
\end{equation}
these two expressions must be identical for the definition of the Hodge-star to be consistent. We can use again Eq.~\eqref{ChiDefinition} to bring the coefficient $ f(\x)$ in Eq.~\eqref{ConsistencyBilateralLinearity1} to the right-hand side of the differentials:
\begin{equation}
\varepsilon^\mu{}_{\nu\rho\sigma} \, f(\x)\, d\x^\nu \wedge d\x^\rho \wedge d\x^\sigma  
=\varepsilon^\mu{}_{\nu\rho\sigma} \, d\x^\alpha \wedge d\x^\beta \wedge d\x^\gamma   \, \chi^\sigma{}_\gamma\chi^\rho{}_\beta\chi^\nu{}_\alpha \triangleright f(\x) \,,
\end{equation}
which is equal to~\eqref{ConsistencyBilateralLinearity2} because $\chi^\mu{}_\nu$ is a Lorentz matrix, and therefore it automatically satisfies the identity
\begin{equation}
\varepsilon^\mu{}_{\nu\rho\sigma} \,  \chi^\sigma{}_\gamma\chi^\rho{}_\beta\chi^\nu{}_\alpha  = \varepsilon^\nu{}_{\alpha\beta\gamma} \,  \, \chi^\mu{}_\nu  \,.
\end{equation}

The consistency in the cases of two- and three-forms can be treated analogously. Let us now verify the inner product property. Given two one-forms $\omega =  d\x^\mu \, \omega_\mu$ and $\rho = d\x^\mu \, \rho_\mu $ (recall that  $\omega^* =   (d\x^\mu \, \omega_\mu)^* =   (\omega_\mu)^* \, d\x^\mu$),
\begin{equation}
 \omega^*   \wedge \bm{\ast} (\rho) = {\sfrac 1 {3!}} \, \varepsilon^\nu{}_{\rho\sigma\lambda} \, (\omega_\mu)^* \, d\x^\mu \wedge d \x^\rho \wedge d \x^\sigma \wedge d \x^\lambda   \, \rho_\nu
= 
  \eta^{\mu\nu} \, (\omega_\mu)^*  \, \rho_\nu\, d^4 \x \,.
\end{equation}
the hermiticity of the sesquilinear form follows from the involutive property of the integral~\eqref{IntegralInvolutive}:
\begin{equation}
\int  \omega^*   \wedge \bm{\ast} (\rho)  =    \eta^{\mu\nu} \int \omega^*_\mu(\x)  \, \rho_\nu (\x)\, d^4\x =   \eta^{\mu\nu}  \overline{\left(  \int \rho^*_\mu(\x)  \, \omega_\nu(\x) \, d^4\x \right)}\,.
\end{equation}

In the case of two two-forms  $\omega =  d\x^\mu \wedge d \x^\nu \, \omega_{\mu\nu}$ and $\rho =  d\x^\mu \wedge d \x^\nu \, \rho_{\mu\nu}$:
\begin{equation}
 \omega^*   \wedge \bm{\ast} (\rho) = {\sfrac 1 {2!}} \, \varepsilon^{\rho\sigma}{}_{\lambda\kappa} \, \omega^*_{\mu\nu}(\x) \, d\x^\mu \wedge d \x^\nu \wedge d \x^\lambda \wedge d \x^\kappa \, \rho_{\rho\sigma}
=  \eta^{\mu\rho} \eta^{\nu\sigma} \, \omega^*_{\mu\nu}(\x)  \, \rho_{\rho\sigma}\, d^4 \x  \,,
\end{equation}
while for one-forms and four-forms the Hermiticity property is trivial:
\begin{equation}
 \left[ f(\x) \right]^*\, \bm{\ast} \left[g(\x) \right] = f(\x)^*  \, g(\x) \,  d^4 \x  \,, \qquad
  \left[d^4 \x    \, f(\x) \right]^*\, \bm{\ast} \left[d^4 \x    \, g(\x) \right] = f(\x)^*  \, g(\x) \,  d^4 \x  \,.
 \end{equation}

We could have used the opposite convention in the definition of the inner product between forms:
\begin{equation}
\int  \, \omega   \wedge \bm{\ast} (\rho^*)  \,,
\end{equation}
but this is identical to (complex conjugate of) the  inner product defined in~\eqref{InnerProductBetweenForms}. We can prove it explicitly on forms of all degrees. For one-forms, the following identities hold:
\begin{equation}\label{SecondOrderingForInnerProduct}
\begin{aligned}
\omega   \wedge \bm{\ast} (\rho^*) &=   d\x^\mu \, \omega_\mu  (\x) \wedge \bm{\ast} (\rho^*_\nu(\x) d\x^\nu  ) 
 =  {\sfrac 1 {3!}} \varepsilon^\nu{}_{\rho\sigma\lambda} \, d\x^\mu \, \omega_\mu (\x)\, \rho^*_\nu(\x) \wedge  d\x^\rho \wedge d\x^\sigma \wedge  d \x^\lambda 
 \\
&=  \eta^{\nu\kappa} \, \chi^\mu{}_\kappa \triangleright \left( \omega_\mu(\x) \, \rho^*_\nu(\x) \right)  \, d^4\x \,,
\end{aligned}
\end{equation}
and the integral is such that:
\begin{equation}
\int \chi^\mu{}_\nu \triangleright f(\x) \, d^4\x = \chi^\mu{}_\nu(o) \,\tilde{f}_\st{L}(o)= \delta^\mu{}_\nu  \,\tilde{f}_\st{L}(o) \,,
\end{equation}
so the integral of the expression~\eqref{SecondOrderingForInnerProduct} coincides with $(\rho,\omega) = \overline{(\omega,\rho)}$ from Eq,~\eqref{InnerProductBetweenForms}. The same proof goes through in the case of two- and three-forms, where an increasing number of operators $\chi^\mu{}_\nu$ act on the whole product between the coefficients of $\omega$ and $\rho$, just to be eliminated by the integral.
Had we chosen to put the Hodge-$\bm{\ast}$ onto the form on the left, because of the following identities (in the case of one-forms):
\begin{equation}
 \bm{\ast} (\omega^*)   \wedge \rho  =    \eta^{\mu\nu}  \omega_\mu^*(\x) \,\rho_\nu(\x) \, d^4 \x \,,
\qquad
 \bm{\ast} (\omega)   \wedge \rho^*  =   \eta^{\mu\nu} \chi^\rho{}_\mu \triangleright \left( \omega_\rho(\x) \,\rho^*_\nu(\x) \right) \, d^4 \x \,,
\end{equation}
one gets that the inner product they define is identical to~\eqref{InnerProductBetweenForms}:
\begin{equation}
\int  \bm{\ast} (\omega^*)   \wedge \rho  = \int  \omega^*   \wedge \bm{\ast} (\rho)  \,, \qquad
\int  \bm{\ast} (\omega)   \wedge \rho^* = \int  \omega  \wedge \bm{\ast} (\rho^*) \,,
\end{equation}
and these identities are easy to prove also for forms of degrees other than one.
The conclusion is that there is only two independent ways of writing the inner products between two forms of the same degree:
\begin{equation}
\begin{gathered}
\int \omega^* \wedge \bm{\ast} (\rho)  = (\omega , \rho ) = \int \bm{\ast} (\omega^*) \wedge \rho \,, 
\\
\int \omega \wedge \bm{\ast} (\rho^*) = (\rho,\omega) = \int \bm{\ast} (\omega) \wedge \rho^*\,,
\end{gathered}
\end{equation}
which are the complex conjugate of each other, $(\omega , \rho )  = \overline{(\rho , \omega )}$. 

Finally, notice that the Hodge-$\bm{\ast}$ commutes with the involution:
 \begin{equation}
\begin{aligned}
 &\bm{\ast}\left[\omega^*(\x)\right]
 =  \omega^*(\x) \, d^4 \x 
  =  \left[\bm{\ast}(\omega[\x]) ]\right)^* \,,
\\
 &\bm{\ast}\left[(d\x^\mu \, \omega_\mu[\x] )^*\right]
 =  {\sfrac 1 {3!}} \varepsilon^\mu{}_{\nu\rho\sigma} \,\omega_\mu^*(\x)\, d\x^\nu \wedge d\x^\rho \wedge d\x^\sigma =  \left[\bm{\ast}(d\x^\mu\, \omega_\mu(\x) ]\right)^* \,,
\\
 &\bm{\ast}\left[(d\x^\mu \wedge d\x^\nu \, \omega_{\mu\nu}[\x] )^*\right]
  =
  {\sfrac 1 {2!}}  \varepsilon^{\mu\nu}{}_{\rho\sigma} \,\omega_{\mu\nu}^*(\x)\, d\x^\rho \wedge d\x^\sigma  =  \left[\bm{\ast}(d\x^\mu\wedge d\x^\nu \, \omega_{\mu\nu}(\x) ]\right)^* \,,
  \\
&\bm{\ast} \left[(d\x^\mu \wedge d\x^\nu \wedge d\x^\rho\, \omega_{\mu\nu\rho}[\x] )^*\right]   =\varepsilon^{\mu \nu\rho}{}_\sigma  \, \omega_{\mu\nu\rho}^*(\x) \, d \x^\sigma =
\bm{\ast} \left[d\x^\mu \wedge d\x^\nu \wedge d\x^\rho\, \omega_{\mu\nu\rho}(\x)\right]^* \,, 
\\
  &\bm{\ast} \left[(d\x^\mu \wedge d\x^\nu \wedge d\x^\rho \wedge d \x^\sigma\, \omega_{\mu\nu\rho\sigma}[\x] )^*\right] 
  = \varepsilon^{\mu\nu\rho\sigma} \, \omega^*_{\mu\nu\rho\sigma}(\x)
 \\
 &=\bm{\ast} \left[d\x^\mu \wedge d\x^\nu \wedge d\x^\rho \wedge d \x^\sigma\, \omega_{\mu\nu\rho\sigma}(\x)\right]^* \,. 
  \end{aligned}
 \end{equation}

\subsection{Lie and inner derivatives}

Following~\cite{Radko_1997,Brzezinski:1992gv}, we can define a notion of Lie derivative if we consider the translation Hopf subalgebra of $U_\ell[iso(3,1)]$, generated by the momentum operators $P^\ell_\mu$. The Lie derivative associates a generic function of the momenta $X$ to an operator 
\begin{equation} 
\pounds_X : \Gamma^n_\ell \to \Gamma^n_\ell \,,
\end{equation}
such that
\begin{equation}
\pounds_X \triangleright (d\x^{\mu_1} \wedge \dots \wedge d\x^{\mu_n} \, \omega_{\mu_1  \dots  \mu_n}) = i\,  d\x^{\mu_1} \wedge \dots \wedge d\x^{\mu_n} \, (X \triangleright \omega_{\mu_1  \dots  \mu_n} ) \,,
\end{equation}
where $X \triangleright f(\x) =  \int d \mu^\st{L}  (p)\, X(p) \, \tilde{f}_\st{L} (p) \, \E[p]$. Notice that the definition of $\pounds_X$ above, compared to the usual expression of the Lie derivative of an n-form in differential geometry, is missing the part that contains the derivatives of the components of the vector field. Therefore $X$ is a noncommuative generalization of the notion of constant vector field. 
Because the momenta commute, one can show that
\begin{equation}
\begin{aligned}
&\pounds_X \triangleright (\omega \wedge \rho) = 
\pounds_X \triangleright (d\x^{\mu_1} \wedge \dots \wedge d\x^{\mu_n} \, \omega_{\mu_1  \dots  \mu_n} \, \wedge
d\x^{\nu_1} \wedge \dots \wedge d\x^{\nu_m} \, \rho_{\nu_1  \dots  \nu_m}) 
\\
&\qquad=
\pounds_X \triangleright [ d\x^{\mu_1} \wedge \dots \wedge d\x^{\mu_n}  \, \wedge
d\x^{\rho_1} \wedge \dots \wedge d\x^{\rho_m} \, (\chi^{\nu_1}{}_{\rho_1} \dots  \chi^{\nu_m}{}_{\rho_m} \triangleright \omega_{\mu_1  \dots  \mu_n})\, \rho_{\nu_1  \dots  \nu_m} ]
\\
&\qquad=
i\, d\x^{\mu_1} \wedge \dots \wedge d\x^{\mu_n}  \, \wedge
d\x^{\rho_1} \wedge \dots \wedge d\x^{\rho_m} \,  X \triangleright [(\chi^{\nu_1}{}_{\rho_1} \dots  \chi^{\nu_m}{}_{\rho_m} \triangleright \omega_{\mu_1  \dots  \mu_n})\, \rho_{\nu_1  \dots  \nu_m} ]
\\
&\qquad=
 i\,  d\x^{\mu_1} \wedge \dots \wedge d\x^{\mu_n}  \, \wedge
d\x^{\rho_1} \wedge \dots \wedge d\x^{\rho_m} \,    (\chi^{\nu_1}{}_{\rho_1} \dots  \chi^{\nu_m}{}_{\rho_m} X^{(1)} \triangleright \omega_{\mu_1  \dots  \mu_n})\, (X^{(2)} \triangleright\rho_{\nu_1  \dots  \nu_m} )
\\
&\qquad= i\, d\x^{\mu_1} \wedge \dots \wedge d\x^{\mu_n} \,(X^{(1)} \triangleright \omega_{\mu_1  \dots  \mu_n} ) \, \wedge
d\x^{\nu_1} \wedge \dots \wedge d\x^{\nu_m} \, (X^{(2)} \triangleright\rho_{\nu_1  \dots  \nu_m} )
\\
&\qquad= (\pounds_{X^{(1)}} \triangleright \omega) \wedge (\pounds_{X^{(2)}} \triangleright \rho)  \,,
\end{aligned}
\end{equation}
where I used Sweedler's notation for the coproduct~\cite{majid_1995}.
So the Lie derivative does not satisfy the (graded) Leibniz rule, but rather a twisted version of it that depends on the coproduct of $X$.
For the same reason (the commutativity of momenta), it is easy to show that the Lie derivative and the differential commute:
\begin{equation}
\pounds_X \circ d = d \circ \pounds_X \,.
\end{equation} 
A set of four inner derivations can be defined as follows:
\begin{equation}
i_\mu : \Gamma^n_\ell \to \Gamma^{n-1}_\ell  \,,
\end{equation}
where
\begin{equation}
i_\mu (d\x^{\mu_1} \wedge \dots \wedge d\x^{\mu_n} \, \omega_{\mu_1  \dots  \mu_n}) = \delta_\mu^{[\mu_1} \, d\x^{\mu_2} \wedge \dots \wedge d\x^{\mu_n]} \, \omega_{\mu_1  \dots  \mu_n} \,,
\end{equation}
and, again, the operators $i_\mu$ do not satisfy the graded Leibniz rule. However, they satisfy a version of the Cartan formula:
\begin{equation}\label{CartanIdentity}
\pounds_{\X_\mu} = d \circ i_\mu + i_\mu \circ d \,,
\end{equation}
where $\X_\mu$ is the covariant momentum operator~\eqref{ExteriorDifferentialExplicitForm}. It is easy to verify this relation, for example, on one-forms:
\begin{equation}
\begin{aligned}
&\omega = d\x^\mu \, \omega_\mu \,,
\\
&d \circ i_\mu (\omega) = d \omega_\mu = i\,  dx^\nu \, \X_\nu \triangleright \omega_\mu \,,
\\
&i_\mu \circ d (\omega) = - i \,i_\mu  (d \x^\rho \wedge d \x^\nu \, \X_\nu \triangleright \omega_\rho ) = -  i \, d \x^\nu \, \X_\nu \triangleright \omega_\mu +
 d \x^\rho  \, \X_\mu \triangleright \omega_\rho \,,
\\
&d \circ i_\mu (\omega) + i_\mu \circ d (\omega) =  i \, d \x^\rho  \, \X_\mu \triangleright \omega_\rho  = \pounds_{\X_\mu} (\omega) \,,
\end{aligned}
\end{equation}
and two-forms:
\begin{equation}
\begin{aligned}
&\omega = d\x^\mu  \wedge d\x^\nu  \, \omega_{\mu\nu} \,,
\\
&d \circ i_\mu (\omega) = d (  d\x^\nu  \, \omega_{\mu\nu} - d\x^\nu    \, \omega_{\nu\mu})= - 2 \, i \, d\x^\nu \wedge d \x^\rho \, \X_\rho \triangleright \omega_{\mu\nu}  
\,,
\\
&i_\mu \circ d (\omega) =  i\,  i_\mu  (d \x^\nu \wedge d \x^\rho  \wedge d \x^\sigma\, \X_\sigma \triangleright \omega_{\nu\rho} ) = 2 \, i \,  d \x^\rho  \wedge d \x^\sigma\, \X_\sigma \triangleright \omega_{\mu\rho} + i\, d \x^\nu \wedge d \x^\rho \, \X_\mu \triangleright \omega_{\nu\rho}  \,,
\\
&d \circ i_\mu (\omega) + i_\mu \circ d (\omega) =   i\, d \x^\nu \wedge d \x^\rho \, \X_\mu \triangleright \omega_{\nu\rho}   = \pounds_{\X_\mu} (\omega) \,.
\end{aligned}
\end{equation}
Finally, the inner derivations satisfy the standard anticommutativity:
\begin{equation}
i_\mu [i_\nu (\omega) ] = - i_\nu [ i_\mu (\omega) ] \,,
\end{equation}
which is trivial to prove.

\newpage

\section{Classical field theories}\label{Sec:FieldTheory}

I will begin with some preliminary remarks on the general structure of quadratic terms in the action of a field theory. Consider, for instance, a mass term for a complex scalar field $\phi(\x)$. There are two possibilities for writing such a term. The first is:
\begin{equation}
\int \phi^* \, \phi \, d^4 \x =  (2\pi)^4 \int \widetilde{\phi^*}_\st{L}(p) \tilde{\phi}_\st{L} [S(p)]  \, d\mu^\st{L}(p) = (2\pi)^4 \int \overline{\tilde{\phi}_\st{R}[S(p)]} \, \tilde{\phi}_\st{L} [S(p)]\, d\mu^\st{L}(p) \,,
\end{equation}
where I used relation~\eqref{FourierTransformStarredFunction} for the Fourier transform of $\phi^*(\x)$. Transforming $p \to S(p)$ and using the relation~\eqref{LeftRightHaarMeasureRelation} between the left and right measures:
\begin{equation}
\int \phi^* \, \phi \, d^4 \x = (2\pi)^4 \int \overline{\tilde{\phi}_\st{R}(p)} \, \tilde{\phi}_\st{L} (p)\, d\mu^\st{R}(p)\,,
\end{equation}
and, finally, using Eq.~\eqref{LeftRightFourierTransformRelation} relating the left and right Fourier transforms:
\begin{equation}\label{Eq:MassTerm1}
\int \phi^* \, \phi \, d^4 \x =  (2\pi)^4 \int  \left| \tilde{\phi}_\st{L} (p) \right|^2 \, d\mu^\st{L}(p) \,,
\end{equation}
which is an expression that highlights the fact that the integral is real and positive.
The second choice for a mass term exchanges $\phi(\x)$ and $\phi^*(\x)$:
\begin{equation}
\int \phi \, \phi^* \, d^4 \x =  (2\pi)^4 \int \tilde{\phi}_\st{L}(p) \widetilde{\phi^*}_\st{L} [S(p)]\, d\mu^\st{L}(p) = (2\pi)^4 \int \tilde{\phi}_\st{L}(p) \, \overline{\tilde{\phi}_\st{R}(p)}\, d\mu^\st{L}(p) \,,
\end{equation}
again, using Eq.~\eqref{LeftRightFourierTransformRelation} to turn 
$\tilde{\phi}_\st{L}(p) \, d\mu^\st{L}(p)$ into $\tilde{\phi}_\st{R}(p) \, d\mu^\st{R}(p)$ lets us write the expression as a quadratic form:
\begin{equation}
\int \phi \, \phi^* \, d^4 \x =  (2\pi)^4 \int \left| \tilde{\phi}_\st{R}(p) \right|^2 \, d\mu^\st{R}(p) \,,
\end{equation}
however, to compare it to Eq.~\eqref{Eq:MassTerm1} we need to express everything in terms of either the left or the right Fourier transform, \textit{e.g.:}
\begin{equation}
\int \phi \, \phi^* \, d^4 \x =  (2\pi)^4 \int \left| \tilde{\phi}_\st{L}(p) \right|^2 \, \mathscr{T}^{-1}(p) \, d\mu^\st{L}(p) \,.
\end{equation}
The two type of terms are equivalent only in the unimodular case, when 
$\mathscr{T}(p) =  1$, otherwise we have, in principle, to include both in the action. Another potential issue is that, in general, nothing ensures that the modular function is Lorentz-invariant, so, in models where it is not, one could be possibly facing Lorentz-breaking choices of terms in the Lagrangian, which might have to be excluded.

Again, recall  that the map  $k \to S(k)$, which I used in the calculations above, might not be bijective, or there might be other global issues in performing this transform which require a separate discussion in each T-Minkowski model. As I already remarked, here I will ignore these issues and focus on the general features of all T-Minkowski models.

\subsection{Complex scalar field}

The previous section clarifies that we have two ways of writing a Poincar\'e-invariant kinetic functional that is quadratic in $d\phi$:
\begin{equation}
(d\phi,d\phi) = \int  d\phi^*   \wedge \bm{\ast} (d\phi)  = -
\int \left[ \X^\mu \triangleright \phi (\x) \right]^*  \, \X_\mu \triangleright \phi (\x)\, d^4\x
 =
 \int S(\X)^\mu \triangleright \phi^* (\x) \, \X	_\mu \triangleright \phi (\x)\, d^4\x \,,
\end{equation}
\begin{equation}
(d\phi^*,d\phi^*) = \int  d\phi \wedge \bm{\ast} (d\phi^*)  =
 -  \int  \X^\mu \triangleright \phi (\x)  \, \left[\X_\mu \triangleright \phi (\x) \right]^* \, d^4\x
 = \int \X^\mu \triangleright \phi (\x) \, S(\X)_\mu \triangleright \phi^* (\x)\, d^4\x \,,
\end{equation}
where I used the Minkowski metric to raise and lower indices.
In terms of the Fourier transform of the scalar field:
\begin{equation}
\begin{aligned}
&\widetilde{(\X_\mu \triangleright \phi)}_\st{L} (k) =  
\X_\mu(k) \tilde{\phi}_\st{L}(k) \,,&
&\widetilde{(S(\X)_\mu \triangleright \phi)}_\st{L} (k) =  
S(\X)_\mu(k) \tilde{\phi}_\st{L}(k) \,,
\\
&\widetilde{(\X_\mu \triangleright \phi^*)}_\st{L} (k) =  
\X_\mu(k) \overline{\tilde{\phi}_\st{R}[S(k)]} \,,&
&\widetilde{(S(\X)_\mu \triangleright \phi^*)}_\st{L} (k) =  
S(\X)_\mu(k)\overline{\tilde{\phi}_\st{R}[S(k)]} \,,
\end{aligned}
\end{equation}
where we used expression~\eqref{FourierTransformStarredFunction} for the Fourier transform of the conjugate of a field. Using the expression for the integral of a product of two functions, Eq.~\eqref{IntegralTwoFunctionsFourierTransform}, we get
\begin{align}
&\begin{aligned}
(d\phi,d\phi) =&   (2\pi)^4 \int \,S(\X)^\mu(k) S(\X)_\mu(k)  \, \overline{\tilde{\phi}_\st{R} [S(k)]} \tilde{\phi}_\st{L}[S(k)] \, d\mu^\st{L}(k) 
\\
=&  -(2\pi)^4 \int \, \square_\ell(k) \, \overline{\tilde{\phi}_\st{R} [S(k)]} \tilde{\phi}_\st{L}[S(k)] \, d\mu^\st{L}(k)
=
- (2\pi)^4 \int \, \square_\ell(k) \, \left| \tilde{\phi}_\st{L} (k) \right|^2 \, d\mu^\st{L}(k)
 \,,
\end{aligned}
\\
&\begin{aligned}
(d\phi^*,d\phi^*) =&  (2\pi)^4 \int \,\X^\mu(k) \X_\mu(k)  \, \tilde{\phi}_\st{L}(k) \overline{\tilde{\phi}_\st{R}(k)} \, d\mu^\st{L}(k) 
\\
=&  - (2\pi)^4 \int \,\square_\ell(k)  \, \tilde{\phi}_\st{L}(k) \overline{\tilde{\phi}_\st{R}(k)} \, d\mu^\st{L}(k)
=
- (2\pi)^4 \int \, \square_\ell(k) \, \left| \tilde{\phi}_\st{R} (k) \right|^2 \, d\mu^\st{R}(k)
 \,,
\end{aligned}
\end{align}
the momentum-space function:
\begin{equation}
\square_\ell(p) = - \eta^{\mu\nu} \X_\mu(p)\X_\nu(p) = - \eta^{\mu\nu} S(\X)_\mu(p)S(\X)_\nu(p) \,,
\end{equation}
is Lorentz-invariant, and reduces to the D'Alembert operator of Minkowski spacetime in the $\ell \to 0$ limit (since $\X_\mu(p) \xrightarrow[\ell \to 0]{}  p_\mu$). This is our deformed mass Casimir, which determines the mass shells of the theory.

The most general T-Poincar\'e-invariant action for a complex scalar field, which reduces to the correct expression in the commutative limit is then 
\begin{equation}\label{ScalarFieldAction}
\mathcal{S} =  {\sfrac 1 2} \left[ \alpha \, (d\phi,d\phi) + (1-\alpha) \, (d\phi^*,d\phi^*)  \right] - {\sfrac{m^2}{2}}\left[ \beta \, (\phi,\phi)+(1-\beta) \, (\phi^*,\phi^*)  \right] \,. 
\end{equation}
In Fourier transform, the action takes the form
\begin{equation}\label{Eq:ScalarAction_FourierTransform}
\mathcal{S}  =  (2\pi)^4 \int  \, \mathcal{C}(p) \, \left| \tilde{\phi}_\st{L}(k)
 \right|^2 \, d\mu^\st{L}(k) \,,
\end{equation}
where
\begin{equation}
\mathcal{C}(p) = \square_\ell(k) \bigg{[} {\sfrac \alpha 2} 
+ {\sfrac {1-\alpha} 2} \, \mathscr{T}^{-1}(k)\bigg{]}  - m^2 \bigg{[} {\sfrac \beta 2}  
+ {\sfrac {1-\beta} 2} \, \mathscr{T}^{-1}(k)\bigg{]}  \,.
\end{equation}
The action~\eqref{Eq:ScalarAction_FourierTransform} is now perfectly equivalent to a commutative field theory, written in Fourier transform. We can extremize it by varying with respect to $\tilde{\phi}_\st{L}(k)$ and its complex conjugate, as we would do in the commutative case:
\begin{equation}
\delta \mathcal{S}  = (2\pi)^4 \int  \mathcal{C}(k) \, \left(\delta  \overline{\tilde{\phi}_\st{L} (k)} \, \tilde{\phi}_\st{L}(k) +  \overline{\tilde{\phi}_\st{L} (k)} \,\delta  \tilde{\phi}_\st{L}(k)  \right) \, d\mu^\st{L}(k) \,,
\end{equation}
which gives the equations of motion
\begin{equation}\label{Eq:ScalarEOM}
\mathcal{C}(k) \, \overline{\tilde{\phi}_\st{L} (k)}
= \mathcal{C}(k) \, \tilde{\phi}_\st{L} (k) = 0\,.
\end{equation}
The extremizing field configurations take the form
\begin{equation}\label{Eq:OnShellScalarFields}
 \tilde{\phi}_\st{L}(q)
 =   K_\st{L}(q) \, \delta \left[ \mathcal{C}(q) \right]  \,,
\qquad
 \overline{\tilde{\phi}_\st{R} (q)} \,,
  =  \overline{K_\st{L}(q)} \, \delta\left[ \mathcal{C}(q) \right]   \,,
\end{equation}
\textit{i.e.} on-shell fields, whose Fourier transform has support on the mass-shell $\mathcal{C}(p)= 0$. Since $\mathscr{T} \geq 0$, the mass-shell $\mathcal{C}(p) = 0$ collapses to $\square_\ell(p) + m^2 = 0$ in the symplifying case $\alpha = \beta$. If there are models where the modular function is not Lorentz-invariant, it may be necessary to make the assumption $\alpha = \beta$, or even $\alpha = \beta = 1$ in order to define a Lorentz-invariant model.

Notice that this result could also have been obtained without resorting to the Fourier transform. The action $\mathcal{S} : C_\ell[\mathbbm{R}^{3,1}] \to \mathbbm{R}$ is a real functional of the noncommutative field $\phi (\x)$, and it makes sense to ask on which field configurations (\textit{i.e.} elements of $C_\ell[\mathbbm{R}^{3,1}]$) it is extremized. The notion of functional variation exists for noncommutative fields too: one considers the following map:
\begin{equation}
\phi(\x) \to \phi(\x)  + \epsilon \, \delta \phi(\x) \,,
\end{equation}
where $\delta \phi(\x) \in C_\ell[\mathbbm{R}^{3,1}]$ and $\epsilon \in \mathbbm{R}$. This is a map from $C_\ell[\mathbbm{R}^{3,1}]$ to itself. Under this map, the action functional transforms to:
\begin{equation}
 \mathcal{S}[\phi] \to \mathcal{S}[\phi  + \epsilon \, \delta \phi] \,,
\end{equation}
and it can be expanded up to first order in $\epsilon$:
\begin{equation}
\mathcal{S}[\phi  + \epsilon \, \delta \phi] \sim \mathcal{S}[\phi]  + \epsilon \, \delta \mathcal{S}[\phi] \,,
\end{equation}
where $\delta \mathcal{S}[\phi] $ is yet another real functional on $C_\ell[\mathbbm{R}^{3,1}]$. 
Asking which field configurations annihilate $ \delta \mathcal{S}[\phi]$ gives the same answer as the Fourier-transform procedure outlined above. To see this, let us write the variation of the action as the integral of a zero-form:
\begin{equation}
\begin{aligned}
\delta \mathcal{S} =
{\sfrac 1 2 } \int \bigg{\{}&   \alpha \, [ S(\X)^\mu \triangleright \delta \phi^* (\x) ]\, [\X_\mu \triangleright \phi (\x)] + (1-\alpha) \, [\X^\mu \triangleright \phi (\x)] \,[ S(\X)_\mu  \triangleright \delta\phi^* (\x) ]
\\
& + \alpha \,  [S(\X)^\mu \triangleright \phi^* (\x) ]\, [\X_\mu \triangleright \delta \phi (\x)] + (1-\alpha) \, [\X^\mu \triangleright \delta  \phi (\x) ]\,[ S(\X)_\mu \triangleright \phi^* (\x) ]
\\
& -\beta \, m^2 \, \delta \phi^* (\x) \phi - (1-\beta) m^2 \, \phi(\x)\delta \phi^* (\x)  \bigg{\}} 
\\
& -\beta \, m^2 \, \phi^* (\x) \delta \phi - (1-\beta) m^2 \, \delta \phi(\x)\phi^* (\x)  \bigg{\}} d^4\x \,,
\end{aligned}
\end{equation}
using the coproduct and counit of $\X_\mu$, Eq.~\eqref{ExteriorDifferentialExplicitForm}, we can integrate by parts the $\X_\mu$ operators that act on $\delta \phi$ 	and $\delta \phi^*$:
\begin{equation}
\begin{aligned}
&\int f(\x) \left[ \X_\mu \triangleright g(\x) \right] d^4\x 
=\cancel{\int \X_\nu \triangleright \left\{ \left[(\chi^{-1})^\nu{}_\mu \triangleright f(\x) \right] \, g(\x) 
\right\} \, d^4 \x} -  \int \left[(\chi^{-1})^\nu{}_\mu \,  \X_\nu \triangleright f(\x) \right] \, g(\x) \, d^4 \x\,,
\\
&\int \left[ \X_\mu \triangleright f(\x) \right] g(\x) \,d^4\x 
=\cancel{\int \X_\nu \triangleright \left[   f(\x) \, g(\x) 
\right] \, d^4 \x} -  \int \left[\chi^\nu{}_\mu \, \triangleright f(\x) \right] \left[  \X_\nu  \triangleright g(\x) \right] d^4 \x \,,
\end{aligned}
\end{equation}
and similarly for the antipode of $\X_\mu$. So the kinetic part of the variation takes the form
\begin{equation}
\begin{aligned}
\delta \mathcal{S}_\st{Kin} =&
- {\sfrac 1 2 } \int \bigg{\{}   \alpha \,   \delta \phi^* (\x) \, \chi^\nu{}_\mu S(\X)^\nu \X_\mu \triangleright \phi (\x) + (1-\alpha) \left[ S(\X)_\nu   \X^\mu \triangleright \phi (\x) \right] \left[ (\chi^{-1})^\nu{}_\mu \delta \phi^* (\x)  \right]
\\
& + \alpha \left[ (\chi^{-1})^\nu{}_\mu \,  \X_\nu S(\X)^\mu \triangleright \phi^* (\x) \right]   \delta \phi (\x) + (1-\alpha) \, [ \chi^\mu{}_\nu \triangleright \delta  \phi (\x) ] \, [\X_\nu S(\X)_\mu \triangleright \phi^* (\x) ] \bigg{\}} d^4\x \,,
\end{aligned}
\end{equation}
and, integrating by part also the $\chi^\mu{}_\nu$ operators (using their group-like coproduct):
\begin{equation}
\begin{aligned}
\delta \mathcal{S}_\st{Kin}  =&
- {\sfrac 1 2 } \int \bigg{\{}   \alpha \,   \delta \phi^* (\x) \, \chi^\nu{}_\mu S(\X)_\nu \X^\mu \triangleright \phi (\x) + (1-\alpha) \left[ \chi^\nu{}_\mu  S(\X)_\nu   \X^\mu \triangleright \phi (\x) \right]  \delta \phi^* (\x) 
\\
& + \alpha \left[ (\chi^{-1})^\nu{}_\mu \,  \X_\nu S(\X)^\mu \triangleright \phi^* (\x) \right]   \delta \phi (\x) + (1-\alpha) \,   \delta  \phi (\x)   \, [(\chi^{-1})^\nu{}_\mu \X_\nu S(\X)^\mu \triangleright \phi^* (\x) ] \bigg{\}} d^4\x \,,
\end{aligned}
\end{equation}
this is identical to
\begin{equation}
\begin{aligned}
 \delta \mathcal{S}_\st{Kin}  =&
{\sfrac 1 2 } \int \bigg{\{}    \alpha \,   \delta \phi^* (\x) \, \X^\mu \X_\mu \triangleright \phi (\x) + (1-\alpha) \left[\X_\mu   \X^\mu \triangleright \phi (\x) \right]  \delta \phi^* (\x) 
\\
& +\alpha \left[ S(\X)_\mu S(\X)^\mu \triangleright \phi^* (\x) \right]   \delta \phi (\x) +(1-\alpha) \,   \delta  \phi (\x)   \, [S(\X)_\mu S(\X)^\mu \triangleright \phi^* (\x) ] \bigg{\}} d^4\x \,.
\end{aligned}
\end{equation}
The whole variation then reads
\begin{equation}
\begin{aligned}
&\delta \mathcal{S} =-
{\sfrac 1 2 } \int \bigg{\{}   \delta \phi^* (\x) \left[ \alpha \,   \square_\ell\triangleright \phi (\x) + \beta \, m^2 \, \phi (\x)  \right] + \left[ (1-\alpha) \, \square_\ell \triangleright \phi (\x) + (1-\beta) \, m^2 \, \phi(\x) \right]  \delta \phi^* (\x) 
\\
&+ \left[ \alpha \, \square_\ell \triangleright \phi^* (\x)  + \beta \, m^2 \, \phi^*(\x) \right]   \delta \phi (\x) + \delta  \phi (\x)   \, \left[ (1-\alpha) \,  \square_\ell \triangleright \phi^* (\x)  + (1-\beta) m^2 \, \phi^* (\x) \right]  \bigg{\}} d^4\x \,.
\end{aligned}
\end{equation}
Using the modular function~\eqref{Eq:TwistedCiclicity}, we can reorder all the variations on the same side and group their coefficients together:
\begin{equation}
\begin{aligned}
&\delta \mathcal{S} = -
{\sfrac 1 2 } \int \bigg{\{}   \delta \phi^* (\x) \left[ \alpha \,   \square_\ell\triangleright \phi (\x) + \beta \, m^2 \, \phi (\x)   +  \mathscr{T}^{-1} \triangleright \left( (1-\alpha) \, \square_\ell \triangleright \phi (\x) + (1-\beta) \, m^2 \, \phi(\x) \right) \right]   
\\
&+ \left[ \alpha \, \square_\ell \triangleright \phi^* (\x)  + \beta \, m^2 \, \phi^*(\x)  +  \mathscr{T} \triangleright  \left( (1-\alpha) \,  \square_\ell \triangleright \phi^* (\x)  + (1-\beta) m^2 \, \phi^* (\x) \right) \right] \delta  \phi (\x)  \bigg{\}} d^4\x \,,
\end{aligned}
\end{equation}
and the extremizing field configurations then satisfy
\begin{equation}
\begin{gathered}
\alpha \,   \square_\ell\triangleright \phi (\x) + \beta \, m^2 \, \phi (\x)   +  \mathscr{T}^{-1} \triangleright \left( (1-\alpha) \, \square_\ell \triangleright \phi (\x) + (1-\beta) \, m^2 \, \phi(\x) \right)  = 0 \,,
\\
 \alpha \, \square_\ell \triangleright \phi^* (\x)  + \beta \, m^2 \, \phi^*(\x)  +  \mathscr{T} \triangleright  \left( (1-\alpha) \,  \square_\ell \triangleright \phi^* (\x)  + (1-\beta) m^2 \, \phi^* (\x) \right)   = 0 \,,
\end{gathered}
\end{equation}
which, written in Fourier transform, are equivalent to~\eqref{Eq:ScalarEOM}.

\subsection{Dirac field}

In order to develop the noncommutative version of a Dirac action, it is necessary to recall the differential-geometric formulation of Dirac fields. I will do this directly in the noncommutative case, which follows closely the commutative one, which can be found for example in Ref.~\cite{Mercuri:2006um}.
In order to find an intrinsic equivalent of the slashed derivative, we can introduce a one-form of  gamma matrices as follows
\begin{equation}
\hat \gamma = d\x^\mu \, \gamma_\mu  \,,
\end{equation}
where the $\gamma_\mu$ are the standard (commutative) gamma matrices associated to the mostly-positive Minkowski metric, with covariant indices:
\begin{equation}
\gamma_\mu = \left\{ \, 
{\small \begin{pmatrix}-1&0&0&0\\0&1&0&0\\0&0&-1&0\\0&0&0&-1\end{pmatrix}}, \, 
{\small \begin{pmatrix}0&0&0&-1\\0&0&1&0\\0&-1&0&0\\1&0&0&0\end{pmatrix}}, \,
{\small \begin{pmatrix}0&0&0&i\\0&0&i&0\\0&i&0&0\\i&0&0&0\end{pmatrix}}, \,
{\small \begin{pmatrix}0&0&-1&0\\0&0&0&-1\\1&0&0&0\\0&1&0&0\end{pmatrix}} \, \right\} \,.
\end{equation}
The standard form of the Dirac action in the intrinsic formulation can be generalized straightforwardly to the noncommutative case:
\begin{equation}\label{Eq:Basic_Dirac_action}
\mathcal{S} =  \int \left(i \, \bar{\psi} \, \hat{\gamma} \wedge \bm{\ast} \left( d\psi \right) - m \, \bar{\psi} \, \psi  \right) \,,
\end{equation}
where $\psi_\alpha(\x) \in C_\ell[\mathbbm{R}^{3,1}]$ is a four-components spinor that transforms linearly under quantum Poincar\'e transformations:
\begin{equation}
\psi'_\alpha (\x') = ( S^\beta{}_\alpha(\La) \otimes \text{\it{id}})  \, \psi_\beta(\x') \,,
\end{equation}
$S^\beta{}_\alpha(\La)$ is a standard spinorial representation of the Lorentz group. The barred spinors are defined as in the commutative case:
\begin{equation}
\bar \psi  = \psi^* \, \gamma^0 \,.
\end{equation}
One can see Dirac spinors as elements of the tensor product $C_\ell[\mathbbm{R}^{3,1}] \otimes \mathbbm{C}^4$, where $ \mathbbm{C}^4$ is the complex vector space of four-component spinors. In the non-cyclic cases, the action~\eqref{Eq:Basic_Dirac_action} is not the most generic one. At the expense of the elegance of the notation, I can write another Lagrangian 4-form that reduces to the correct one in the commutative limit, however, I can only do so with the spinor indices explicited. Call $\alpha,\beta,\delta , \dots =  \{1,2,3,4\}$. The kinetic term in  the action~\eqref{Eq:Basic_Dirac_action} then reads:
\begin{equation}\label{Eq:Basic_Dirac_action_w/indices}
i \, \bar{\psi} \, \hat{\gamma} \wedge \bm{\ast} \left( d\psi \right)
=
- \psi^*_\alpha (\x) \,(\gamma_0)_{\alpha\beta} (\gamma_\mu)_{\beta\delta} \left[  \X^\mu \triangleright \psi_\delta (\x)  \right] \,,  
\end{equation}
but we can also write a term like this:
\begin{equation}
 \left[  \X^\mu \triangleright \psi_\delta (\x)  \right] \, \psi^*_\alpha (\x)  \,(\gamma_0)_{\alpha\beta} (\gamma_\mu)_{\beta\delta} \,,  
\end{equation}
which coincides with the one above only in the cyclic cases. A similar choice is present also for the mass term:
\begin{equation}
m \int  \bar{\psi} \, \psi  \, d^4\x = m \int \psi^*_\alpha(\x)  \, (\gamma^0)_{\alpha\beta} \psi_\beta(\x)   \, d^4\x \neq 
 m \int (\gamma^0)_{\alpha\beta} \psi_\beta(\x)   \,  \psi^*_\alpha(\x)   \, d^4\x \,.
\end{equation}
In the present paper, I will focus only on the choice~\eqref{Eq:Basic_Dirac_action} for the action, for illustrative purposes.
In Fourier transform, the action reads
\begin{equation} 
\begin{aligned}
\mathcal{S} =&  \int \left(  - \X^\mu[S(k)]  \, \overline{\tilde{\psi}^\st{R}_\alpha [S(k)]} \,(\gamma_0)_{\alpha\beta} (\gamma_\mu)_{\beta\delta}  \, \tilde{\psi}^\st{L}_\delta [S(k)]    - m \,  \overline{\tilde{\psi}^\st{R}_\alpha [S(k)]} \, (\gamma_0)_{\alpha\beta} \,  \tilde{\psi}^\st{L}_\beta [S(k)] \right) d\mu^\st{L}(k) \,,  
\\
=& - \int   \overline{\tilde{\psi}^\st{L}_\alpha (k)} \,(\gamma_0)_{\alpha\beta} \left[  \X^\mu(k) (\gamma_\mu)_{\beta\delta} + m \, \delta_{\beta\delta} \right]  \, \tilde{\psi}^\st{L}_\delta (k)   \,  d\mu^\st{L}(k) \,.  
\end{aligned}
\end{equation}
where, again ignoring global issues on momentum space, I transformed $k \to S(k)$ and used the properties of the left- and right-invariant measures. The above action is extremized by the solutions of the noncommutative Dirac equation:
\begin{equation}
i \, \slashed{\partial} \triangleright \psi (\x) - m \, \psi(\x) = 0 \,,
\end{equation}
where the slashed derivative is a linear operator $\slashed{\partial} : C_\ell[\mathbbm{R}^{3,1}] \otimes  \mathbbm{C}^4 \to C_\ell[\mathbbm{R}^{3,1}] \otimes \mathbbm{C}^4$ defined by the following integral kernel:
\begin{equation}
(\slashed{\partial} \triangleright \psi(\x) )_\alpha = i \, \int  \X_\mu(p) \, (\gamma^\mu)_{\alpha\beta} \, \tilde{\psi}^\st{L}_\beta  (p) \, d\mu^\st{L}(p) \,. 
\end{equation}

\subsection{Maxwell field}

The vector potential is a one-form $A = d\x^\mu \, A_\mu(\x)$, which changes under gauge transformations  as follows
\begin{equation} \label{GaugeTransformExpression}
A' = U^* \, A \, U +  i\, U^* \, d \, U \,,
\end{equation}
where
\begin{equation}
U(\x) = e^{i \, \rho (\x) } \,, \qquad \rho(\x) \in C_\ell[\mathbbm{R}^{3,1}] \,, \qquad \rho^*(\x) = \rho(\x) \,, 
\end{equation}
which implies 
\begin{equation}
U^*(\x) U(\x) = U (\x) U^*(\x) = \1 \,.
\end{equation}
Notice that, unless we are in a particular case in which $[d \rho, \rho] =0$, we cannot assume that $ U^* \, d \, U  = i \, d \rho$. The general expression for the inhomogeneous part of the gauge transform in~\eqref{GaugeTransformExpression} will not be linear in $\rho$. Similarly, unless for some reason $[\rho, A] =0$, the homogeneous part of~\eqref{GaugeTransformExpression} too will depend nonlinearly on $\rho$. 
Eq~\eqref{GaugeTransformExpression}  can be written in components as
\begin{equation}
A'_\mu(\x) = [\chi^\nu{}_\mu \triangleright U^*(\x)] \, A_\nu(\x) \, U(\x) - [\chi^\nu{}_\mu \triangleright U^*(\x)] \, [\X_\nu \triangleright U(\x)]\,.
\end{equation}
So far, the one-form $A$ was not necessarily supposed to be real. Taking the star of the transformation law we get:
\begin{equation}
A'^* = U^* \, A^* \, U - i \,  (d \, U)^*  \, U \,,
\end{equation}
but the complex conjugate of $dU$ can be written as
\begin{equation}
\begin{aligned}
(d U)^* &= [i \, d\x^\mu \,\X_\mu \triangleright U(\x)]^* =  - i \,[\X_\mu \triangleright U(\x)]^* \, d \x^\mu = - i \, S(\X)_\mu \triangleright U^*(\x) \, d \x^\mu \\
&=  i \,(\chi^{-1})^\nu{}_\mu \, \X_\nu \triangleright U^*(\x) \, d \x^\mu = i \,  d\x^\nu \, \X_\nu \triangleright U^*(\x) 
=  d (U^*)\,,
\end{aligned}
\end{equation}
and we can use the unitarity condition $U^*U=\1$ and the Leibniz rule~\eqref{Eq:Exterior_differential_def} to prove
\begin{equation}
0 = d (U^* \, U) = d (U^*) \, U  + U^* \, dU  \,, \qquad \Rightarrow \qquad 
d (U^*) \, U  = -  U^* \, dU   \,,
\end{equation}
so the complex conjugate of $A'$ is the same as the gauge transformation of $A^*$ by $U$:
\begin{equation}
A'^* = U^* \, A^* \, U - i\, (d \, U)^*  \, U
= U^* \, A^* \, U + i \,  U^* \, dU  \,,
\end{equation}
this reassures us on the gauge-covariance of the involution.
I can therefore impose a reality condition on the vector potential
\begin{equation}
A^* =  A \,, \qquad  A^* = A^*_\mu \, d\x^\mu  = d\x^\mu \,  A_\mu ~~ \Rightarrow ~~ A^*_\mu = (\chi^{-1})^\nu{}_\mu \triangleright A_\nu \,,
\end{equation}
this is not an ultralocal expression as in the commutative case. It depends on ``derivatives'' of the fields, and therefore its gauge-covariance is far from obvious. However, the proof above implies the covariance of the reality condition. We can see this in components:
\begin{equation}
\begin{aligned}
[A'_\mu(\x)]^* &= U^*(\x)\, A^*_\nu(\x) \, [\chi^\nu{}_\mu \triangleright U^*(\x)]^* -  [\X_\mu \triangleright U(\x)]^*\, [\chi^\nu{}_\mu \triangleright U^*(\x)]^* 
\\
&=  U^*(\x)\, A^*_\nu(\x) \, [(\chi^{-1})^\nu{}_\mu \triangleright U(\x)]  + [(\chi^{-1})^\rho{}_\nu \X_\rho \triangleright U^*(\x)]\, [(\chi^{-1})^\nu{}_\mu \triangleright U(\x)]
\\
&=  
U^*(\x)\, [(\chi^{-1})^\rho{}_\nu \triangleright A_\rho(\x)] \, [(\chi^{-1})^\nu{}_\mu \triangleright U(\x)]
 + (\chi^{-1})^\nu{}_\mu  \triangleright \left( [\X_\nu \triangleright U^*(\x)] \, U(\x)  \right)
  \\
&=  
 U^*(\x)\,(\chi^{-1})^\rho{}_\mu \triangleright  [A_\rho(\x) \,   U(\x)]
  + (\chi^{-1})^\nu{}_\mu  \triangleright \left( [\X_\nu \triangleright U^*(\x)] \, U(\x)  \right)
    \\
&=  
(\chi^{-1})^\nu{}_\mu \triangleright \left( [\chi^\rho{}_\nu \triangleright U^*(\x)]\, A_\rho(\x) \,   U(\x) 
+ [\X_\nu \triangleright U^*(\x)] \, U(\x)  \right)    \\
&=  
(\chi^{-1})^\nu{}_\mu \triangleright \left( [\chi^\rho{}_\nu \triangleright U^*(\x)]\, A_\rho(\x) \,   U(\x) 
- [\chi^\rho{}_\nu \triangleright U^*(\x)] \, [\X_\rho \triangleright U(\x)]  \right)
\\
&= (\chi^{-1})^\nu{}_\mu \triangleright A'_\nu (\x) \,.
\end{aligned}
\end{equation}
where I used the action~\eqref{Eq:Conjugate_xi_chi} of $\chi^\mu{}_\nu$ and $\X_\rho$  on conjugate functions, and their coproducts~\eqref{ExteriorDifferentialExplicitForm} in order to integrate them by part, as for example in:
\begin{equation}
[\X_\nu \triangleright U^*(\x)] \, U(\x) 
= \cancel{\X_\nu \triangleright [U^*(\x) U(\x)]} - [\chi^\rho{}_\nu \triangleright U^*(\x)] \, [\X_\rho \triangleright U(\x)] \,.
\end{equation}

I can now introduce a Faraday two-form:
\begin{equation}
F = dA + A \wedge A \,,
\end{equation}
where the wedge product between $A$ and itself is not automatically zero because of noncommutativity:
\begin{equation}
A \wedge  A  = d\x^\mu \, A_\mu(\x) \wedge d\x^\nu \, A_\nu(\x)
= d\x^\mu \wedge d\x^\rho \, [ \chi^\nu{}_\rho \triangleright A_\mu(\x) ] A_\nu(\x) \,.
\end{equation}
With this definition, it is easy to prove that $F$ is covariant under gauge transformations:
\begin{equation}
F' = U^* \, F \, U  \,.
\end{equation}
The Faraday form is real if the potential is:
\begin{equation}
F^* = (dA)^* + A^* \wedge A^* = d (A^*) + A^* \wedge A^* =  d A  + A \wedge A = F \,.
\end{equation}
One can write the following two real functionals with $F$: the deformed Maxwell and  Pontryagin actions (modulo normalization constants):
\begin{equation}
\mathcal{S}_\st{Maxwell} = \int F \wedge \bm{\ast} F \,,
\qquad
\mathcal{S}_\st{Pontryagin} = \int F \wedge F  \,,
\end{equation}
these functionals are T-Poincar\'e-invariant, but they are not gauge invariant unless the integral is cyclic. In fact,
\begin{equation}
\begin{aligned}
&F' \wedge \bm{\ast} F' = U^* \, F \, U \wedge U^* \,  \bm{\ast} (F) \, U  = U^* \, F  \wedge  \bm{\ast} (F) \, U  \,,
\\
&F' \wedge F' = U^* \, F \, U \wedge U^* \,  F \, U  = U^* \, F  \wedge  F \, U  \,,
\end{aligned}
\end{equation}
one would like to bring the $U^*$ and $U$ factors that surround around the two Lagrangians in contact by using the cyclicity of the integral:
\begin{equation}
\begin{aligned}
&\int F' \wedge \bm{\ast} F' = \int  U^* \, F  \wedge  \bm{\ast} (F) \, U  = \int  U \, U^* \, F  \wedge  \bm{\ast} (F)= \int  F  \wedge  \bm{\ast} (F) \,,
\\
&\int F' \wedge  F' = \int  U^* \, F  \wedge  F \, U  = \int  U \, U^* \, F  \wedge F= \int  F  \wedge F \,,
\end{aligned}
\end{equation}
however this is of course possible only if the integral is, indeed, cyclic. In the noncyclic cases, alternative (and more complicated) routes need to be taken, if we want to describe gauge fields~\cite{Mathieu:2020ccc,Kupriyanov:2020axe}.

\subsection{Noether theorem, conserved currents and energy-momentum tensor}

%

The differential calculus and the exterior, inner and Lie derivative can be used to introduce a covariant notion of conservation laws for (classical) field theories on any T-Minkowski spacetime~\cite{Mercati:2011aa}. I will only discuss the example of a scalar field here. Consider the simplest example of complex scalar   Lagrangian, \textit{i.e.} Eq.~\eqref{ScalarFieldAction} with $\alpha = \beta = 1$. The Lagrangian 4-form is
\begin{equation}
\mathscr{L} =  {\sfrac 1 2 } \, d\phi^*   \wedge \bm{\ast} (d\phi) - {\sfrac 1 2 } \, m^2 \, \phi^* \, \phi \, d^4 \x \,.
\end{equation}	
The following vector-valued 3-form:
\begin{equation}
j_\mu = {\sfrac 1 2 } \, ( \X_\mu  \triangleright \phi^*) \, \bm{\ast} (d \phi) +  {\sfrac 1 2 } \,  \bm{\ast} [d (\chi^\nu{}_\mu \triangleright \phi^*)] \, (\X_\nu \triangleright  \phi )   - i_\mu \triangleright \mathscr{L} \,,
\end{equation} 
is closed on-shell. In fact:
\begin{equation}
\begin{aligned}
d j_\mu =&  {\sfrac 1 2 } \, d ( \X_\mu  \triangleright \phi^*) \wedge \bm{\ast} (d \phi) -  {\sfrac 1 2 } \,  \bm{\ast} [d (\chi^\nu{}_\mu \triangleright \phi^*)] \wedge d (\X_\nu \triangleright  \phi )
\\& + {\sfrac 1 2 } \, ( \X_\mu  \triangleright \phi^*) \, d\bm{\ast}  d \phi  +  {\sfrac 1 2 } \,  d \bm{\ast}  d (\chi^\nu{}_\mu \triangleright \phi^*)\, (\X_\nu \triangleright  \phi )
 - d \circ i_\mu \triangleright \mathscr{L} 
\,,
\end{aligned}
\end{equation}
the first line can be shown to be equivalent to the Lie derivative of the kinetic term in the Lagrangian:
\begin{equation}
\begin{aligned}
&{\sfrac 1 2 } \, d ( \X_\mu  \triangleright \phi^*) \wedge \bm{\ast} (d \phi) -  {\sfrac 1 2 } \,  \bm{\ast} [d (\chi^\nu{}_\mu \triangleright \phi^*)] \wedge d (\X_\nu \triangleright  \phi ) 
\\
&=
{\sfrac 1 2 } \, d ( \X_\mu  \triangleright \phi^*) \wedge \bm{\ast} (d \phi) +  {\sfrac 1 2 } \,  d (\chi^\nu{}_\mu \triangleright \phi^*)  \wedge \bm{\ast}  [d (\X_\nu \triangleright  \phi )]
\\
&=  {\sfrac 1 2 } \, \pounds_{\X_\mu} \triangleright\left[  d\phi^*   \wedge \bm{\ast} (d\phi) \right] \,.
\end{aligned}
\end{equation}
The first two terms in the second line form a total Lie derivative on-shell, when $d \bm{\ast}  d \, \phi = - m^2 \,  \phi  \, d^4 \x$ and $d \bm{\ast}  d \, \phi^* = - m^2 \,  \phi^* \, d^4 \x$:
\begin{equation}
\begin{aligned}
  {\sfrac 1 2 } \, ( \X_\mu  \triangleright \phi^*) \, d\bm{\ast}  d \phi  +  {\sfrac 1 2 } \,  d \bm{\ast}  d (\chi^\nu{}_\mu \triangleright \phi^*)\, (\X_\nu \triangleright  \phi ) =&  - {\sfrac 1 2 } \, m^2 \, \left[ ( \X_\mu  \triangleright \phi^*) \,\phi  +  (\chi^\nu{}_\mu \triangleright \phi^*)\, (\X_\nu \triangleright  \phi ) \right] \\
  =& - {\sfrac 1 2 } \, m^2 \,  \pounds_{\X_\mu} \triangleright \left(  \phi^*   \, \phi \right) \,,
\end{aligned}
\end{equation}
and the remaining term is a total Lie derivative because of the Cartan identity:
\begin{equation}
 d \circ i_\mu \triangleright \mathscr{L} = \pounds_{\X_\mu} \triangleright \mathscr{L} - i_\mu  \circ \cancel{d \mathscr{L} } \,,
\end{equation}
so we may write
\begin{equation}
\begin{aligned}
d j_\mu =&     \pounds_{\X_\mu} \triangleright\left[ {\sfrac 1 2 } \, d\phi^*   \wedge \bm{\ast} (d\phi) - {\sfrac 1 2 } \, m^2 \,\phi^*  \, \phi  - \mathscr{L} \right] 
 = 0 \,.
\end{aligned}
\end{equation}
This is the statement of Noether's theorem for translation symmetries:
there are four current 3-forms $j_0$, $j_1$, $j_2$, $j_3$, whose closure represents the conservation of four corresponding Noether charges. The four components of the current 3-forms $j_\mu$ consitute the components of the energy-momentum tensor of the theory:
\begin{equation}
\begin{aligned}
j_\mu =& {\sfrac 1 2 } \, ( \X_\mu  \triangleright \phi^*) \, \bm{\ast} (d x^\nu)  (\X_\nu \triangleright \phi) +  {\sfrac 1 2 } \,  \bm{\ast} (d x^\nu)   (\X_\nu \chi^\rho{}_\mu \triangleright \phi^*)  \, (\X_\rho \triangleright  \phi )   - \bm{\ast}(d x^\nu) \,\eta_{\mu\nu} \, \mathscr{L}
\\
=& {\sfrac 1 2 } \,  \bm{\ast} (d x^\nu)  \, ( \chi^\rho{}_\nu \, \X_\mu  \triangleright \phi^*) \, (\X_\rho \triangleright \phi) +  {\sfrac 1 2 } \,  \bm{\ast} (d x^\nu)   (\X_\nu \chi^\rho{}_\mu \triangleright \phi^*)  \, (\X_\rho \triangleright  \phi )   - \bm{\ast}(d x^\nu) \,\eta_{\mu\nu} \, \mathscr{L}
\\
=&    \bm{\ast} (d x^\nu) \,  \left[  ( \X_{(\mu} \, \chi^\rho{}_{\nu)}  \triangleright \phi^*) \, (\X_\rho \triangleright \phi)- \eta_{\mu\nu} \, \mathscr{L} \right]  =    \bm{\ast} (d x^\nu) \,  T_{\mu\nu} \,.
\end{aligned}
\end{equation}
The energy-momentum tensor:
\begin{equation}
T_{\mu\nu} (\x) = ( \X_{(\mu} \, \chi^\rho{}_{\nu)}  \triangleright \phi^*) \, (\X_\rho \triangleright \phi)- \eta_{\mu\nu} \, \mathscr{L}   \,,
\end{equation}
can be calculated on-shell, for example, on a plane-wave solution of the equations of motion:
\begin{equation}\label{OnShellPlanewaves}
\phi(\x) = N \, \E[k] \,, \qquad \phi^*(\x) = \bar{N} \, \E[S(k)] \,, \qquad \X_\mu(k)\X^\mu(k) = - m^2\,,
\end{equation}
the plane wave $\E[k]$ is an eigenfunction of the operators  $\X_\mu$, with eigenvalues $\X_\mu(k)$. Then:
\begin{equation}
\begin{aligned}
T_{\mu\nu} (\x) &= \left[ \X_{(\mu}[S(k)] \, \chi^\rho{}_{\nu)}[S(k)] \, \X_\rho(k)  + \eta_{\mu\nu} \, \left(\X_{\mu}(k) \X^{\mu}(k) + m^2 \right)  \right] \, |N|^2  
\\
&= - (\chi^{-1})^\sigma_{(\mu}(k) \, (\chi^{-1})^\rho{}_{\nu)}(k) \, \X_{\sigma}(k)\, \X_\rho(k)   \, |N|^2 = - S[\X_{(\mu}(k)]\, S[\X_{\nu)}(k) ]  \, |N|^2  \,,
\end{aligned}
\end{equation}
Had we chosen the other simplest choice of Lagrangian, Eq.~\eqref{ScalarFieldAction} with $\alpha = \beta = 0$:
\begin{equation}
\mathscr{L} =  {\sfrac 1 2 } \, d\phi   \wedge \bm{\ast} (d\phi^*) - {\sfrac 1 2 } \, m^2 \, \phi  \, \phi^* \, d^4 \x \,,
\end{equation}
the conserved current 3-form would have been:
\begin{equation}
\begin{aligned}
j_\mu =&  {\sfrac 1 2 } \, ( \X_\mu  \triangleright \phi) \, \bm{\ast} (d \phi^*) +  {\sfrac 1 2 } \,  \bm{\ast} [d (\chi^\nu{}_\mu \triangleright \phi)] \, (\X_\nu \triangleright  \phi^* )   - i_\mu \triangleright \mathscr{L} \,,
\\
=&    \bm{\ast} (d x^\nu) \,  \left[  ( \X_{(\mu} \, \chi^\rho{}_{\nu)}  \triangleright \phi) \, (\X_\rho \triangleright \phi^*)- \eta_{\mu\nu} \, \mathscr{L} \right]   \,,
\end{aligned}
\end{equation}
which, on the on-shell plane-waves~\eqref{OnShellPlanewaves}, corresponds to the following energy-momentum tensor:
\begin{equation}
\begin{aligned}
T_{\mu\nu} (\x) &=   \X_{(\mu}(k) \, \chi^\rho{}_{\nu)}(k) \, \X_\rho[S(k)]  \, |N|^2   = - \X_{(\mu}(k) \, \chi^\rho{}_{\nu)}(k) \, (\chi^{-1})^\sigma{}_\rho (k) \X_\sigma (k)   \, |N|^2  
\\
&=   -  \X_{(\mu}(k) \,  \X_{\nu)}(k)    \, |N|^2  \,.
\end{aligned}
\end{equation}
 
One can also introduce a conserved current associated to the invariance of the action~\eqref{ScalarFieldAction} (for any value of $\alpha$ and $\beta$) under global changes of phase, $\phi \to e^{i \, \theta} \, \phi$, $\phi^* \to e^{-i \, \theta} \, \phi^*$ where $\theta \in \mathbbm{R}$. The following 3-form:
 \begin{equation}
 j =  \bm{\ast} \left(  \phi^* \, d \phi - d \phi^* \, \phi \right) 
 \end{equation}
is closed on-shell, because:
\begin{equation}
dj =   d \phi^* \wedge  \bm{\ast} \left(d \phi \right)  +  \bm{\ast} \left( d \phi^*\right)  \wedge   d \phi  +   \phi^* \,   \left(d \bm{\ast} d \phi \right)  -  \left(d \bm{\ast} d \phi^*\right)  \, \phi \,,
\end{equation} 
the last two terms cancel each other on the solutions of the equations of motion, while the first two can be written as:
\begin{equation}
\begin{aligned}
d \phi^* \wedge  \bm{\ast} \left(d \phi \right) =& d\x^\mu  \,  \X_\mu \triangleright \phi^* \wedge  \bm{\ast} \left( d\x^\nu  \right) \, \X_\nu \triangleright \phi
 = d\x^\mu  \wedge   \bm{\ast} \left(  \X_\mu \triangleright \phi^* d\x^\nu  \right) \, \X_\nu \triangleright \phi
\\
=& d\x^\mu   \wedge  \bm{\ast} \left(  d\x^\rho  \right)  \,  \chi^\nu{}_\rho \X_\mu \triangleright \phi^* \, \X_\nu \triangleright \phi \,,
\\
\bm{\ast} \left( d \phi^*\right)  \wedge   d \phi =&
\bm{\ast} \left( d\x^\mu \right) \,  \X_\mu \triangleright \phi^* \wedge   d\x^\nu  \,  \X_\nu \triangleright \phi
\\
=&
\bm{\ast} \left( d\x^\mu \right)\wedge    d\x^\rho \,  \chi^\nu{}_\rho \X_\mu \triangleright \phi^*   \,  \X_\nu \triangleright \phi \,,
\end{aligned}
\end{equation}
and, just like in the commutative case, $d\x^\mu   \wedge  \bm{\ast} \left(  d\x^\rho  \right)  = - \bm{\ast} \left( d\x^\mu \right)\wedge    d\x^\rho $.

\section{Conclusions} \label{Sec:Conclusions}

In this paper I focused on the general features of field theory on T-Minkowski noncommutative spacetimes, in a model-independent fashion. These, as introduced in Part I~\cite{Mercati:2023apu}, fall into 11 categories of qualitatively different models (although these split, in turn, into 17 automorphism classes), but they all share some basic features that allow one to say a lot about the general properties of field theories that can be built upon them, without having to specialize to any particular model. The most important such characteristic is the fact that the algebra of coordinates~$C_\ell[\mathbbm{R}^{3,1}]$ is a Lie algebra (possibly with a central extension), which implies that the ordered exponentials, which form a basis for the noncommutative functions thanks to the Poincar\'e--Birkhoff--Witt property, form a Lie group $\mathcal{G}_\ell$, whose group manifold I identify with the momentum space of the theory. In Sec.~\ref{Sec:Fourier}, a complete Fourier theory, for the noncommutative analogous of Schwarzian functions, was developed, following the notion of generalized Weyl systems introduced in~\cite{Agostini:2002de}. In this way, noncommutative functions can be mapped to commutative Fourier transforms, and noncommutative field theory reformulated, in Fourier transform, as a commutative field theory written in general (non necessarily linear/Cartesian) coordinates in momentum space. A notion of integral can be introduced, which is invariant under T-Poincar\'e transformations, and independent of the ordering choice for representing noncommutative polynomials (which is the same as invariance under changes of coordinates on momentum space). The natural structures on the Lie group $\mathcal{G}_\ell$, like the Haar measure and the invariant vector fields, play structural roles in the properties of the algebra of noncommutative functions. For example, the unimodularity of $\mathcal{G}_\ell$ implies the ciclicity of the integral. Moreover, the convolution product between Fourier transforms is related to the group product of $\mathcal{G}_\ell$. 
These group-theoretical techniques allowed me, in Sec.~\ref{Sec:T-Poincare_algebra},  to write the action of finite T-Poincar\'e transformations on general T-Minkowski functions, and to study their infinitesimal version, which leads to the Hopf-algebra deformation of the Poincar\'e algebra that is dual to the T-Poincar\'e group (written in a bicrossproduct-like basis). In the following Section~\ref{Sec:DifferentialGeometry} I developed all the basic tools of differential geometry: the differential calculus, the exterior derivative, the exterior algebra, Hodge star, Lie and inner derivative, all covariant under T-Poincar\'e transformations and explicitly calculable, once a model is specified.
Armed with this technology, in the last Section, number~\ref{Sec:FieldTheory}, I was able to study a complex scalar field and derive its equations of motion in two versions, \textit{i.e.} in Fourier transform and in direct space, and to show their equivalence. I could also study Dirac fields and, in the unimodular case, Abelian gauge theories. The last result concerns the Noether theorem, whose validity I could verify in the case of a scalar field and for T-Poincar\'e symmetry, thereby deriving the energy-momentum tensor of the theory.

All the results above only concern `classical' (as opposed to quantum) noncommutative field theories. The reader may find the concept of a classical field theory on a noncommutative (\textit{i.e.} `quantum') spacetime puzzling, and, indeed, it is not clear whether such a concept makes physical sense. There might be a regime in which quantum field theory effects are negligible, that is, one is not sensitive to effects controlled by the Planck constant $\hbar$ (``tree level'' QFT), but remnants of the underlying noncommutative structure of spacetime are detectable in classical field theory processes, because one has some sensitivity to the noncommutativity scale $\ell$. However, the conjectured existence of this `classical-noncommutative' regime is not the main motivation of the present paper. Rather, I studied these classical theories as a necessary stepping stone in order to prepare the ground for the study on noncommutative quantum field theories. Paradoxically, these admit, in my opinion, a more straightforward physical interpretation than their classical counterparts: as shown in~\cite{Mercati:2023apu}, the braided algebras of $N$ points~\eqref{xaxb_comm_rel} admits a commutative subalgebra generated by the \textit{differences between coordinates of different points,} \textit{i.e.} $\x^\mu_a - \x^\mu_b$, $a\neq b = 1,\dots,N$. Then, all translation-invariant (and therefore all  T-Poincar\'e-invariant) functions of $N$ points are \textit{commutative.} This implies that the $N$-point functions, which in commutative QFT encode the entire physical content of the theory, are all commutative and do not require special care for their interpretation. They can simply be compared to their commutative counterparts to see whether there are any physical effects caused by the noncommutativity. This is an exceptional opportunity, offered by the special algebraic properties of T-Minkowski spacetimes, to draw physical conclusions without incurring in crippling arbitrariness and uncertainty in the interpretation of the results.
The ultimate goal is, then, the study of noncommutative QFTs. An appropriate formalism for this could be, for example, the \textit{braided QFT} approach initiated by Oeckl~\cite{Oeckl:1999zu,Oeckl:2000eg,Sasai:2007me}, more recently studied in~\cite{Bogdanovic:2023izt,DimitrijevicCiric:2023hua,Giotopoulos:2021ieg,DimitrijevicCiric:2021jea} (also the related works~\cite{Fiore:2007vg,Fiore:2007zz,Fiore:2008ta} are worth mentioning). In the present paper I obtained a series of important results towards that goal, which puts us now in a position to develop a formulation of QFT on T-Minkowski spacetimes, as we can define all types of Standard Model fields (scalar, Dirac and gauge),  each of which is rigorously defined as element of a specific algebra ($C_\ell[\mathbbm{R}^{3,1}]$, $C_\ell[\mathbbm{R}^{3,1}] \otimes \mathbbm{C}^4$ and $\Gamma_\ell$) and has good transformation properties under T-Poincar\'e, and we can write action functionals that are T-Poincar\'e invariant. Moreover, these actions can be expressed in Fourier transform, and the whole theory mapped to a commutative field theory on a nontrivial momentum space. The functional integral can then be defined  in principle in the standard way in Fourier transform, and the N-point functions can presumably be deduced from it, and with them the physical predictions of the theory. This will be the objective of future works.

\section*{Acknowledgements}

Many thanks to Giuseppe Fabiano for his insightful comments and suggestions, which helped improve the quality of this manuscript. 
I acknowledge  support by the Agencia Estatal de Investigaci\'on (Spain) under grants CNS2023-143760 and PID2023-148373NB-I00 funded by MCIN/AEI/10.13039/50110001103,
and by the Q-CAYLE Project funded by the Regional Government of Castilla y Le\'on (Junta de Castilla y Le\'on) and by the Ministry of Science and Innovation MICIN through NextGenerationEU (PRTR C17.I1).

\section*{Appendices}

\appendix

\section{Calculation of the Lorentz transform of momenta}\label{Appendix:Lorentz_transform_momenta}

Noncommutative exponentials, introduced in~\eqref{GeneralOrderedExponentials} through an ordering prescription, can be written as a product of $n$ exponentials, each depending on a single linear combination of the coordinates, as in
\begin{equation}
\E[k] = e^{i k^1_\mu \, \x^\mu} e^{i k^2_\mu \, \x^\mu} \dots e^{i k^n_\mu \, \x^\mu} \,,
\end{equation}
for example, the lexicographic order is obtained with $n=4$ and $k^1_\mu = k_0 \, \delta^0\mu$, $k^2_\mu = k_1 \, \delta^1\mu$, $k^3_\mu = k_2 \, \delta^2\mu$, $k^4_\mu = k_3 \, \delta^3\mu$. The Weyl ordering is simply $n = 1$ and $k^1_\mu = k_\mu$, and an ordering in which the $0$ coordinate is symmetrized with respect to the other three, who are in Weyl ordering, is obtained with $n=3$ and $k^1_\mu = {\sfrac 1 2} k_0 \, \delta^0_\mu$, $k^2_\mu = \sum_{i=1}^3 k_i \, \delta^i_\mu$, $k^3_\mu = {\sfrac 1 2} k_0 \, \delta^0_\mu$.
Now that we have an explicit parametrization for any arbitrarily-ordered exponential, we can study its Poincar\'e transformation:
\begin{equation}\label{Eq:ProofPoincareTransformMomenta_0}
\E'[k] = e^{i k^{(1)}_\mu \, \La^\mu{}_\nu \otimes \x^\nu + i k^{(1)}_\mu \, \A^\mu \otimes \1}  e^{i k^{(2)}_\mu \, \La^\mu{}_\nu \otimes \x^\nu + i k^{(2)}_\mu \, \A^\mu \otimes \1} \dots  e^{i k^{(n)}_\mu \, \La^\mu{}_\nu \otimes \x^\nu + i k^{(n)}_\mu \, \A^\mu \otimes \1} \,,
\end{equation}
each of the terms in the above product can be split separately into a product of exponentials of the form:
\begin{equation}
e^{i k^{(a)}_\mu \, \La^\mu{}_\nu \otimes \x^\nu + i k^{(a)}_\mu \, \A^\mu \otimes \1}
=
e^{i \, \theta^{\mu\nu} \, Z_{\mu\nu}(\La,k^{(a)})\otimes \1 } \, e^{i \, J_\mu ( \La, k^{(a)} ) \otimes \x^\mu} \, e^{ i k^{(a)}_\mu \, \A^\mu \otimes \1} \,,
\end{equation}
it is not hard to convince ourselves that the expression above holds. It is a consequence of the commutation relations~\eqref{T-PoincareGroup_commutators} and~\eqref{x_comm_rel}. We can see this explicitly by considering the Zassenhaus theorem, which provides an inverse for the BCH formula:
\begin{equation}
e^{\hat{X}+\hat{Y}}  = e^{\hat{X}} \, e^{\hat{Y}} \, \prod_{n=2}^\infty e^{C_n(\hat{X},\hat{Y})}  \,, 
\end{equation}
where $C_n(\hat{X},\hat{Y})$ are Lie polynomials of degree $n$, \textit{i.e.}, they are a sum of commutators of $n$ generators each, in a specific order.\footnote{The Zassenhaus theorem, as well as the Campbell-Baker-Haussdorff-Dynkin theorem hold is usually formulated inb the context of unital Banach algebras over the complex or real numbers (although there are many proofs of the BCHD theorem in different contexts~\cite{bonfiglioli2011topics}), where the convergence of the infinite sums that appear in the theorems can be discussed. Here I am content to work with formal power series and ignote convergence issues.} For example:
\begin{equation}
C_2(\hat{X},\hat{Y}) = -\sfrac {1}{2} \, [\hat{X},\hat{Y}] \,, \qquad C_3(\hat{X},\hat{Y}) = \sfrac{1}{6} \left( 2 \, [\hat{Y},[\hat{X},\hat{Y}]] + [\hat{X},[\hat{X},\hat{Y}]] \right) \,.
\end{equation}
The main point is then that, for $\hat{X} =  i k^{(a)}_\mu \, \A^\mu \otimes \1$ and $\hat{Y}=i k^{(a)}_\mu \, \La^\mu{}_\nu \otimes \x^\nu$, the nested commutators $C_n(\hat{X},\hat{Y})$  can only produce terms of the following two kinds:
\begin{equation}\label{Eq:General_form_Lie_polynomials}
F(\La) \otimes \1 \,, \qquad  G_\mu(\La) \otimes \x^\mu \,,
\end{equation}
where $F$ and $G_\mu$ are functions of $\La^\mu{}_\nu$. The fact that the expressions above do not involve $\A^\mu$ is a straightforward consequence of the fact that $\A^\mu$ does not appear on the right-hand sides of the commutation relatios~\eqref{T-PoincareGroup_commutators} and~\eqref{x_comm_rel}. What remains to be proven is that the possible expressions are at most linear in $\x^\mu$. This, again, is a direct consequence of the commutation relations. The basic commutator is of the form~\eqref{Eq:General_form_Lie_polynomials}:
\begin{equation}
[\hat{X},\hat{Y}] = [i k^{(a)}_\rho \, \A^\rho \otimes \1 \, , \, i k^{(a)}_\mu \, \La^\mu{}_\nu \otimes \x^\nu  ]
=i \, f^{\alpha\beta}{}_\gamma \, k^{(a)}_\mu \,  k^{(a)}_\rho\left( \La^\mu{}_\alpha \, \La^\rho{}_\beta \, \delta^\gamma{}_\nu -  \delta^\mu{}_\alpha \, \delta^\rho{}_\beta \, \La^\gamma{}_\nu \right)   \otimes \x^\nu  \,,
\end{equation}
and, acting on the above with the adjoint actions $[\hat{X}, \, . \, ]$ and  $[\hat{Y}, \, . \, ]$ an arbitrary amount of times always gives a polynomial composed of monomials of the general form~\eqref{Eq:General_form_Lie_polynomials}. In fact, $[\hat{X}, \, . \, ]$ just changes the function of $\La^\mu{}_\nu$ that is on the left-hand side of the tensor product, while $[\hat{Y}, \, . \, ]$, on the other hand,  acts on objects of the form~\eqref{Eq:General_form_Lie_polynomials} by commuting the respective $\x^\mu$ coordinates on the right-hand side of the tensor product, which, by~\eqref{x_comm_rel}, gives an expression at most linear in $\x^\mu$. Finally, notice that the terms of the form $F(\La) \otimes \1$ are linear in $\theta^{\mu\nu}$, because they can only arise as a result of the commutator between two  $\x^\mu$  coordinates, and any further commutator by  $\x^\mu$ will erase the term proportional to the identity. Therefore, these terms can only have arisen as the result of one last commutator by $\hat{X}$, and must be linear in $\theta^{\mu\nu}$.
Because of all the above, the Zassenhaus formula for our exponential takes the following form:
\begin{equation}
e^{i k^{(a)}_\mu \, \La^\mu{}_\nu \otimes \x^\nu + i k^{(a)}_\mu \, \A^\mu \otimes \1}
=
e^{i k^{(a)}_\mu \, \A^\mu \otimes \1} \, e^{i k^{(a)}_\mu \, \La^\mu{}_\nu \otimes \x^\nu} \, \prod_{n=0}^\infty e^{i \, \theta^{\mu\nu} \, Z^n_{\mu\nu}(\La,k^{(a)})\otimes \1 + i \, H^n_\mu ( \La, k^{(a)} ) \otimes \x^\mu} 
\,,
\end{equation}
which, using the group law of algebra~\eqref{x_comm_rel}, can be rearranged into:
\begin{equation}\label{Eq:ProofPoincareTransformMomenta_1}
e^{i k^{(a)}_\mu \, \La^\mu{}_\nu \otimes \x^\nu + i k^{(a)}_\mu \, \A^\mu \otimes \1}
=
e^{i k^{(a)}_\mu \, \A^\mu \otimes \1} \, e^{i \, \theta^{\mu\nu} \, Z_{\mu\nu}(\La,k^{(a)})\otimes \1 + i \, H_\mu ( \La, k^{(a)} ) \otimes \x^\mu} \,.
\end{equation}
Recall now the definition of the backreaction of momenta on Lorentz matrices~\eqref{Eq:Def_BackReaction}. On an expoential of a single linear combination of generators $\A^\mu$, this reduces to the exponential map~\eqref{Eq:ExponentialMapActionOnLambda}, defined as the solution of Eqs.~\eqref{Eq:ExponentialMapDiffEq} and~\eqref{Eq:ExponentialMapDiffEqInitialCondition}:
\begin{equation}
e^{i q_\mu \, \A^\mu} \, \La^\mu{}_\nu = \Upsilon^\mu{}_\nu(q,\La) \, e^{i q_\mu \, \A^\mu} \,,
\end{equation} 
this implies
\begin{equation}
e^{i q_\mu \, \A^\mu \otimes \1} \, e^{i \, \theta^{\mu\nu} \, Z_{\mu\nu}(\La,k^{(a)})\otimes \1} = e^{i \, \theta^{\mu\nu} \, Z_{\mu\nu}[\Upsilon(\La,q),k^{(a)}]\otimes \1}\, e^{i q_\mu \, \A^\mu \otimes \1} \,,
\end{equation}
and, on the second exponential in Eq.~\eqref{Eq:ProofPoincareTransformMomenta_1},
\begin{equation}\label{Eq:ProofPoincareTransformMomenta_2}
\begin{aligned}
&e^{i q_\mu \, \A^\mu \otimes \1} \, e^{i \, H_\mu ( \La, k^{(a)} ) \otimes \x^\mu} 
= e^{i q_\mu \, \A^\mu \otimes \1} \, \sum_{n=0}^\infty \frac{i^n}{n!} H_{\mu_1}( \La, k^{(a)} ) \dots H_{\mu_n}( \La, k^{(a)} )  \otimes \x^{\mu_1} \dots \x^{\mu_n}
\\
&= \sum_{n=0}^\infty \frac{i^n}{n!} H_{\mu_1}[\Upsilon( \La, q ), k^{(a)} ] \dots H_{\mu_n}[\Upsilon( \La, q ), k^{(a)} ]  \otimes \x^{\mu_1} \dots \x^{\mu_n} \,  e^{i q_\mu \, \A^\mu \otimes \1}
\\
&= e^{i \, H_\mu [\Upsilon( \La, q ), k^{(a)} ] \otimes \x^\mu}\,  e^{i q_\mu \, \A^\mu \otimes \1} \,.
\end{aligned}
\end{equation}
Eq.~\eqref{Eq:ProofPoincareTransformMomenta_1} then reduces to
\begin{equation}\label{Eq:ProofPoincareTransformMomenta_3}
e^{i k^{(a)}_\mu \, \La^\mu{}_\nu \otimes \x^\nu + i k^{(a)}_\mu \, \A^\mu \otimes \1}
=
e^{i \, \theta^{\mu\nu} \, Z_{\mu\nu}[\Upsilon( \La, k^{(a)} ),k^{(a)}]\otimes \1 + i \, H_\mu [\Upsilon( \La, k^{(a)} ), k^{(a)} ] \otimes \x^\mu} \, e^{i k^{(a)}_\mu \, \A^\mu \otimes \1}  \,.
\end{equation}
We can use these results to reorder the right-hand side of Eq.~\eqref{Eq:ProofPoincareTransformMomenta_0} with all the $\A^\mu$'s to the right. The first step is to rewrite each term of~\eqref{Eq:ProofPoincareTransformMomenta_0} as in Eq.~\eqref{Eq:ProofPoincareTransformMomenta_1}:
\begin{equation}
\E'[k] = \prod_{a=1}^n  e^{i k^{(a)}_\mu \, \La^\mu{}_\nu \otimes \x^\nu + i k^{(a)}_\mu \, \A^\mu \otimes \1}   =
\prod_{a=1}^n e^{i k^{(a)}_\mu \, \A^\mu \otimes \1} \, e^{i \, \theta^{\mu\nu} \, Z_{\mu\nu}(\La,k^{(a)})\otimes \1 + i \, H_\mu ( \La, k^{(a)} ) \otimes \x^\mu}  \,,
\end{equation}
and then, using Eq.~\eqref{Eq:ProofPoincareTransformMomenta_2} and  Eq.~\eqref{Eq:ProofPoincareTransformMomenta_3} repeatedly:
\begin{equation}
\begin{aligned}
\E'[k] =& 
  \exp \left( i \, \theta^{\mu\nu} \sum_{a=1}^n  Z_{\mu\nu}\left\{
\Upsilon \left[ k^{(1)} ,\Upsilon \left[ k^{(2)} , \dots , \Upsilon \left[ k^{(a-1)} , \Upsilon(\La,k^{(a)}) \right] \dots \right]\right] ,k^{(a)}\right\} \otimes \1 \right)
\\
&\prod_{b=1}^n  e^{i \, H_\mu \left\{
\Upsilon \left[ k^{(1)} ,\Upsilon \left[ k^{(2)} , \dots , \Upsilon \left[ k^{(b-1)} ,
\Upsilon(\La,k^{(b)}) \right] \dots \right]\right] ,k^{(a)}\right\} \otimes \x^\mu } 
\prod_{c=1}^n e^{i k^{(c)}_\mu \, \A^\mu \otimes \1}  \,,
\end{aligned}
\end{equation}
which, by reordering the exponentials of $\x^\mu$, can be recasted in the form
\begin{equation}
\E'[k] =  e^{i \, \theta^{\mu\nu} \, \sigma_{\mu\nu} (k,\La) \otimes \1} ~: \, e^{i \, \lambda_\mu (k,\La) \otimes \x^\mu} \, :  \, : \, e^{i \, k_\mu \, \A^\mu  \otimes \1} \,:\,.
\end{equation}

\small

\providecommand{\href}[2]{#2}\begingroup\raggedright\endgroup


\begin{thebibliography}{10}

\bibitem{Mercati:2023apu}
F.~Mercati, ``{T-Minkowski noncommutative spacetimes I: Poincar\'e groups,
  differential calculi and braiding},''
  \href{http://arxiv.org/abs/2311.16249}{{\ttfamily arXiv:2311.16249
  [hep-th]}}.

\bibitem{wess1999qdeformed}
J.~Wess, ``q-Deformed Heisenberg Algebras,''
  \href{http://arxiv.org/abs/math-ph/9910013}{{\ttfamily arXiv:math-ph/9910013
  [math-ph]}}.

\bibitem{AmelinoCamelia:2000mn}
G.~Amelino-Camelia, ``{Relativity in space-times with short distance structure
  governed by an observer independent (Planckian) length scale},''
  \href{http://dx.doi.org/10.1142/S0218271802001330}{{\em Int. J. Mod. Phys.}
  {\bfseries D11} (2002) 35--60},
  \href{http://arxiv.org/abs/gr-qc/0012051}{{\ttfamily arXiv:gr-qc/0012051
  [gr-qc]}}.

\bibitem{Woronowicz:1989}
S.~Woronowicz, ``Differential calculus on compact matrix pseudogroups (quantum
  groups),'' \href{http://dx.doi.org/10.1007/BF0122141}{{\em Commun. Math.
  Phys.} {\bfseries 122} (1989) 125–170}.

\bibitem{Meier:2023kzt}
T.~Meier and S.~J. van Tongeren, ``{Quadratic Twist-Noncommutative Gauge
  Theory},'' \href{http://dx.doi.org/10.1103/PhysRevLett.131.121603}{{\em Phys.
  Rev. Lett.} {\bfseries 131} no.~12, (2023) 121603},
  \href{http://arxiv.org/abs/2301.08757}{{\ttfamily arXiv:2301.08757
  [hep-th]}}.

\bibitem{Meier:2023lku}
T.~Meier and S.~J. van Tongeren, ``{Gauge theory on twist-noncommutative
  spaces},'' \href{http://dx.doi.org/10.1007/JHEP12(2023)045}{{\em JHEP}
  {\bfseries 12} (2023) 045}, \href{http://arxiv.org/abs/2305.15470}{{\ttfamily
  arXiv:2305.15470 [hep-th]}}.

\bibitem{Ballesteros_1995}
A.~Ballesteros, E.~Celeghini, F.~J. Herranz, M.~A. del Olmo, and M.~Santander,
  ``Universal R-matrices for non-standard (1+1) quantum groups,''
  \href{http://dx.doi.org/10.1088/0305-4470/28/11/015}{{\em J. Phys. A}
  {\bfseries 28} no.~11, (Jun, 1995) 3129--3138}.

\bibitem{BALLESTEROS_1996}
A.~Ballesteros and F.~J. Herranz, ``(2+1) null-plane quantum Poincaré group
  from a factorized universal R-matrix,''
  \href{http://dx.doi.org/10.1142/s0217732396001739}{{\em Mod. Phys. Lett. A}
  {\bfseries 11} no.~21, (Jul, 1996) 1745--1755}.

\bibitem{Ballesteros:1996awf}
A.~Ballesteros, F.~J. Herranz, and C.~M. Pere\~na, ``{Null-plane quantum
  universal $R$-matrix},''
  \href{http://dx.doi.org/10.1016/S0370-2693(96)01435-9}{{\em Phys. Lett. B}
  {\bfseries 391} (1997) 71--77},
  \href{http://arxiv.org/abs/q-alg/9607009}{{\ttfamily arXiv:q-alg/9607009}}.

\bibitem{majid_1995}
S.~Majid, \href{http://dx.doi.org/10.1017/CBO9780511613104}{{\em Foundations of
  Quantum Group Theory}}.
\newblock Cambridge University Press, 1995.

\bibitem{Aschieri:2020ifa}
P.~Aschieri, ``{Cartan structure equations and Levi-Civita connection in
  braided geometry},'' \href{http://arxiv.org/abs/2006.02761}{{\ttfamily
  arXiv:2006.02761 [math.QA]}}.

\bibitem{Aschieri:2005zs}
P.~Aschieri, M.~Dimitrijevic, F.~Meyer, and J.~Wess, ``{Noncommutative geometry
  and gravity},'' \href{http://dx.doi.org/10.1088/0264-9381/23/6/005}{{\em
  Class. Quant. Grav.} {\bfseries 23} (2006) 1883--1912},
  \href{http://arxiv.org/abs/hep-th/0510059}{{\ttfamily arXiv:hep-th/0510059}}.

\bibitem{Chari}
A.~N.~P. Vyjayanthi~Chari, {\em A guide to quantum groups}.
\newblock Cambridge University Press, 1994.
\newblock
  \url{https://www.google.com/books/edition/A_Guide_to_Quantum_Groups/_fRhQgAACAAJ?hl=en}.

\bibitem{Zakrzewski_1997}
S.~Zakrzewski, ``Poisson Structures on the Poincar{\'e} Group,''
  \href{http://dx.doi.org/10.1007/s002200050091}{{\em Commun. Math. Phys.}
  {\bfseries 185} no.~2, (May, 1997) 285--311},
  \href{http://arxiv.org/abs/q-alg/9602001v1}{{\ttfamily q-alg/9602001v1}}.

\bibitem{Lizzi:2021rlb}
F.~Lizzi and F.~Mercati, ``{$\kappa$-Poincar\'e-comodules, Braided Tensor
  Products and Noncommutative Quantum Field Theory},''
  \href{http://dx.doi.org/10.1103/PhysRevD.103.126009}{{\em Phys. Rev. D}
  {\bfseries 103} (2021) 126009},
  \href{http://arxiv.org/abs/2101.09683}{{\ttfamily arXiv:2101.09683
  [hep-th]}}.

\bibitem{DiLuca:2022idu}
M.~G. Di~Luca and F.~Mercati, ``{New class of plane waves for
  $\kappa$-noncommutative quantum field theory},''
  \href{http://dx.doi.org/10.1103/PhysRevD.107.105018}{{\em Phys. Rev. D}
  {\bfseries 107} no.~10, (2023) 105018},
  \href{http://arxiv.org/abs/2211.11627}{{\ttfamily arXiv:2211.11627
  [hep-th]}}.

\bibitem{Fabiano:2023xke}
G.~Fabiano and F.~Mercati, ``{Multiparticle states in braided lightlike
  $\kappa$-Minkowski noncommutative QFT},''
  \href{http://dx.doi.org/10.1103/PhysRevD.109.046011}{{\em Phys. Rev. D}
  {\bfseries 109} no.~4, (2024) 046011},
  \href{http://arxiv.org/abs/2310.15063}{{\ttfamily arXiv:2310.15063
  [hep-th]}}.

\bibitem{tolstoy2007twisted}
V.~N. Tolstoy, ``Twisted Quantum Deformations of Lorentz and Poincar\'{e}
  algebras,'' in {\em Invited talk at the VII International Workshop `Lie
  Theory and its Applications in Physics',18--24 June 2007, Varna, Bulgaria}.
\newblock 2007.
\newblock \href{http://arxiv.org/abs/0712.3962}{{\ttfamily arXiv:0712.3962
  [math.QA]}}.

\bibitem{Lizzi:2022hcq}
F.~Lizzi, L.~Scala, and P.~Vitale, ``{Localization and observers in
  \ensuremath{\varrho}-Minkowski spacetime},''
  \href{http://dx.doi.org/10.1103/PhysRevD.106.025023}{{\em Phys. Rev. D}
  {\bfseries 106} no.~2, (2022) 025023},
  \href{http://arxiv.org/abs/2205.10862}{{\ttfamily arXiv:2205.10862
  [hep-th]}}.

\bibitem{Fabiano:2023uhg}
G.~Fabiano, G.~Gubitosi, F.~Lizzi, L.~Scala, and P.~Vitale, ``{Bicrossproduct
  vs. twist quantum symmetries in noncommutative geometries: the case of
  \ensuremath{\varrho}-Minkowski},''
  \href{http://dx.doi.org/10.1007/JHEP08(2023)220}{{\em JHEP} {\bfseries 08}
  (2023) 220}, \href{http://arxiv.org/abs/2305.00526}{{\ttfamily
  arXiv:2305.00526 [hep-th]}}.

\bibitem{Lukierski:2005fc}
J.~Lukierski and M.~Woronowicz, ``{New Lie-algebraic and quadratic deformations
  of Minkowski space from twisted Poincare symmetries},''
  \href{http://dx.doi.org/10.1016/j.physletb.2005.11.052}{{\em Phys. Lett. B}
  {\bfseries 633} (2006) 116--124},
  \href{http://arxiv.org/abs/hep-th/0508083}{{\ttfamily arXiv:hep-th/0508083}}.

\bibitem{Majid:1994cy}
S.~Majid and H.~Ruegg, ``{Bicrossproduct structure of kappa Poincare group and
  noncommutative geometry},''
  \href{http://dx.doi.org/10.1016/0370-2693(94)90699-8}{{\em Phys. Lett. B}
  {\bfseries 334} (1994) 348--354},
  \href{http://arxiv.org/abs/hep-th/9405107}{{\ttfamily arXiv:hep-th/9405107}}.

\bibitem{Woronowicz1987_1}
S.~L. Woronowicz, ``Twisted SU(2) Group. An Example of a Non-Commutative
  Differential Calculus,''
  \href{http://dx.doi.org/10.2977/PRIMS/1195176848}{{\em PRIMS} {\bfseries 23}
  no.~1, (1987) 117--181}.

\bibitem{Woronowicz1987_2}
S.~L. Woronowicz, ``Compact matrix pseudogroups,''
  \href{http://dx.doi.org/10.1007/BF01219077}{{\em Commun. Math. Phys.}
  {\bfseries 111} (1987) 613--665}.

\bibitem{Agostini:2002de}
A.~Agostini, F.~Lizzi, and A.~Zampini, ``{Generalized Weyl systems and kappa
  Minkowski space},'' \href{http://dx.doi.org/10.1142/S021773230200871X}{{\em
  Mod. Phys. Lett. A} {\bfseries 17} (2002) 2105--2126},
  \href{http://arxiv.org/abs/hep-th/0209174}{{\ttfamily arXiv:hep-th/0209174}}.

\bibitem{enwiki:1214961199}
{Wikipedia contributors}, ``Gimbal lock --- {Wikipedia}{,} The Free
  Encyclopedia,'' 2024.
\newblock
  \url{https://en.wikipedia.org/w/index.php?title=Gimbal_lock&oldid=1214961199}.
  [Online; accessed 11-April-2024].

\bibitem{achilles2012early}
R.~Achilles and A.~Bonfiglioli, ``The early proofs of the theorem of Campbell,
  Baker, Hausdorff, and Dynkin,''
  \href{http://dx.doi.org/10.1007/s00407-012-0095-8}{{\em Archive for history
  of exact sciences} {\bfseries 66} (2012) 295--358}.

\bibitem{deitmar2014_Harmonic_Analysis_book}
A.~Deitmar and S.~Echterhoff,
  \href{http://dx.doi.org/10.1007/978-3-319-05792-7}{{\em Principles of
  harmonic analysis}}.
\newblock Springer, 2014.

\bibitem{Hersent:2020lsr}
K.~Hersent, ``{On the UV/IR mixing of Lie algebra-type noncommutatitive
  \ensuremath{\phi}$^{4}$-theories},''
  \href{http://dx.doi.org/10.1007/JHEP03(2024)023}{{\em JHEP} {\bfseries 24}
  (2020) 023}, \href{http://arxiv.org/abs/2309.08917}{{\ttfamily
  arXiv:2309.08917 [hep-th]}}.

\bibitem{Majid:2006xn}
S.~Majid, ``{Algebraic approach to quantum gravity. II. Noncommutative
  spacetime},'' \href{http://arxiv.org/abs/hep-th/0604130}{{\ttfamily
  arXiv:hep-th/0604130}}.

\bibitem{Arzano:2020jro}
M.~Arzano, A.~Bevilacqua, J.~Kowalski-Glikman, G.~Rosati, and J.~Unger,
  ``{$\kappa$-deformed complex fields and discrete symmetries},''
  \href{http://dx.doi.org/10.1103/PhysRevD.103.106015}{{\em Phys. Rev. D}
  {\bfseries 103} no.~10, (2021) 106015},
  \href{http://arxiv.org/abs/2011.09188}{{\ttfamily arXiv:2011.09188
  [hep-th]}}.

\bibitem{Gubitosi:2011hgc}
G.~Gubitosi and F.~Mercati, ``{Relative Locality in $\kappa$-Poincar\'e},''
  \href{http://dx.doi.org/10.1088/0264-9381/30/14/145002}{{\em Class. Quant.
  Grav.} {\bfseries 30} (2013) 145002},
  \href{http://arxiv.org/abs/1106.5710}{{\ttfamily arXiv:1106.5710 [gr-qc]}}.

\bibitem{Podles:1995qy}
P.~Podles, ``{Solutions of Klein-Gordon and Dirac equations on quantum
  Minkowski spaces},'' \href{http://dx.doi.org/10.1007/BF02101287}{{\em Commun.
  Math. Phys.} {\bfseries 181} (1996) 569--586},
  \href{http://arxiv.org/abs/q-alg/9510019}{{\ttfamily arXiv:q-alg/9510019}}.

\bibitem{Radko_1997}
O.~V. Radko and A.~A. Vladimirov, ``On the algebraic structure of differential
  calculus on quantum groups,'' \href{http://dx.doi.org/10.1063/1.531952}{{\em
  Journal of Mathematical Physics} {\bfseries 38} no.~10, (Oct., 1997)
  5434–5446}. \url{http://dx.doi.org/10.1063/1.531952}.

\bibitem{Brzezinski:1992gv}
T.~Brzezinski, ``{Remarks on bicovariant differential calculi and exterior Hopf
  algebras},'' \href{http://dx.doi.org/10.1007/BF00777376}{{\em Lett. Math.
  Phys.} {\bfseries 27} (1993) 287--300}.

\bibitem{Mercuri:2006um}
S.~Mercuri, ``{Fermions in Ashtekar-Barbero connections formalism for arbitrary
  values of the Immirzi parameter},''
  \href{http://dx.doi.org/10.1103/PhysRevD.73.084016}{{\em Phys. Rev. D}
  {\bfseries 73} (2006) 084016},
  \href{http://arxiv.org/abs/gr-qc/0601013}{{\ttfamily arXiv:gr-qc/0601013}}.

\bibitem{Mathieu:2020ccc}
P.~Mathieu and J.-C. Wallet, ``{Gauge theories on \ensuremath{\kappa}-Minkowski
  spaces: twist and modular operators},''
  \href{http://dx.doi.org/10.1007/JHEP05(2020)112}{{\em JHEP} {\bfseries 05}
  (2020) 112}, \href{http://arxiv.org/abs/2002.02309}{{\ttfamily
  arXiv:2002.02309 [hep-th]}}.

\bibitem{Kupriyanov:2020axe}
V.~G. Kupriyanov, M.~Kurkov, and P.~Vitale, ``{$\kappa$-Minkowski-deformation
  of U(1) gauge theory},''
  \href{http://dx.doi.org/10.1007/JHEP01(2021)102}{{\em JHEP} {\bfseries 01}
  (2021) 102}, \href{http://arxiv.org/abs/2010.09863}{{\ttfamily
  arXiv:2010.09863 [hep-th]}}.

\bibitem{Mercati:2011aa}
F.~Mercati, ``{Quantum $\kappa$-deformed differential geometry and field
  theory},'' \href{http://dx.doi.org/10.1142/S021827181650053X}{{\em Int. J.
  Mod. Phys. D} {\bfseries 25} no.~05, (2016) 1650053},
  \href{http://arxiv.org/abs/1112.2426}{{\ttfamily arXiv:1112.2426 [math.QA]}}.

\bibitem{Oeckl:1999zu}
R.~Oeckl, ``{Braided quantum field theory},''
  \href{http://dx.doi.org/10.1007/s002200100375}{{\em Commun. Math. Phys.}
  {\bfseries 217} (2001) 451--473},
  \href{http://arxiv.org/abs/hep-th/9906225}{{\ttfamily arXiv:hep-th/9906225}}.

\bibitem{Oeckl:2000eg}
R.~Oeckl, ``{Untwisting noncommutative $R^d$ and the equivalence of quantum
  field theories},''
  \href{http://dx.doi.org/10.1016/S0550-3213(00)00281-9}{{\em Nucl. Phys. B}
  {\bfseries 581} (2000) 559--574},
  \href{http://arxiv.org/abs/hep-th/0003018}{{\ttfamily arXiv:hep-th/0003018}}.

\bibitem{Sasai:2007me}
Y.~Sasai and N.~Sasakura, ``{Braided quantum field theories and their
  symmetries},'' \href{http://dx.doi.org/10.1143/PTP.118.785}{{\em Prog. Theor.
  Phys.} {\bfseries 118} (2007) 785--814},
  \href{http://arxiv.org/abs/0704.0822}{{\ttfamily arXiv:0704.0822 [hep-th]}}.

\bibitem{Bogdanovic:2023izt}
D.~Bogdanovi\'c, M.~Dimitrijevi\'c~\'Ciri\'c, V.~Radovanovi\'c, and R.~J.
  Szabo, \href{http://dx.doi.org/10.22323/1.436.0338}{``{BV quantization of
  braided scalar field theory},''} in {\em {CORFU2022: 22th Hellenic School and
  Workshops on Elementary Particle Physics and Gravity}}.
\newblock 4, 2023.
\newblock \href{http://arxiv.org/abs/2304.14073}{{\ttfamily arXiv:2304.14073
  [hep-th]}}.

\bibitem{DimitrijevicCiric:2023hua}
M.~Dimitrijevi\'c~\'Ciri\'c, N.~Konjik, V.~Radovanovi\'c, and R.~J. Szabo,
  ``{Braided quantum electrodynamics},''
  \href{http://dx.doi.org/10.1007/JHEP08(2023)211}{{\em JHEP} {\bfseries 08}
  (2023) 211}, \href{http://arxiv.org/abs/2302.10713}{{\ttfamily
  arXiv:2302.10713 [hep-th]}}.

\bibitem{Giotopoulos:2021ieg}
G.~Giotopoulos and R.~J. Szabo, ``{Braided symmetries in noncommutative field
  theory},'' \href{http://dx.doi.org/10.1088/1751-8121/ac5dad}{{\em J. Phys. A}
  {\bfseries 55} no.~35, (2022) 353001},
  \href{http://arxiv.org/abs/2112.00541}{{\ttfamily arXiv:2112.00541
  [hep-th]}}.

\bibitem{DimitrijevicCiric:2021jea}
M.~Dimitrijevi\'c~\'Ciri\'c, G.~Giotopoulos, V.~Radovanovi\'c, and R.~J. Szabo,
  ``{Braided $L_{\infty}$-algebras, braided field theory and noncommutative
  gravity},'' \href{http://dx.doi.org/10.1007/s11005-021-01487-x}{{\em Lett.
  Math. Phys.} {\bfseries 111} no.~6, (2021) 148},
  \href{http://arxiv.org/abs/2103.08939}{{\ttfamily arXiv:2103.08939
  [hep-th]}}.

\bibitem{Fiore:2007vg}
G.~Fiore and J.~Wess, ``{On full twisted Poincare' symmetry and QFT on
  Moyal-Weyl spaces},''
  \href{http://dx.doi.org/10.1103/PhysRevD.75.105022}{{\em Phys. Rev. D}
  {\bfseries 75} (2007) 105022},
  \href{http://arxiv.org/abs/hep-th/0701078}{{\ttfamily arXiv:hep-th/0701078}}.

\bibitem{Fiore:2007zz}
G.~Fiore, ``{Can QFT on Moyal-Weyl spaces look as on commutative ones?},''
  \href{http://dx.doi.org/10.1143/PTPS.171.54}{{\em Prog. Theor. Phys. Suppl.}
  {\bfseries 171} (2007) 54--60},
  \href{http://arxiv.org/abs/0705.1120}{{\ttfamily arXiv:0705.1120 [hep-th]}}.

\bibitem{Fiore:2008ta}
G.~Fiore, \href{http://dx.doi.org/10.1142/9789812833556_0005}{``{On the
  consequences of twisted Poincare' symmetry upon QFT on Moyal noncommutative
  spaces},''} in {\em {Quantum field theory and beyond: Essays in honor of
  Wolfhart Zimmermann's 80th birthday (Ringberg Symposium)}}, pp.~64--84.
\newblock 9, 2008.
\newblock \href{http://arxiv.org/abs/0809.4507}{{\ttfamily arXiv:0809.4507
  [hep-th]}}.

\bibitem{bonfiglioli2011topics}
A.~Bonfiglioli and R.~Fulci,
  \href{http://dx.doi.org/10.1007/978-3-642-22597-0}{{\em Topics in
  noncommutative algebra: the theorem of Campbell, Baker, Hausdorff and
  Dynkin}}, vol.~2034.
\newblock Springer, 2011.

\end{thebibliography}
\end{document}